\documentclass[twocolumn]{aastex631}

\usepackage{xcolor}
\usepackage{graphicx}
\usepackage{hyperref}
\usepackage{amsmath}
\usepackage{txfonts}

\begin{document}

\title{JWST NIRCam simulations and observations of AGN ionization cones in cosmic noon galaxies}

\author[0009-0002-0710-194X]{S. Lebowitz}
\affiliation{Steward Observatory, University of Arizona, 933 N. Cherry Ave., Tucson, AZ 85721, USA}

\author[0000-0003-4565-8239]{K. Hainline}
\affiliation{Steward Observatory, University of Arizona, 933 N. Cherry Ave., Tucson, AZ 85721, USA}

\author[0000-0002-0000-2394]{S. Juneau}
\affiliation{NSF NOIRLab, 950 N. Cherry Ave., Tucson, AZ, 85719, USA}

\author[0000-0002-6221-1829]{J. Lyu}
\affiliation{Steward Observatory, University of Arizona, 933 N. Cherry Ave., Tucson, AZ 85721, USA}

\author[0000-0003-2919-7495]{C. C. Williams}
\affiliation{NSF NOIRLab, 950 N. Cherry Ave., Tucson, AZ, 85719, USA}
\affiliation{Steward Observatory, University of Arizona, 933 N. Cherry Ave., Tucson, AZ 85721, USA}

\author[0000-0002-8909-8782]{S. Alberts}
\affiliation{Steward Observatory, University of Arizona, 933 N. Cherry Ave., Tucson, AZ 85721, USA}

\author[0000-0003-3310-0131]{X. Fan}
\affiliation{Steward Observatory, University of Arizona, 933 N. Cherry Ave., Tucson, AZ 85721, USA}

\author[0000-0002-7893-6170]{M. Rieke}
\affiliation{Steward Observatory, University of Arizona, 933 N. Cherry Ave., Tucson, AZ 85721, USA}

\begin{abstract}
The extended narrow line region (NLR) of Active Galactic Nuclei (AGN) provides a valuable laboratory for exploring the relationship between AGN and their host galaxies, often appearing as an “ionization cone” that can extend out to the very edge of the galaxy. We use new James Webb Space Telescope (JWST) NIRCam imaging to study the morphologies and sizes of NLRs traced by [\ion{O}{3}] at cosmic noon ($z\sim2-3$). To determine NIRCam's ability to probe the properties of the NLR in cosmic noon galaxies, we present simulated images of AGN at $z=2-3$ created with archival data cubes from the Multi Unit Spectroscopic Explorer (MUSE) of nine nearby ($z<0.05$) AGN host galaxies with previously confirmed extended NLRs. Our simulated images show that NIRCam is able to resolve the morphologies of NLRs at $z=2-3$ with narrow- and medium-band observations. We then search for extended NLRs with NIRCam medium-band observations targeting the [\ion{O}{3}]$+\mathrm{H}\beta$ emission in twenty-seven previously identified AGN at $z=2.4-3.4$ in the Great Observatories Origins Deep Survey South (GOODS-S) field. We detect six galaxies in our observed sample with [\ion{O}{3}]$+\mathrm{H}\beta$ morphologies consistent with AGN ionization cones with characteristic NLR sizes of $1-2.5$ kpc. Thanks to our simulated data, we can predict the effects of cosmological surface brightness dimming and instrument noise on the NLR size measurements at high redshift, which has the biasing effect of lowering the NLR size-AGN luminosity trend that is observed at low redshift by a factor of $\sim 2$. 
\end{abstract}

\section{Introduction} \label{sec:intro}
A major challenge in extragalactic astronomy is understanding how star formation is regulated and suppressed over the evolutionary history of a galaxy. One proposed mechanism for modulating star formation involves the interactions between an actively accreting supermassive black hole, known as an Active Galactic Nucleus (AGN), and its host galaxy through the internal regulation of the galaxy's gaseous environment. Cosmological simulations show that AGN ``feedback'' is essential for suppressing star formation in massive galaxies and reproducing the observed galaxy color bi-modality and deficiency of massive galaxies at $z=0$ \citep{Sijacki2007, Croton2006, Scharre2024, Puchwein2013}. AGN feedback can be induced by radiation pressure generated from the accretion of matter onto the black hole (“quasar mode”) and/or by charged particles accelerated from the AGN’s magnetic fields in radio-emitting jets (“radio mode”) \citep{Fabian2012}.

Quasar, or radiative-mode, feedback is associated with high luminosity AGN powered by accretion disks radiating up to $10^{47}$ erg s$^{-1}$, typically fueled by star-forming environments rich in cold, dense gas \citep{Laha2021, Heckman&Best2014, Morganti2017, Rupke2005, Martin2005}. These AGN are able to influence the evolution of the host galaxy by ionizing, heating, or expelling gas through multi-scale, multi-phase outflows, which can lead to quenched star formation \citep{Fabian2012, Morganti2017, Heckman&Best2014, DiMatteo2005}. In particular, radiatively-driven, ionized outflows have been observed across spatial scales ranging from $1$ parsec to tens of kiloparsecs \citep{Revalski2021, Cresci2023}, often residing in a highly ionized, low density region generated by ionizing UV photons radiated from the black hole's accretion disk \citep{Antonucci1993, Joh2021}. This region is commonly known as the ``narrow line region'' (NLR), named after the spectroscopically observed narrow emission lines, kinematically broadened with typical widths of $10^{2-3}$ km s$^{-1}$ \citep{Bennert2002}. The anisotropic escape and collimation of this ionizing radiation generates a NLR ($0.1-10$ kpc) that protrudes bilaterally from the AGN, frequently described as an AGN ``ionization bicone'' \citep{Antonucci1993, Bennert2002, Netzer2015}. 

The NLR is a useful tool for studying the co-evolution of AGN and their host galaxies as it can be spatially resolved observationally \citep{Joh2021}. In particular, the warm, ionized phase ($T \sim 10^{4}K$) of the NLR is well-traced by the rest-frame optical [\ion{O}{3}]$\lambda5007$ line, a forbidden transition occurring only in extremely hot, low-density environments \citep{Heckman&Best2014, Morganti2017, Peimbert2017}. Studies targeting the [\ion{O}{3}] emission in nearby AGN have revealed extended NLRs with sizes as large as 20 kpc, signifying that AGN are able to ionize gas out to the edge of their host galaxy \citep{Sun2017, Hainline2013, Haineline2014}. NLR sizes have also been found to strongly correlate with AGN luminosity at low redshifts ($z<0.6$), indicating that more luminous AGN are able to exert a greater influence in ionizing the available gas within their host galaxies \citep{Liu2014, Haineline2014, Dempsey2018}. While the properties of the NLR of AGN host galaxies have been well-studied at low redshift, it remains uncertain how the size of the NLR scales as a function of luminosity at higher redshift, particularly at cosmic noon ($z=2-4$), when galaxy growth and AGN accretion activity are at its peak \citep{Madau2014}.

Cosmic noon marks a pivotal era in galaxy evolution that traces the formation of today's most massive galaxies \citep{Madau2014}. During this epoch, black hole growth and star formation rates are at their highest, with galaxies assembling half of their current stellar mass \citep{Schreiber2020}. Additionally, the quasar luminosity function reaches its maximum at $z=2-3$ \citep{Hopkins2007, Pei1995}, suggesting a greater number density of luminous AGN (L$_{\mathrm{bol}}>10^{45}$ erg s$^{-1}$) capable of producing extended NLRs. Models also indicate that AGN-driven outflows may be more prevalent and more effective at exerting feedback at this epoch \citep{Heckman&Best2014, Fischer2010, Veilleux2001}. Several low-redshift studies have linked AGN feedback to ionized outflows in the NLR by using integral field unit (IFU) spectroscopy to probe the kinematics of the [\ion{O}{3}] gas \citep{Lopez-coba, Durre&Mould2018, Gonzalez-Martin, Ventur&Marconi2021}. 
While IFU spectroscopy is well-suited for kinematic studies of individual objects, wide-field imaging surveys targeting [\ion{O}{3}]$+\mathrm{H}\beta$ offer the opportunity to characterize the morphologies and sizes of NLRs for large samples of AGN host galaxies, and identify extended NLRs which can then be followed up spectroscopically to search for ionized outflows.

However, resolving the rest-frame optical [\ion{O}{3}] emission at higher redshifts has been a long-standing observational challenge. At $z=2-4$, [\ion{O}{3}] is shifted into the near-infrared (NIR), making ground-based observations difficult due to atmospheric absorption of NIR photons. The James Webb Space Telescope (JWST), with its unparalleled resolution and sensitivity in the infrared, enables a groundbreaking exploration of AGN ionization cones in cosmic noon galaxies using its primary imager, NIRCam \citep{Rieke2023a, Rigby2023}. In this paper, we explore whether NIRCam is able to resolve the morphologies of AGN ionization cones at cosmic noon by simulating narrow- and medium-band NIRCam images for AGN at $z=2-3$, using archival data cubes from the Multi Unit Spectroscopic Explorer (MUSE) of nine nearby ($z<0.05$) AGN host galaxies with previously confirmed extended NLRs. Upon demonstrating that such observations are feasible, we then search for extended NLRs at cosmic noon through NIRCam medium-band observations targeting the [\ion{O}{3}]$+\mathrm{H}\beta$ emission in twenty-seven previously identified AGN at $z=2.4-3.4$ in the Great Observatories Origins Deep Survey south \citep[GOODS-S, ][]{Giavalisco2004} field. We analyze the [\ion{O}{3}]$+\mathrm{H}\beta$ morphologies of this sample and investigate how their NLR sizes vary with AGN luminosity. Thanks to our simulated images, we can predict the impact of observational biases—such as instrumental noise, cosmological surface brightness dimming, and wide-band imaging—on the measurement of NLR sizes at cosmic noon. 

We described our simulated and observed samples in Section \ref{sec:data}. In Section \ref{sec:methods}, we describe our methods for creating the simulated images and constructing [\ion{O}{3}]$+\mathrm{H}\beta$ maps from our observed images. In Sections \ref{sec:simulation_results} and \ref{sec:observation_results}, we respectively present the results from our simulated images and observations. Our findings are discussed in Section \ref{sec:discussion} and our conclusions are detailed in Section \ref{sec:conclusion}. Throughout the paper, we adopt the standard $\Lambda CDM$ cosmology with $H_{o}=70$ km s$^{-1}$ Mpc$^{-1}$, $\Omega_{\mathrm{m}}=0.3$, and $\Omega_{\Lambda}=0.7$.

\section{Data} \label{sec:data}
To predict what [\ion{O}{3}] ionization cones would look
like at cosmic noon observed by NIRCam, we selected nine low-z galaxies ($z<0.05$) with previously confirmed extended narrow line regions and used their archival MUSE datacubes to construct simulated NIRCam images at $z=2-3$. We then constructed an observed sample of twenty-seven previously confirmed AGN at $z=2.4-3.4$, chosen to leverage publicly available deep photometry in the GOODS-S field. We describe our archival sample selection in Section \ref{subsection:muse}. The selection of our NIRCam observations is described in Section \ref{subsection:nircam}.

\subsection{MUSE}\label{subsection:muse}
The Multi Unit Spectroscopic Explorer (MUSE) is an integral field spectrograph (IFU) installed on the Very Large Telescope at the European Southern Observatory (ESO) \citep{Bacon2010}. The MUSE instrument's wide field of view ($1\arcmin\times1\arcmin$) and pixel resolution of $0.2\arcsec$ enables high spatial resolution at $0.464-0.93$ $\mu$m for exploring the [\ion{O}{3}] morphologies of nearby galaxies. In addition, IFU spectroscopy provides spectral data for each spectral pixel of the image. These capabilities allow for spatial mapping of [\ion{O}{3}] ionization cones in nearby galaxies ($z<0.6$). 

All MUSE data cubes used in the simulation were obtained from the ESO Archive Science Portal \citep{Romaniello2022, Alberto2019, Romaniello2023} and processed using the standard MUSE data reduction pipeline available in the ESO archive \citep{Weilbacher2020}. Our MUSE sample was selected from the AMUSING++ Nearby Galaxy Compilation presented in \cite{Lopez-coba}, with the exception of Mrk 573 which was selected from \cite{Gonzalez-Martin} due to its prominent bicone. The Amusing++ Nearby Galaxy Compilation is a collection of 635 galaxies with data from MUSE IFU from $0.0002<z<0.1$. From this compilation, we selected nearby ($z<0.05$) galaxies publicly available in the ESO archive with visually observed significant [\ion{O}{3}] features, with emphasis given to galaxies that showed conic [\ion{O}{3}] morphologies. These objects do not represent a complete sample of low-z galaxies with observed AGN ionization cones. The MUSE datacubes used in the simulation were observed in Wide Field Mode. A summary of our simulated sample can be found in Table \ref{table:MUSE} and specific details of each object can be found in Appendix \ref{section:Appendix_A}. 

\begin{deluxetable*}{l l c c l c c c}
\tablecolumns{8}
\tablecaption{MUSE AGN Sample \label{table:MUSE}}
\tablehead{Galaxy & Program ID & $z$ & $f$ & Log[$\mathrm{L}_{\mathrm{bol}}$] & NLR radius & NLR radius & NLR radius \\ 
\colhead {} & \colhead {} & \colhead {} & \colhead {} & [erg s$^{-1}$] & w/o noise (kpc) & w/ noise (kpc) & w/o noise (kpc)\\
\colhead {} & \colhead {} & \colhead {} & \colhead {} & \colhead {} & Medium-band & Medium-band & Narrow-band}
\startdata
NGC 2992 & 094.B-0321(A) & 0.007 & 0.74 & 43.36 & 3.8 & 1.2 & 2.0\\
NGC 7582 & 095.B-0934(A) & 0.005 & 0.80 & 43.16 & 3.4 & 0.9 & 0.9\\ 
IC 5063 & 60.A-9339(A) & 0.011 & 0.66 & 44.24 & 4.1 & 1.4 & 3.5\\   
ESO 428-14 & 097.D-0408(A) & 0.006 & 0.72 & 42.30 & 2.2 & 0.9 & 1.3\\
NGC 5728 & 097.B-0640(A) & 0.009 & 0.81 & 44.16 & 3.3 & 1.0 & 1.6\\
Mrk 573 & 106.21C7.001 & 0.017 & 0.74 & 44.71 & 3.6 & 1.9 & 4.2\\
IC 1657 & 097.D-0408(A) & 0.013 & 0.70 & 43.54 & 5.6 & 2.8 & 3.8\\
IC 1481 & 106.21C7.002 & 0.020 & 0.71 & $>$43.01 & 4.5 & 1.3 & 1.8\\ 
Mrk 926 & 095.B-0015(A) & 0.047 & 1.00 & 45.62 & 4.6 & 2.4 & 3.6
\enddata
\tablecomments{See Appendix \ref{section:Appendix_A} for notes on individual galaxies. Object redshifts are spectroscopic. $f$ is the scaling factor applied to the continuum subtraction to account for the slope of the continuum.}
\end{deluxetable*}

\subsection{NIRCam} \label{subsection:nircam}
The Near Infrared Camera (NIRCam) is the primary imager on board the James Webb Space Telescope (JWST) \citep{Rieke2005, Rieke2003, Beichman2012, Rieke2023a}. NIRCam has two modules (A and B) which can simultaneously view two adjacent ($2.2\arcmin\times2.2\arcmin$) fields of view across a short wavelength ($0.6-2.3$ $\mu$m, with a pixel resolution of $0.031\arcsec$) channel and a long wavelength ($2.4-5.0$ $\mu$m, with a pixel resolution of $0.063\arcsec$) channel. 

We selected our objects from the GOODS-S pre-JWST AGN catalog presented in \cite{Lyu2022}. This catalog consists of $\sim 900$ AGN in the 3D-HST GOODS-S footprint. AGN are classified according to nine selection techniques. These include mid-IR SED, mid-IR color type, X-ray luminosity, X-ray-radio luminosity relation, radio loudness, radio slope index, optical spectrum, optical SED, and variability \citep{Lyu2022}. We refer the reader to \cite{Lyu2022} for more details on the AGN selection. To leverage the available 1.82 $\mu$m (F182M) and 2.10 $\mu$m (F210M) medium-band photometry in the First Reionization Epoch Spectroscopically Complete Observations \citep[FRESCO, ][]{Oesch2023} and the JWST Extragalactic Medium Survey \citep[JEMS, ][]{Williams2023}, we selected AGN in the redshift ranges of $z=2.44-3.39$. For objects at $2.44 < z < 2.93$, [\ion{O}{3}] can be targeted by F182M, and for objects at $2.98 < z < 3.39$, [\ion{O}{3}] can be targeted by F210M. The average $5\sigma$ depth of these images are 0.035 MJy sr$^{-1}$ (in F182M) and 0.044 MJy sr$^{-1}$ (in F210M), determined by measuring the $5\sigma$ standard deviation in the $15x15$ corner pixels. The NIRCam sample and the filters used to target [\ion{O}{3}] or continuum emission are summarized in Table \ref{table:NIRCam}. 

Our NIRCam objects were observed with medium-band NIRCam data as part of FRESCO \citep{Oesch2023} and JEMS \citep{Williams2023} in the GOODS-S field \citep{Giavalisco2004}. [\ion{O}{3}]$+\mathrm{H}\beta$ maps of our sample were constructed using medium-band F182M and F210M photometry from JADES, JEMS, and FRESCO. We also use wide-band photometry in F444W (JADES+FRESCO) and F150W (JADES)) to construct RGB images for our sample. Photometric redshifts and EAZY fit models \citep{Brammer2008} were obtained from the JWST Advanced Deep Extragalactic Survey (JADES) \citep{Hainline2023, Rieke2023b, Eisenstein2023a, Eisenstein2023b}. Photometric redshifts were derived from fits to Kron aperture photometry for each source, and represent the redshift at the minimum chi-square from the EAZY template fits. Spectroscopic redshifts are provided when available from the literature. After redshift cuts, we compose a sample of twenty-seven AGN. The JADES IDs and object redshifts are reported in Table \ref{table:NIRCam}. Specific details of each object can be found in Appendix \ref{section:Appendix_C}. 

\begin{deluxetable*}{l l c c c c c l}
\centering
\tablecolumns{8}
\tablewidth{0pt}
\tablecaption{NIRCam AGN Sample  \label{table:NIRCam}}
\tablehead{JADES ID & $z$ & $f$ & R.A. & Dec. & [\ion{O}{3}] & Continuum & [\ion{O}{3}]$+\mathrm{H}\beta$ \\
\colhead {} & \colhead {} & \colhead {} & (deg) & (deg) & Filter & Filter &  Morphology}
\startdata
    111187 & 2.76 & 0.80 & 53.092843 & -27.80131 & F182M & F210M & Extended \\
    144830 & 2.90 & 1.01 & 53.108582 & -27.757321 & F182M & F210M & Cone \\
    147220 & 2.73* & 0.85 & 53.108139 & -27.754065 & F182M & F210M & Compact \\
    182930 & 3.03* & 1.11 & 53.161524 & -27.85607 & F210M & F182M & Cone \\
    187160 & 2.92* & 0.65 & 53.063882 & -27.843767 & F182M & F210M & Extended \\
    188085 & 2.90 & 1.02 & 53.137716 & -27.845074 & F182M & F210M & Companion \\
    188852 & 2.73 & 0.89 & 53.173974 & -27.843388 & F182M & F210M & Cone \\ 
    194373 & 2.51 & 0.86 & 53.140062 & -27.826514 & F182M & F210M & Compact \\
    196290 & 2.58* & 0.78 & 53.148855 & -27.821178 & F182M & F210M & Extended \\
    197581 & 2.64 & 0.60 & 53.141682 & -27.816629 & F182M & F210M & Extended\\
    199505 & 2.58* & 0.75 & 53.185844 & -27.810029 & F182M & F210M & Compact \\
    199773 & 2.66 & 0.81 & 53.163241 & -27.809057 & F182M & F210M & Companion \\
    201715 & 3.11 & 1.31 & 53.104026 & -27.802457 & F210M & F182M & Extended \\ 
    201902 & 2.71 & 0.70 & 53.093815 & -27.801388 & F182M & F210M & Companion \\
    202086 & 2.61* & 0.70 & 53.110846 & -27.800666 & F182M & F210M & Extended \\
    202380 & 2.65 & 0.93 & 53.129801 & -27.799204 & F182M & F210M & Extended \\
    202630 & 2.63 & 0.87 & 53.145598 & -27.798975 & F182M & F210M & Compact \\
    207632 & 2.57 & 0.77 & 53.134037 & -27.781039 & F182M & F210M & Cone \\
    208000 & 2.50* & 0.75 & 53.146160 & -27.779939 & F182M & F210M & Extended \\
    209117 & 2.73 & 0.88 & 53.160609 & -27.776249 & F182M & F210M & Cone \\
    214145 & 2.76 & 1.01 & 53.181181 & -27.760363 & F182M & F210M & No [\ion{O}{3}] \\
    216241 & 2.91 & 0.65 & 53.121861 & -27.752775 & F182M & F210M & Extended \\ 
    217830 & 2.61 & 0.80 & 53.162737 & -27.744344 & F182M & F210M & Cone \\  
    219594 & 3.22 & 1.18 & 53.174470 & -27.733386 & F210M & F182M & No [\ion{O}{3}] \\
    226033 & 3.14 & 1.06 & 53.153855 & -27.840255 & F210M & F182M & Extended \\
    243761 & 3.08 & 1.07 & 53.178092 & -27.861805 & F210M & F182M & Extended \\
    245063 & 2.93* & 0.95 & 53.082562 & -27.755328 & F182M & F210M & No [\ion{O}{3}] 
\enddata
\tablecomments{See Appendix \ref{section:Appendix_C} for notes on individual galaxies. Spectroscopic redshifts are denoted by $z$*. All other redshifts are photometric. $f$ is the scaling factor applied to the continuum subtraction to account for the slope of the continuum.}
\end{deluxetable*}

\section{Methods} \label{sec:methods}
We describe our methods for creating simulated NIRCam images based on MUSE data in Section \ref{subsection:simulation_methods}. In Section \ref{subsection:observation_methods}, we discuss our methods for constructing [\ion{O}{3}]$+\mathrm{H}\beta$ maps for our observed NIRCam sample by performing continuum subtractions.

\subsection{Simulation} \label{subsection:simulation_methods}
We construct simulated images to explore the morphologies of [\ion{O}{3}] ionization cones in cosmic noon galaxies as observed by NIRCam.
Our simulation uses MUSE datacubes to produce artificial narrow- and medium-band NIRCam images at $z=2-3$. The simulation is pre-programmed with filter-set data for NIRCam, the MUSE pixel resolution ($0.2\mathrm{\arcsec}$), and the NIRCam pixel resolution for the short ($0.031\mathrm{\arcsec}$) and long ($0.063\mathrm{\arcsec}$) wavelength channels. The simulation takes in the datacube file for the MUSE object, its redshift, the "test" redshift to be simulated, and the NIRCam filter choice. In this paper, we show examples of simulated images constructed with a variety of NIRCam filters (F162M, F162M, F182M, and F210M, F323N, and F360M) to demonstrate the applicability of various filters for targeting [\ion{O}{3}] at different redshifts and with narrow- versus medium-band imaging. The process for producing simulated NIRCam images can be divided into four steps:

(1) Redshift datacube. The simulation first redshifts the MUSE datacube to the "test" redshift. This is done using the astropy package with FlatLambdaCDM cosmology to redshift the wavelength and flux for each spaxel \citep{astropy:2013, astropy:2018, astropy:2022}. 

(2) Filter integration. The simulation integrates each spaxel of the redshifted datacube with the selected NIRCam filter. The filter integration is performed with the NIRCam $+$ OTE Module A filter transmission curves as Module A has a higher efficiency than Module B due to its double-sided anti-reflection coating. In order for the integration to be performed, the spectral wavelength range at the simulated wavelength must extend to the left and right edges of the filter curve. The simulation checks that the filter curve covers the full wavelength range of the redshifted spectrum. If this check is false, the spectrum is extended to the filter edge using a 3$\sigma$ clipped linear fit to approximate the flux values for the extended spectrum. This extra step is only needed when integrating our sample with F162M and F182M as the redshifted spectrum does not fully extend to the left edge of these medium-band filters when simulated at $z=2.3$ and $z=2.7$, respectively. The simulation integrates the redshifted spectrum (F$_{\mathrm{\lambda}}$) with the NIRCam filter throughput using a rectangular integration. This step returns the average flux through the NIRCam filter curve for each MUSE pixel at the spatial resolution of MUSE. 

(3) Convolution with Webb PSF. To account for the NIRCam point spread function (PSF) at the simulated wavelength, the simulation convolves the simulated image at the MUSE resolution with a theoretical PSF generated using WebbPSF, a tool created for JWST instruments that simulates JWST PSFs \citep{Perrin2014}. The PSFs for JWST are computed based on a library of optical path difference (OPD) maps consistent with JWST's optical error budget. This convolution is justified because, at $z\sim2-3$, MUSE offers significantly higher resolution compared to NIRCam. For instance, at $z=2.3$, MUSE's pixel resolution of $0.2\arcsec$ corresponds to an effective spaxel size of $0.02$ kpc, while NIRCam's pixel resolution of $0.031\arcsec$ corresponds to an effective pixel size of $0.25$ kpc at the same redshift.

(4) Resolution adjustment. The convolved image is then degraded to the resolution of NIRCam. This resolution adjustment is performed by calculating the number of MUSE pixels (n$_{\mathrm{MUSE}}$) that correspond to one NIRCam pixel (n$_{\mathrm{NIRCam}}$) length at the test redshift. This is the ratio of the of the proper distance per NIRCam pixel at the test redshift to the proper distance per MUSE pixel at the object redshift. For simplicity, this value is rounded down to avoid fractional pixel sizes. The flux of the new NIRCam pixel is equal to the sum of the MUSE pixel fluxes in a box with an area of n$_{\mathrm{MUSE}}$ $\times$ n$_{\mathrm{MUSE}}$. This step returns the final simulated NIRCam image at the test redshift for the selected filter. An oversampling factor can also be input to simulate the effects of using a primary dither pattern during observations to remove artifacts and improve the image quality (see Section \ref{subsection:oversample} for further discussion of the oversampling factor). This increases the number of pixels that the light is spread over, improving the resolution of the final image. The NIRCam pixel resolution is then the ratio of the proper distance per NIRCam pixel at the test redshift to the proper distance per MUSE pixel at the object redshift, divided by the oversampling factor. 

To isolate the [\ion{O}{3}] emission, we need to subtract the contribution from the stellar continuum. This can be done by constructing two simulated images using NIRCam filters that separately target the [\ion{O}{3}] line and the adjacent continuum. The continuum image can then be subtracted from the [\ion{O}{3}]$+\mathrm{continuum}$ image. This method generates a map of isolated [\ion{O}{3}] emission when a narrow band is used to target [\ion{O}{3}], and a map of combined [\ion{O}{3}]$+\mathrm{H}\beta$ emission when a medium band is used. Depending on the redshift of the object, medium-band filters may isolate [\ion{O}{3}] from $+\mathrm{H}\beta$ if the [\ion{O}{3}] line lies near the filter's edge. However, this situation is suboptimal due to the reduced throughput at the filter's boundaries. For the continuum subtractions performed in this study, we also include a scaling factor, $f$, to account for the slope of the continuum. We calculate $f$ by taking the ratio of the average continuum flux in F182M to the average continuum flux in F210M over the central $10\times10$ pixels. The $f$ factor is then multiplied by the pixel fluxes of the continuum image in the continuum subtraction (i.e. $\text{flux(F182M)}-f \times \text{flux(F210M)}$). The scaling factors for each object in our simulated sample are reported in Table \ref{table:MUSE}. 

In Sections \ref{subsection:MUSE_noise}-\ref{subsection:MUSE_NLR_size}, we convert our images to surface brightness units by multiplying the average flux per pixel by the width of the selected filter and dividing by the pixel area.  We then add noise to the final image to simulate the noise of the NIRCam instrument. To simulate instrument noise in our MUSE sample, we constructed simulated noise images with the same noise properties as our NIRCam observations to enable direct comparison between our simulated and observed samples. We simulate the instrumental noise of NIRCam only when constructing simulated medium-band images at $z=2.7$ using the F182M and F210M filters, so as to use the measured noise level from our F182M and F210M observed images. We measure the average noise level for our observations with detected NLRs (10 objects) by averaging the standard deviation of the $15 \times 15$ pixel corner regions across this sample (see Section \ref{subsection:NIRCam_morphologies} for details about the classification of our observed NLR sample and Table \ref{table:NLR_sizes} for a summary of these objects). The simulated noise is then added using a Gaussian distribution with a width equivalent to the measured mean noise level. From the observed NLR sample, the average measured noise levels of the F182M and F210M images are $\sigma_{NC}=3.7 \times 10^{-17}$ erg s$^{-1}$ cm$^{-2}$ arcsec$^{-2}$ and $\sigma_{NC}=3.0 \times 10^{-17}$ erg s$^{-1}$ cm$^{-2}$ arcsec$^{-2}$, respectively. These noise levels are measured in surface brightness units to enable calculations of NLR sizes above a limiting surface brightness level in Sections \ref{subsection:MUSE_NLR_size} and \ref{subsection:NLR-lum}. Throughout this paper, all simulated images constructed without added noise are shown in gray-scale and images with simulated NIRCam noise are shown in color. 

\subsection{Observations: Continuum Subtractions} \label{subsection:observation_methods}
\begin{figure*}
        \includegraphics[width=180mm,height=80mm]{"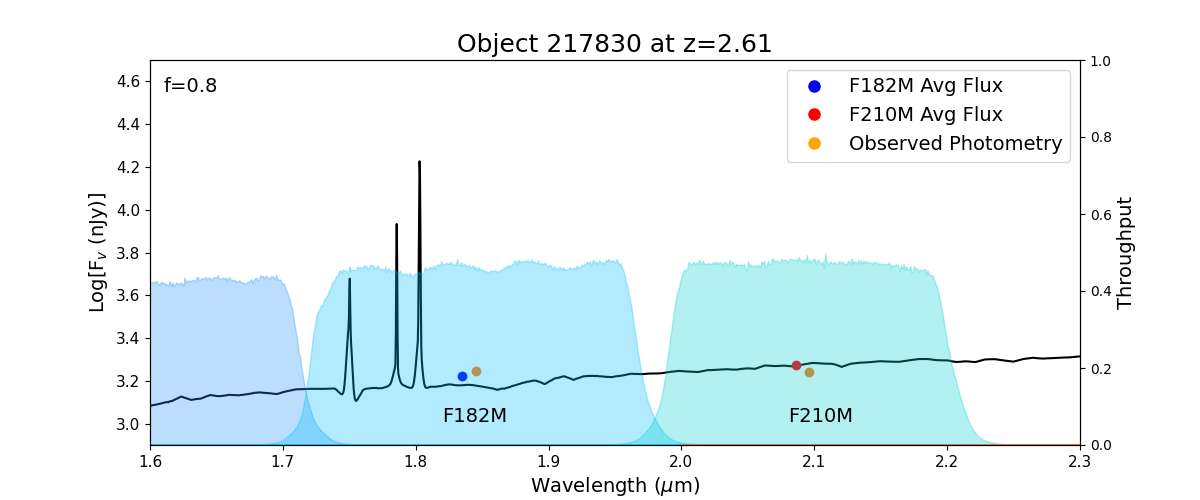"}
        \includegraphics[width=180mm,height=80mm]{"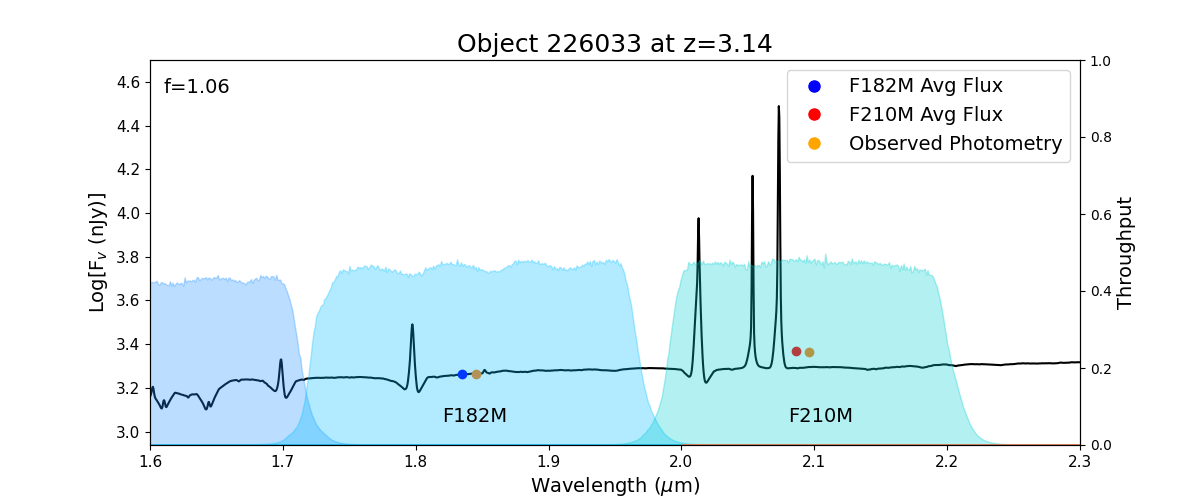"} 
        \caption{Spectral energy distribution (SED) profiles for an object (217830) in the redshift range, $2.44<z<2.93$ (top), and an object (226033) in the redshift range, $2.98<z<3.39$ (bottom). The observed photometric flux through F182M and F210M are represented by a yellow point. The calculated average flux through F182M is represented by a blue point and the calculated average flux through F210M is represented by a red point. The observed photometric fluxes and calculated average fluxes are shown at different wavelengths for clarity. The calculated $f$ factors are shown in the top left corner of the plots}\label{fig:SED}
\end{figure*}

To search for extended NLRs at cosmic noon, we construct [\ion{O}{3}]$+\mathrm{H}\beta$ maps for our NIRCam medium-band observations (F182M and F210M) of AGN at $z = 2.3 - 3.4$. The filters used to target [\ion{O}{3}] and the continuum emission for each object are reported in Table \ref{table:NIRCam}. For objects at $2.44<z<2.93$, F210M is subtracted from F182M. For objects at $2.98<z<3.39$, F182M is subtracted from F210M. To perform an accurate filter subtraction, we also include the same scaling factor, $f$, used for our simulated images to account for the continuum slope (see Figure \ref{fig:SED} for example). For the observed sample, these scaling factors are derived from the best-fitting EAZY template photometry for each source \citep{Hainline2023}, calculated as the ratio of the estimated average continuum flux through the filter targeting [\ion{O}{3}] to the average flux through the filter targeting continuum emission. The average continuum flux through the filter targeting [\ion{O}{3}] is estimated by performing a linear fit to the continuum in small windows left and right of the [\ion{O}{3}] doublet. The $f$ factor is then multiplied by the pixel fluxes of the continuum image in the continuum subtraction. The scaling factors for our observed sample are reported in Table \ref{table:NIRCam}.

We show examples of the the spectral energy distribution (SED) profiles for an object at $z<2.93$ (top) and at $z>2.98$ (bottom) in Figure \ref{fig:SED}. The calculated scaling factors ($f$) are reported in the top left corner of the plots. For six objects (JADES ID = 187160, 197581, 202086, 216241, 201902, 198545), the calculated $f$ factor resulted in an over-subtraction (i.e. negative pixels values in the central $10\times10$ pixels). The SED profiles for these objects have a very red continuum and/or H$\beta$ absorption which may result in the model not fully accounting for the [\ion{O}{3}] emission. For these objects, the $f$ factor is manually adjusted down until the continuum subtraction results in non-negative values in the central $10\times10$ pixels.

\section{Results: Simulated NIRCam Images} \label{sec:simulation_results}
Before continuing the analysis of the NIRCam observations, we turn back to the simulated images to predict how instrumental effects and cosmological dimming influence the observations of NLRs at cosmic noon. For illustration purposes, we explore whether NIRCam is able to resolve the morphologies of [\ion{O}{3}] cones at $z=2-3$ through simulated NIRCam narrow- and medium-band images based on MUSE datacubes of low-z AGN. In this section, we present the results derived from our simulated images. We use the nearby galaxy, NGC 2992, as a reference object in Sections \ref{subsection:oversample}-\ref{subsection:MUSE_NLR_size} due to the clarity and visibility of its [\ion{O}{3}] bicone. NGC 2992 ($z=0.007296$) is spiral galaxy, optically classified as a Seyfert 1.9 with an AGN bolometric luminosity of $10^{43.3}$ erg s$^{-1}$ \citep{Mingozzi, Kakkad2022}. NGC 2992 has an ionization bicone extending approximately $4$ kpc SE and NW of the central engine, detectable in [\ion{O}{3}] and H$\alpha$, and is also reported to host a double-lobed radio structure \citep{Friedrich, Zanchettin2023}.

In Section \ref{subsection:oversample}, we demonstrate how the convolution with WebbPSF and the inclusion of an oversampling factor influences the NIRCam simulated images of NGC 2992. In Section \ref{subsection:filter_choice}, we examine the impact of the filter width (narrow- versus medium-band) on the the [\ion{O}{3}] and [\ion{O}{3}]$+\mathrm{H}\beta$ maps. In Section \ref{subsection:MUSE_noise}, we explore how simulated instrument noise alters the detection of the ionization cone in NGC 2992. In Section \ref{subsection:MUSE_NLR_size}, we presents our calculations of the NLR size for the $z=2.7$ simulated images, with and without simulated instrument noise. In Section \ref{subsection:NLR-lum}, we explore the simulated $z=2.7$ NLR size-AGN luminosity relation for the complete MUSE sample. A montage of the $z=2.7$ simulated NIRCam medium-band images for our MUSE AGN sample, along with information for each object, can be found in Appendix \ref{section:Appendix_A}.

\subsection{Effects of Convolution and Oversample} \label{subsection:oversample}
\begin{figure*}
        \centering
        \includegraphics[width=180mm,trim={1cm 2cm 1cm 3cm},clip]{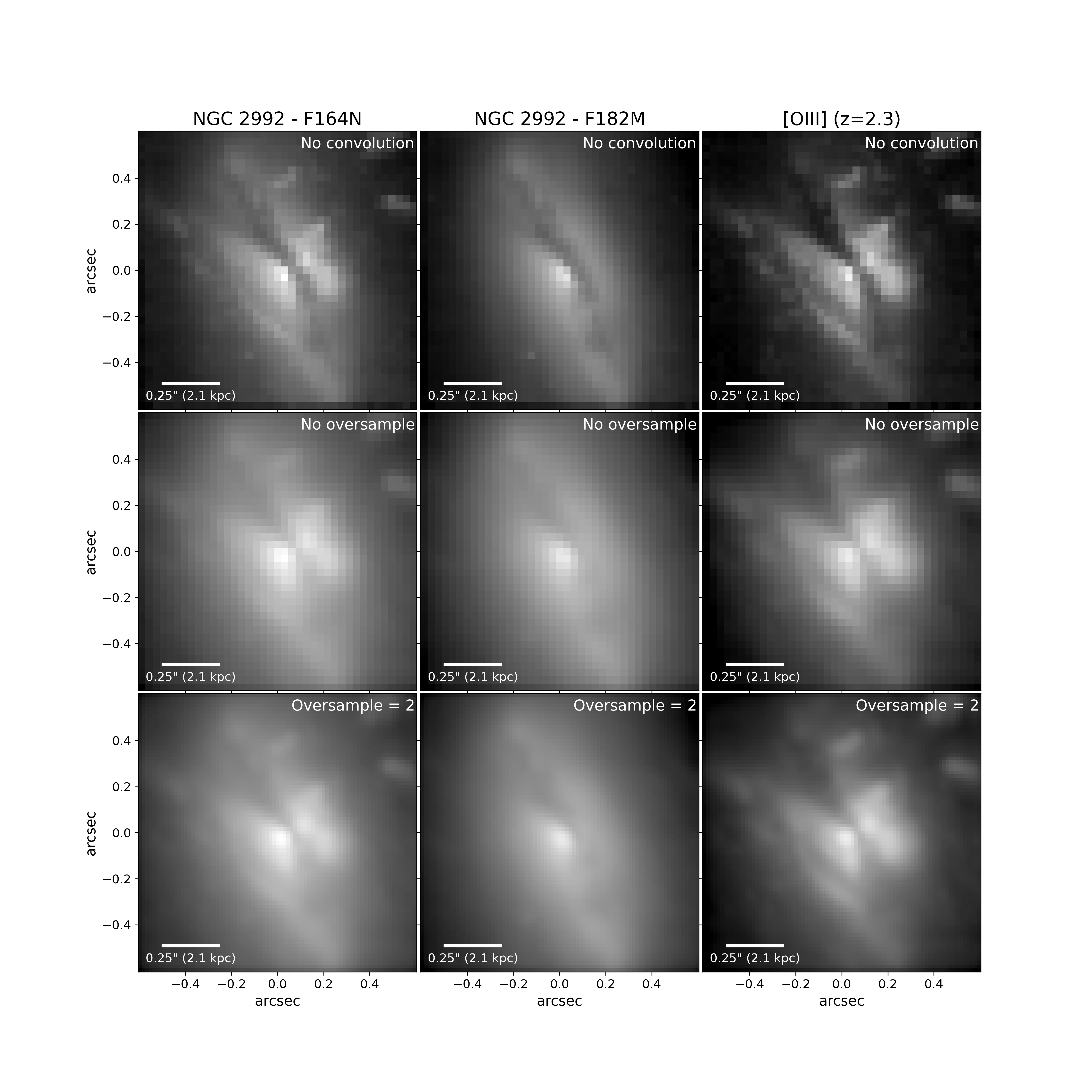}        
        \caption{Simulated NIRCam image of NGC 2992 at z=2.3 without simulated noise}. Left column: F164N ([\ion{O}{3}]). Middle column: F182M (continuum). Right column: Continuum subtracted [\ion{O}{3}] map ($F164N-F182M$). First row from the top: Image is not convolved with WebbPSF and no oversampling factor is applied. Second row: Convolved with WebbPSF, but no oversampling factor is applied. Third row: Convolved with WebbPSF and an oversampling factor of $2$ is applied.\label{fig:Oversample}
\end{figure*}

In Figure \ref{fig:Oversample}, we show the effects of convolving the simulated NIRCam image of NGC 2992 with WebbPSF and the inclusion of an oversampling factor on the detection of the [\ion{O}{3}] cones at $z=2.3$. In this section, we target [\ion{O}{3}] using a narrow-band filter to show simulated [\ion{O}{3}] maps constructed by isolating the [\ion{O}{3}] line, with minimal contamination from continua. At $z=2.3$, [\ion{O}{3}] can be targeted by F164N (first column). We use F182M to target the adjacent continuum emission (middle column), and perform a continuum subtraction ($F164N-F182M$) to construct a map of the isolated [\ion{O}{3}] emission (right column). 

In the first row, we show our simulated image of NGC 2992 at $z=2.3$ with no convolution or added oversampling factor. The [\ion{O}{3}] bicones are clearly visible in the F164N image (left). The [\ion{O}{3}] cones are not visible in the F182M image (middle), which captures the continuum emission. In the continuum-subtracted image (right), the [\ion{O}{3}] cones are clearly detected. The second row shows the simulated image of NGC 2992 at $z=2.3$ convolved with WebbPSF. Despite the lower spatial resolution, the bicone is still visible in the continuum-subtracted [\ion{O}{3}] map (right panel in second row), but appears less defined compared to the non-convolved version in the first row. This outcome is expected, as convolving the image with WebbPSF incorporates instrumental effects that cause the light to spread across its pixels, thereby reducing the image resolution. 

The image clarity can be improved by the addition of a PSF oversampling factor, which increases the image resolution and offsets the effects of the convolution. In the third row, we apply an oversampling factor of $2$ to simulate using a primary dither pattern during observations to remove artifacts and improve the image quality. As shown in the continuum-subtracted image (right), the inclusion of the oversampling factor results in a stronger detection of the bicone. We recommend using a fine dither pattern for observations of AGN ionization cones at $z\geq2.3$ to mitigate the impact of the instrumental PSF on resolution. We also explored extending our simulation to higher redshifts ($z=5.4$) to investigate whether NIRCam could detect AGN ionization cones at earlier epochs in cosmic history. However, to maintain the paper’s focus on cosmic noon, we present our $z=5.4$ simulated images of NGC 2992 in Appendix \ref{subsection:Appendix_B}. The remainder of the simulated images presented in this paper are convolved with the NIRCam PSF without an oversampling factor applied.

\subsection{Effects of Filter Choice} \label{subsection:filter_choice}
\begin{figure*}
        \centering
        \includegraphics[width=180mm,trim={1cm 1cm 1cm 2cm},clip]{"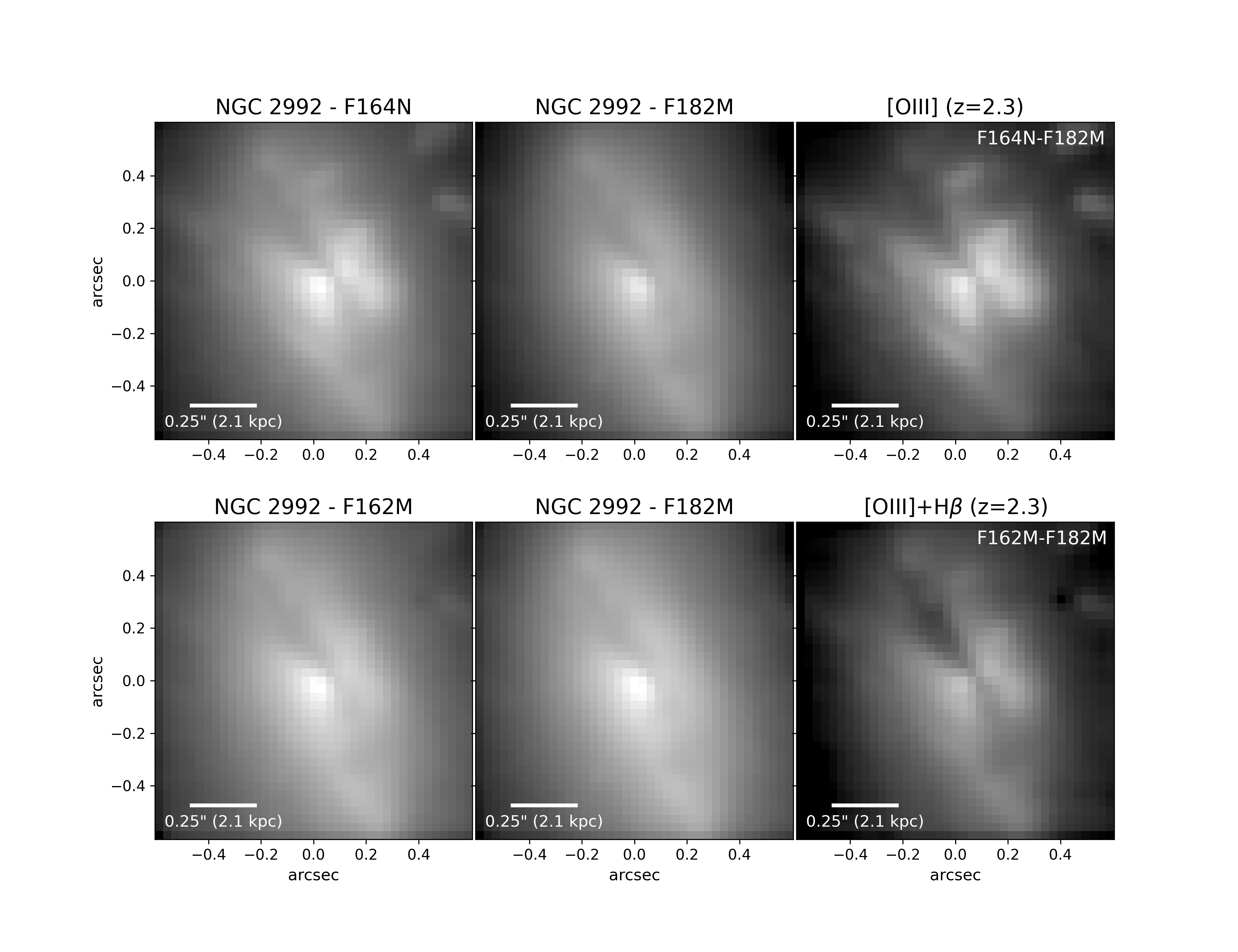"}
        \caption{Simulated NIRCam images of NGC 2992 without noise} at $z=2.3$ using narrow-band and medium-band filters to target [\ion{O}{3}]. First row: F164N (left), F182M (middle), continuum subtraction ($F164N-F182M$) (right). Second row: F162M (left), F182M (middle), continuum subtraction ($F162M-F182M$) (right). \label{fig:Filter Choice}
\end{figure*}

In Figure \ref{fig:Filter Choice}, we compare using narrow- (F164N) versus medium-band (F162M) filters to target the [\ion{O}{3}] emission at $z=2.3$. Ideally, narrow-band filters can focus directly on [\ion{O}{3}] emission with minimal contamination from continuum and other emission or absorption lines such as H$\beta$. However, data from large extragalactic sets are unlikely to use narrow-band imaging due to the smaller redshift ranges that [\ion{O}{3}] can be targeted, and longer exposure times needed to reach the same signal-to-noise as in a wider band. We explore how using medium-band imaging to target [\ion{O}{3}] alters the detection of the [\ion{O}{3}] bicone in the simulated $z=2.3$ image of NGC 2992.

The first row of Figure \ref{fig:Filter Choice} shows simulated images of NGC 2992 using a narrow-band filter (F164N) to target [\ion{O}{3}]. The second row shows simulated images using a medium-band filter (F162M) to target [\ion{O}{3}]. The [\ion{O}{3}] map constructed with the narrow-band filter better delineates the morphology of the [\ion{O}{3}] bicone compared to the [\ion{O}{3}]$+\mathrm{H}\beta$ map constructed with the medium-band filter. However, the [\ion{O}{3}] bicone is still distinctly resolved in the map constructed with medium-band filters. This figure demonstrates that medium-band imaging is adequate for resolving the [\ion{O}{3}] cones at $z=2.3$, but some of the structural detail, such as the dust lanes and shape of the cones, is reduced likely due to contamination from H$\beta$ and continuum emission in the medium-band. This result demonstrates the efficacy of using medium-band imaging to search for AGN ionization cones, which is significant given that extragalactic datasets have so far provided an abundance of medium-band imaging, but limited narrow-band imaging. In the interest of keeping our exploration relevant and applicable to available datasets, the remainder of our simulated images are constructed with medium-band filters, using the same filters (F182M and F210M) that comprise our observational sample. 

\subsection{Simulating Noise} \label{subsection:MUSE_noise}
At higher redshifts, the observed light from distant objects diminishes with the square of their distance from the observer. This effect is compounded by cosmological surface brightness dimming, proportional to $(1+z)^{4}$, which makes distant objects appear fainter due to the expansion of space and curvature \citep{Calvi2014}. As a consequence, high-redshift galaxies and their low-surface-brightness features can be challenging to detect due to instrumental noise. Instrument noise can reduce the overall quality of an image by suppressing fine details, adding grittiness, and creating hot and cold pixels. The dominant sources of noise for NIRCam include readout noise (thermal kTC noise, pixel read noise, and 1/f noise), and dark current noise \citep{Leisenring2016, Robberto2010}. In this section, we investigate how instrumental noise alters the detection of the ionization cone in the simulated image of NGC 2992 at $z=2.7$. 

\begin{figure*}                     
    \centering
    \includegraphics[width=180mm,trim={1cm 1cm 1cm 2cm},clip]{"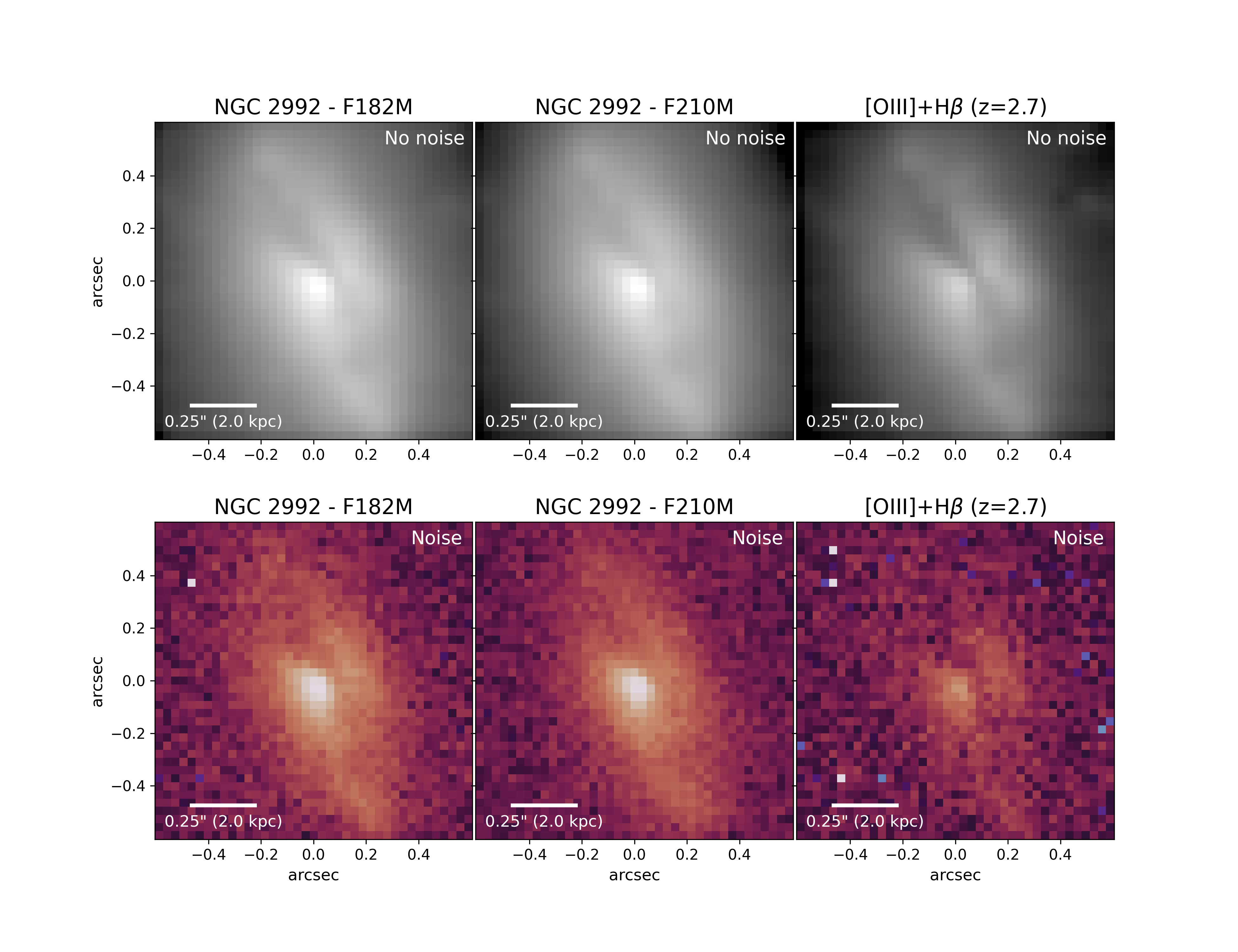"}        
    \caption{NIRCam simulated images of NGC 2992 without noise (top row) and with noise (bottom row) at $z=2.7$. The images are convolved with WebbPSF. Left column: F182M ([\ion{O}{3}]+continuum), Middle column: F210M (continuum), Right column: [\ion{O}{3}]$+\mathrm{H}\beta$ map constructed through the continuum subtraction, $F182M-F210M$.} \label{fig:noise}
\end{figure*}

We simulate NIRCam noise by constructing mock F182M and F210M images with the same noise properties as our NIRCam observations (see Section \ref{subsection:simulation_methods} for details on the construction of simulated noise images). In Figure \ref{fig:noise}, we show the results of simulating NIRCam instrument noise for NGC 2992 at $z=2.7$ with medium-band imaging. The top row shows simulated images of NGC 2992 without instrument noise, in surface brightness units, at $z=2.7$ for F182M (left), F210M (middle), and the continuum-subtracted [\ion{O}{3}]$+\mathrm{H}\beta$ map ($F182M-F210M$, right). The bottom row shows the same images, but with simulated NIRCam noise added. The [\ion{O}{3}]$+\mathrm{H}\beta$ map with noise reveals that some of the galaxy's surface brightness is reduced to levels below the instrument's noise threshold, causing the bicone to appear less extended. Although the bicone of NGC 2992 is still visible above the instrument noise, the reduction of the apparent size of the bicone indicates that instrument noise may have a significant effect on our ability to map the full extent of the NLR in cosmic noon galaxies, especially given the strong dependency of cosmological surface brightness dimming on redshift ($(1+z)^{-4}$). We discuss the measured NLR sizes for NGC 2992, with and without simulated instrument noise, in the following section.

\subsection{Calculating NLR Size} \label{subsection:MUSE_NLR_size}
Studies have shown that AGN at $z < 0.6$ can ionize gas out to the edge of their host galaxy, producing extended NLRs with sizes as large as $20$ kpc \cite{Sun2017, Haineline2014}. However, it is not known if AGN at higher redshifts can similarly affect their host's environment on kpc-scales or if we can detect the NLR at cosmic noon above the noise level of NIRCam. We compare the NLR sizes of NGC 2992 simulated at $z=2.7$ with and without added instrument noise to explore how noise affects the observed size of the NLR. 

Determining the size of the NLR is highly dependent on the spatial resolution of the observations \citep{Ventur&Marconi2021}. One common method of determining the sizes of the NLR is to measure the extent of the [\ion{O}{3}] emission down to a fixed surface brightness level \citep{Sun2017}. We use this method to estimate the radii of the NLRs for our $z=2.7$ simulated sample. Although many low-redshift NLR studies adopt the common definition of the NLR radii as the semi-major axis of the $10^{-15}/(1+z)^{4}$ erg s$^{-1}$ cm$^{-2}$ arcsec$^{-2}$ isophote of the [\ion{O}{3}]$\lambda$5007$\AA$ line \citep{Liu2014, Sun2017, Hainline2013, Haineline2014}, we find this very faint surface brightness level (SB$ \approx 5 \times 10^{-18}$ erg s$^{-1}$ cm$^{-2}$ arcsec$^{-2}$) is comparable to the uncertainty in the residual background of the simulated [\ion{O}{3}]$+\mathrm{H}\beta$ images resulting in the false detection of [\ion{O}{3}]$+H\beta$ emission above the surface brightness threshold across the entire image. Therefore, we choose alternative surface brightness limits for our noiseless and noise samples that allow us to make measurements of the NLR size. 

We find that surface brightness limits equivalent to $3\sigma$ of the mean noise levels allow for NLR size measurements for both sample types. The noise level for each simulated $z=2.7$ image (with and without noise) is calculated by averaging the standard deviation of $15 \times 15$ pixel corner regions in each [\ion{O}{3}]$+\mathrm{H}\beta$ map. These noise levels were averaged across all nine MUSE AGN (with and without noise) to determine the mean noise level for each sample type - noting that the noise level for the [\ion{O}{3}]$+\mathrm{H}\beta$ maps with simulated noise was inherited from the observed images. For the simulated noiseless images, this corresponds to a surface brightness limit of $2.4 \times 10^{-17}$ erg s$^{-1}$ cm$^{-2}$ arcsec$^{-2}$. For the simulated images with noise, this surface brightness limit is $1.4 \times 10^{-16}$ erg s$^{-1}$ cm$^{-2}$ arcsec$^{-2}$. Ideally, the same limiting surface brightness should be used across sample types to facilitate direct comparisons of NLR sizes between the [\ion{O}{3}]$+\mathrm{H}\beta$ maps with and without noise. However, when testing the same surface brightness limits, we found that the limit derived for the noiseless images (SB$=2.4 \times 10^{-17}$ erg s$^{-1}$ cm$^{-2}$) is too faint to measure the NLR sizes in the noisy images. Conversely, the limit derived for the noisy images (SB$=1.4 \times 10^{-16}$ erg s$^{-1}$ cm$^{-2}$) is too high to accurately measure the NLR sizes in the noiseless images.

\begin{figure*}
        \centering
        \includegraphics[width=200mm,trim={4.5cm 0 0 0},clip]{"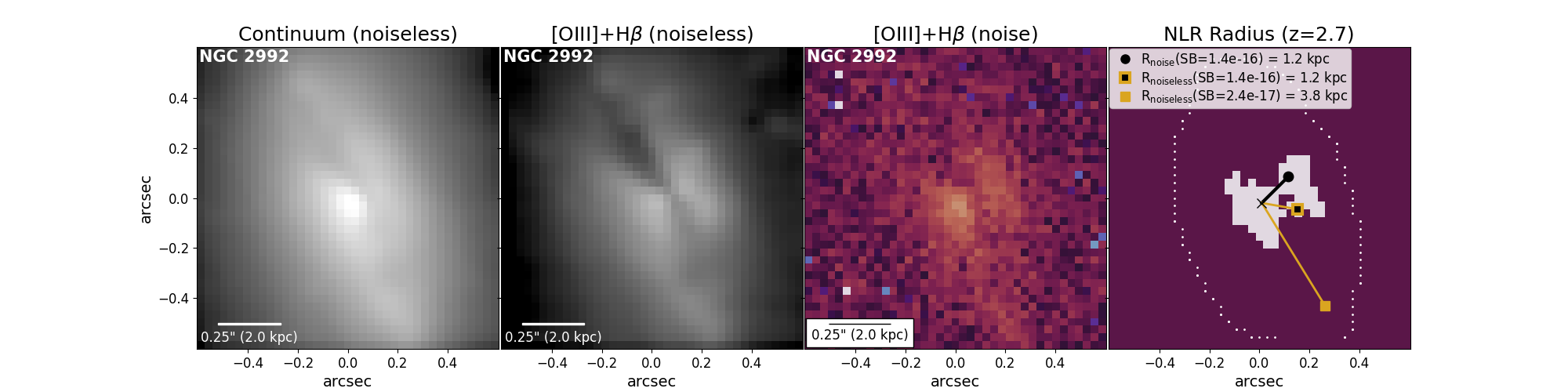"}
        \caption{NIRCam simulated images of our NGC 2992 at $z=2.7$. First column: continuum emission without noise (F210M). Second column: [\ion{O}{3}]$+\mathrm{H}\beta$ map without noise ($F182M - F210M$). Third column: [\ion{O}{3}]$+\mathrm{H}\beta$ map with simulated NIRCam noise (third column). Fourth column: Plot of the NLRs with simulated noise, measured above a fixed surface brightness level of $1.4 \times 10^{-16}$ erg s$^{-1}$ cm$^{-2}$ arcsec$^{-2}$ (gray area), and without simulated noise, measured above a fixed surface brightness level of $2.4\times 10^{-17}$ erg s$^{-1}$ cm$^{-2}$ arcsec$^{-2}$ (white outline). The average NLR radius measured from the [\ion{O}{3}]$+\mathrm{H}\beta$ map with noise is shown by the black circle. The average NLR radius measured from the [\ion{O}{3}]$+\mathrm{H}\beta$ map without noise is shown by the yellow square.}
        \label{fig:NGC2992_montage}
\end{figure*}

In Figure \ref{fig:NGC2992_montage}, we show simulated images of NGC 2992 at $z=2.7$. We show the continuum emission (F210M) and the [\ion{O}{3}]$+\mathrm{H}\beta$ map ($F182M - F210M$) without noise in the first and second columns respectively, and the [\ion{O}{3}]$+\mathrm{H}\beta$ map with simulated instrument noise in the third column. The fourth column shows our measurements of the [\ion{O}{3}]$+\mathrm{H}\beta$ emission above fixed surface brightness limits for the [\ion{O}{3}]$+\mathrm{H}\beta$ maps without noise (traced by the dashed white line) and the [\ion{O}{3}]$+\mathrm{H}\beta$ map with noise (traced by the gray region). We calculate the average NLR radius for these images (noise/noiseless) by measuring the average radial extent of the [\ion{O}{3}]$+\mathrm{H}\beta$ emission above the limiting surface brightness level chosen for each sample type. The average radial extent is computed by taking the median histogram value of the distances from the center to the edge pixels. For simplicity, we do not consider the cone opening angle in this measurement. The average NLR radius for the [\ion{O}{3}]$+\mathrm{H}\beta$ map with noise is shown by the black circle and the average NLR radius for the [\ion{O}{3}]$+\mathrm{H}\beta$ map without noise is shown by the yellow square. To show the dependency of the measured NLR size on the limiting surface brightness level, we also include a measurement of the NLR radius for the [\ion{O}{3}]$+\mathrm{H}\beta$ map without noise using the same surface brightness limit as the [\ion{O}{3}]$+\mathrm{H}\beta$ map with noise (SB$=1.4 \times 10^{-16}$ erg s$^{-1}$ cm$^{-2}$), represented by a black square with a yellow outline. All three points are projected along the maximum distance from the center to the edge pixels to show the radial direction in which the [\ion{O}{3}]$+\mathrm{H}\beta$ emission is most extended the noisy and noiseless [\ion{O}{3}]$+\mathrm{H}\beta$ maps.

In the $z=2.7$ simulated image of NGC 2992 without noise, the calculated average NLR radius is $3.8$ kpc (using SB limit $=2.4 \times 10^{-17}$ erg s$^{-1}$ cm$^{-2}$) and $1.2$ kpc (using SB limit $=1.4 \times 10^{-16}$ erg s$^{-1}$ cm$^{-2}$). The radial extent of the bicone is reduced in the simulated image with noise, with an average radius of $1.2$ kpc (using SB limit $=1.4 \times 10^{-16}$ erg s$^{-1}$ cm$^{-2}$). These size differences highlight the impact of instrumental noise on measuring the NLR size at $z=2.7$, particularly due to the strong dependence of our measurements on the selected surface brightness limit. Additionally, the NLR radius derived from the simulated medium-band image (without noise) may overestimate the intrinsic NLR size due to possible contamination from H$\beta$ and continuum emission in the medium-band. We therefore include the average NLR radii calculated from simulated narrow-band images in Table \ref{table:MUSE} to provide a more accurate reference of the intrinsic NLR sizes for NGC 2992 and the full MUSE sample. These narrow-band NLR sizes are discussed in greater detail in Section \ref{subsection:NLR_sizes}. The next section will further explore the effect of noise on the observed NLR sizes as a function of AGN luminosity.

\subsection{Simulating the NLR Size-AGN Luminosity Relation at Cosmic Noon}
\label{subsection:NLR-lum}
NLR sizes have been found to correlate with AGN luminosity, with higher AGN luminosities corresponding to larger photoionized regions \citep{Bennert2002, Liu2013a, Liu2013b, Liu2014, Hainline2013, Haineline2014}. However, these studies have concentrated on galaxies at $z<0.6$. Our simulated images of NGC 2992 at $z=2.7$ showed that cosmological dimming and instrument noise reduce the observed size of the NLR. We explore how these effects might impact the observed NLR size-AGN luminosity relationship at cosmic noon, by simulating this trend for our full MUSE AGN sample at $z=2.7$. We refer the reader to Section \ref{subsection:muse} for details on our MUSE AGN sample selection. Information for each object can be found in Appendix \ref{section:Appendix_A}, along with a montage of $z=2.7$ simulated NIRCam medium-band images for these objects, with and without simulated NIRCam noise (Figures \ref{fig:Montage_a} and \ref{fig:Montage_b}). The NLR radii for our simulated sample are calculated using the same methods and surface brightness limits discussed in Section \ref{subsection:MUSE_NLR_size}. The average NLR radii for the full MUSE sample, calculated from the medium-band [\ion{O}{3}]$+\mathrm{H}\beta$ maps (with and without noise), are reported in Table \ref{table:MUSE} and shown in Figures \ref{fig:Montage_a} and \ref{fig:Montage_b} (Appendix \ref{section:Appendix_A}). 
 
\begin{figure}
        \centering
        \includegraphics[width=90mm,trim={0.5cm 0 0 1cm},clip]{"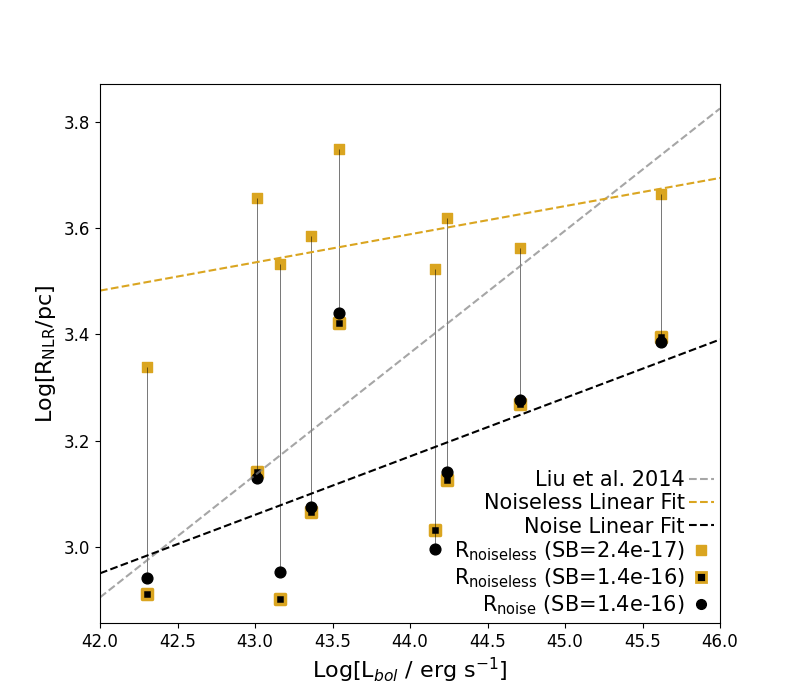"}
        \caption{Plot of average NLR radii for our full MUSE sample both with and without instrument noise at $z=2.7$ versus their AGN bolometric luminosities. The NLR radii are calculated down to a minimum surface brightness of $2.4\times 10^{-17}$ erg s$^{-1}$ cm$^{-2}$ arcsec$^{-2}$ for the noiseless [\ion{O}{3}]$+\mathrm{H}\beta$ maps and $1.4 \times 10^{-16}$ erg s$^{-1}$ cm$^{-2}$ arcsec$^{-2}$ for the [\ion{O}{3}]$+\mathrm{H}\beta$ maps with simulated noise. The average NLR radii for the noiseless [\ion{O}{3}]$+\mathrm{H}\beta$ images are represented by yellow squares and the average NLR radii for the noise [\ion{O}{3}]$+\mathrm{H}\beta$ images are represented by black circles. The gray dashed line shows the observed low-redshift best-fit line from \cite{Liu2014}. The yellow dashed line represents linear fit to the simulated sample without noise (yellow squares). The black dashed line represents linear fit to the simulated sample with noise (black circles).}
        \label{fig:MUSE_NLR_plot}
\end{figure}

In Figure \ref{fig:MUSE_NLR_plot}, we show the calculated average NLR radii for our full simulation sample at $z=2.7$ (with and without instrument noise), plotted against their AGN bolometric luminosities. The NLR sizes are calculated down to a minimum surface brightness level of $2.4\times10^{-17}$ erg s$^{-1}$ cm$^{-2}$ arcsec$^{-2}$ for the [\ion{O}{3}]$+\mathrm{H}\beta$ maps without simulated noise (yellow squares) and $1.4\times10^{-16}$ erg s$^{-1}$ cm$^{-2}$ arcsec$^{-2}$ for the [\ion{O}{3}]$+\mathrm{H}\beta$ maps with simulated noise (black circles). We also overlay measurements of the NLR radius for the [\ion{O}{3}]$+\mathrm{H}\beta$ maps without noise, calculated down to the limiting surface brightness of $1.4 \times 10^{-16}$ erg s$^{-1}$ cm$^{-2}$ (black squares with yellow outlines), to further emphasize the significant influence of the limiting surface brightness level on our NLR measurements. 

Both the simulated noise and noiseless samples show a positive trend between NLR radii and AGN luminosity, in agreement with the NLR size-AGN luminosity trend observed at low redshift. The noiseless sample shows larger scatter compared to the sample with noise, with a median NLR radius of $3.8$ kpc and standard deviation of $0.9$ kpc. The noise sample (black circles) has a median NLR radius of $1.3$ kpc and standard deviation of $0.6$ kpc. The higher standard deviation in NLR radii for the noiseless sample may result from an overestimation of the NLRs in ESO 428-14, IC 1657, Mrk 926, and NGC 5728, likely caused by difficulties in differentiating [\ion{O}{3}] emission from the ionization cone from other sources above the limiting surface brightness. 

We find that the addition of instrument noise results in smaller observed NLR radii across the entire sample. The median difference between the measured average NLR radii for the the sample without simulated noise (SB$=2.4\times10^{-17}$ erg s$^{-1}$ cm$^{-2}$ arcsec$^{-2}$) and with simulated noise (SB$=1.4\times10^{-16}$ erg s$^{-1}$ cm$^{-2}$ arcsec$^{-2}$) is $2.5$ kpc ($0.48$ dex) with a standard deviation of $0.5$ kpc ($0.11$ dex). For reference, when the same limiting surface brightness level (SB$=1.4\times10^{-16}$ erg s$^{-1}$ cm$^{-2}$ arcsec$^{-2}$) is applied to both sample types, we find that the noiseless NLR radii are $-0.01$ dex smaller than the NLR radii of the noisy sample. However, these NLR radii (black squares with yellow outlines) do not reflect accurate NLR measurements for the noiseless [\ion{O}{3}]$+\mathrm{H}\beta$ maps of our MUSE AGN sample (as evidenced by the disparities between these plotted size measurements and the white dashed outlines in the fourth column of Figures \ref{fig:Montage_a} and \ref{fig:Montage_b} in Appendix \ref{section:Appendix_A}. These results indicate that NIRCam is able to detect ionization cones in cosmic noon galaxies above the instrumental noise level, but the measured sizes of these NLRs may be biased lower by instrumental noise and cosmological surface brightness dimming.
 
For comparison with the NLR size-AGN luminosity trend observed at low redshift, we also plot an empirically derived line (gray dashed line) in Figure \ref{fig:MUSE_NLR_plot} using the best-fit relationship from \cite{Liu2014} in terms of the bolometric luminosity:

\begin{equation}
\begin{aligned}
&\log(\frac{R_{\mathrm{NLR}}}{pc}) = 0.23 \times \log(\frac{L_{[\mathrm{OIII}]}}{10^{42} \mathrm{erg s}^{-1}}) + 3.72
\label{eq:R_NLR}
\end{aligned}
\end{equation} 

where $L_{\mathrm{bol}}$, from \cite{Lamastra2009}, is given by:

\begin{equation}
\begin{aligned}
&\frac{L_{\mathrm{bol}}}{L_{[\mathrm{OIII}]}}\simeq 3500
\label{eq:L_bol}
\end{aligned}
\end{equation} 

The best-fit line given by (\ref{eq:R_NLR}) is derived from a sample of $z<0.6$ AGN that include obscured quasars \citep{Liu2013a, Hainline2013, Greene2011, Humphrey2010}, unobscured quasars \citep{Liu2014,Husemann2013}, and Seyfert 2 galaxies \citep{Bennert2006, Fraquelli2003}, where the NLR radii are measured above a limiting surface brightness of $10^{-15}/(1+z)^{4}$ erg s$^{-1}$ cm$^{-2}$ arcsec$^{-2}$. Although we were unable to measure the NLR sizes of our $z=2.7$ simulated medium-band sample using this standard surface brightness threshold, we perform linear fits for the simulated noiseless (yellow squares) and noisy (black circles) samples to compare their slopes to the low-redshift literature trend. The simulated noiseless sample linear fit (yellow dashed line) is given by:

\begin{equation}
\begin{aligned}
&\log(\frac{R_{\mathrm{NLR_{noiseless}}}}{pc}) = 0.053 \times \log(\frac{L_{[\mathrm{OIII}]}}{10^{42} \mathrm{erg s}^{-1}}) + 3.67
\label{eq:R_NLR_noiseless}
\end{aligned}
\end{equation}

The simulated noise sample linear fit (black dashed line) is given by:

\begin{equation}
\begin{aligned}
&\log(\frac{R_{\mathrm{NLR_{noise}}}}{pc}) = 0.11 \times \log(\frac{L_{[\mathrm{OIII}]}}{10^{42} \mathrm{erg s}^{-1}}) + 3.34
\label{eq:R_NLR_noise}
\end{aligned}
\end{equation} 

The slope of the linear fit to the simulated noise sample is a factor of $\sim 2$ lower than the slope ($0.23$) of the derived low-redshift trend. While our simulated noise and noiseless samples show a positive trend between NLR radii and AGN luminosity, we find our simulated samples have shallower slopes compared to the slope of the derived best-fit line at low redshift. We discuss these results further in Section \ref{subsection:NLR_sizes}.
 
\section{Results: NIRCam Observations} \label{sec:observation_results}
In Section \ref{sec:simulation_results}, we showed simulated, $z=2-3$, NIRCam images of [\ion{O}{3}] ionization cones for low-redshift AGN observed by MUSE. Here we present real NIRCam observations targeting the [\ion{O}{3}]$+\mathrm{H}\beta$ emission for $27$ AGN at $z=2.4-3.4$ in the GOODS-S field. Notes on individual objects can be found in Appendix \ref{section:Appendix_C}. In Section \ref{subsection:NIRCam_morphologies}, we summarize the various [\ion{O}{3}]$+\mathrm{H}\beta$ morphologies observed for our objects and our classification scheme. In Section \ref{subsection:NIRCam_size_lum}, we explore the NLR Size-AGN luminosity trend for our observed sample and compare to our $z=2.7$ simulated noise sample.

\subsection{[\ion{O}{3}]$+\mathrm{H}\beta$ Morphologies} \label{subsection:NIRCam_morphologies}
\begin{figure*}
        \centering
        \includegraphics[width=160mm,height=38mm,trim={0 1.5cm 0 2cm},clip]{"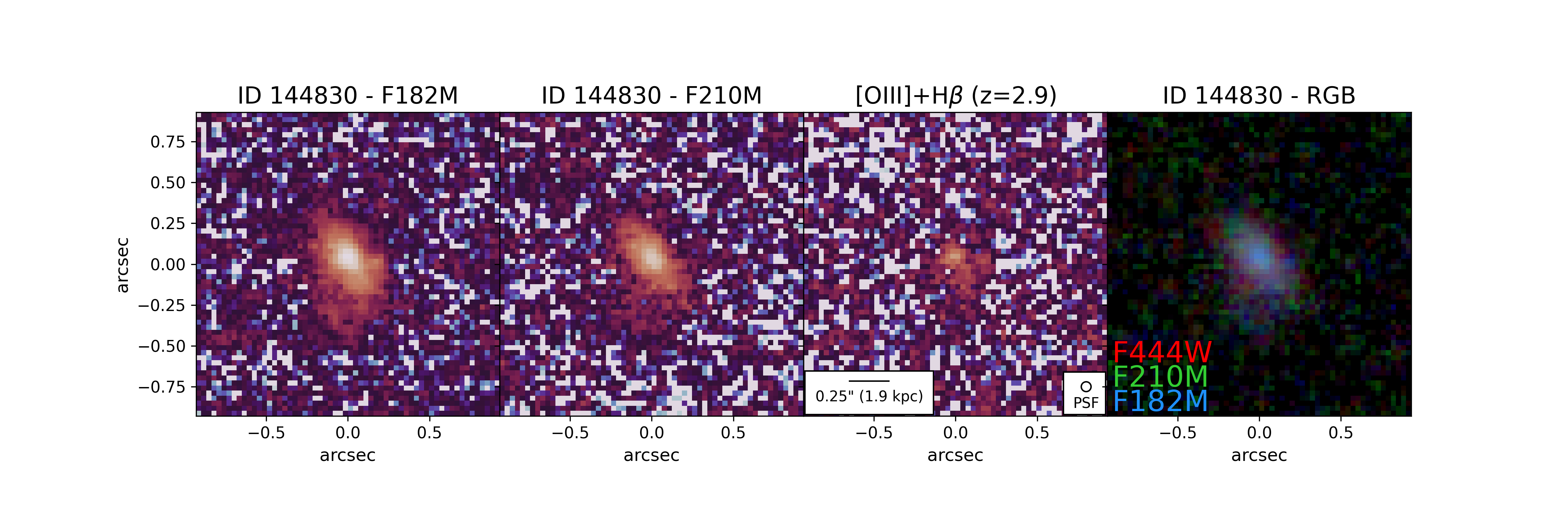"} 
        \includegraphics[width=160mm,height=38mm,trim={0 1.5cm 0 2cm},clip]{"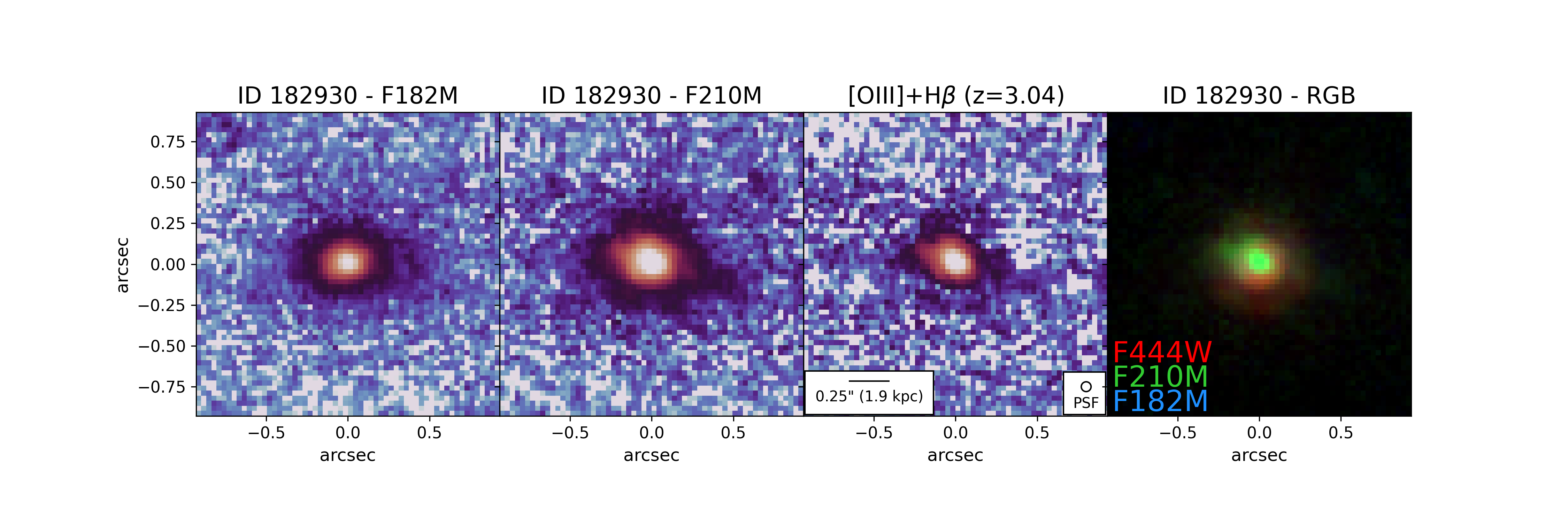"}
        \includegraphics[width=160mm,height=38mm,trim={0 1.5cm 0 2cm},clip]{"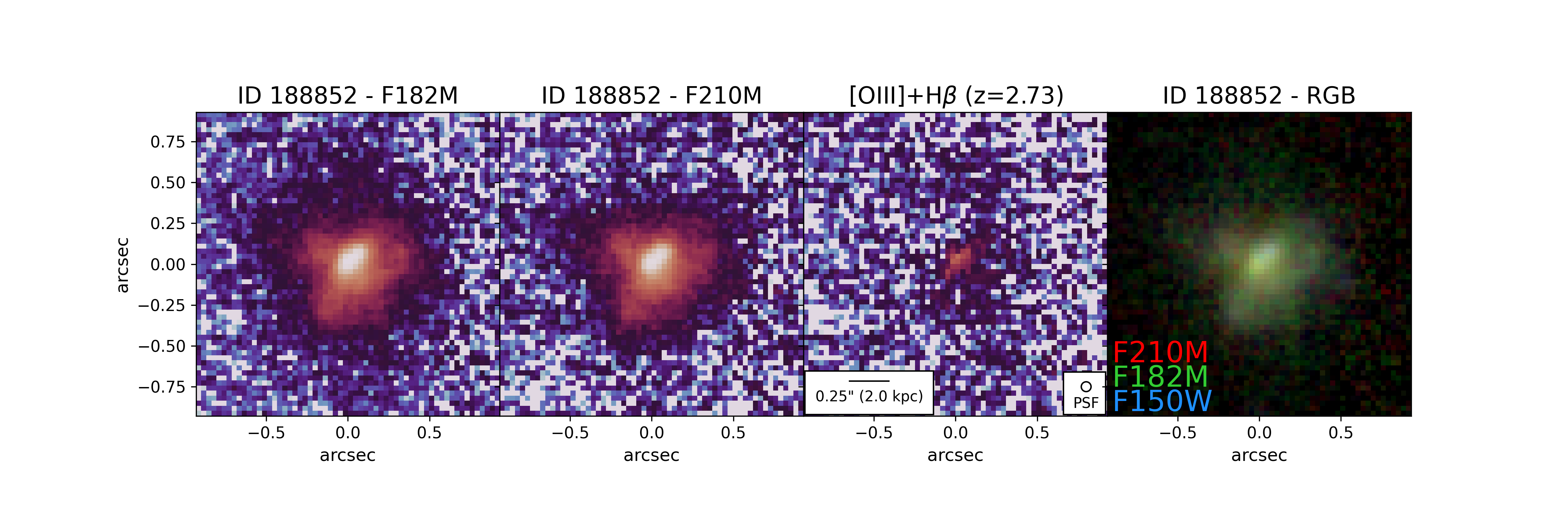"}  
        \includegraphics[width=160mm,height=38mm,trim={0 1.5cm 0 2cm},clip]{"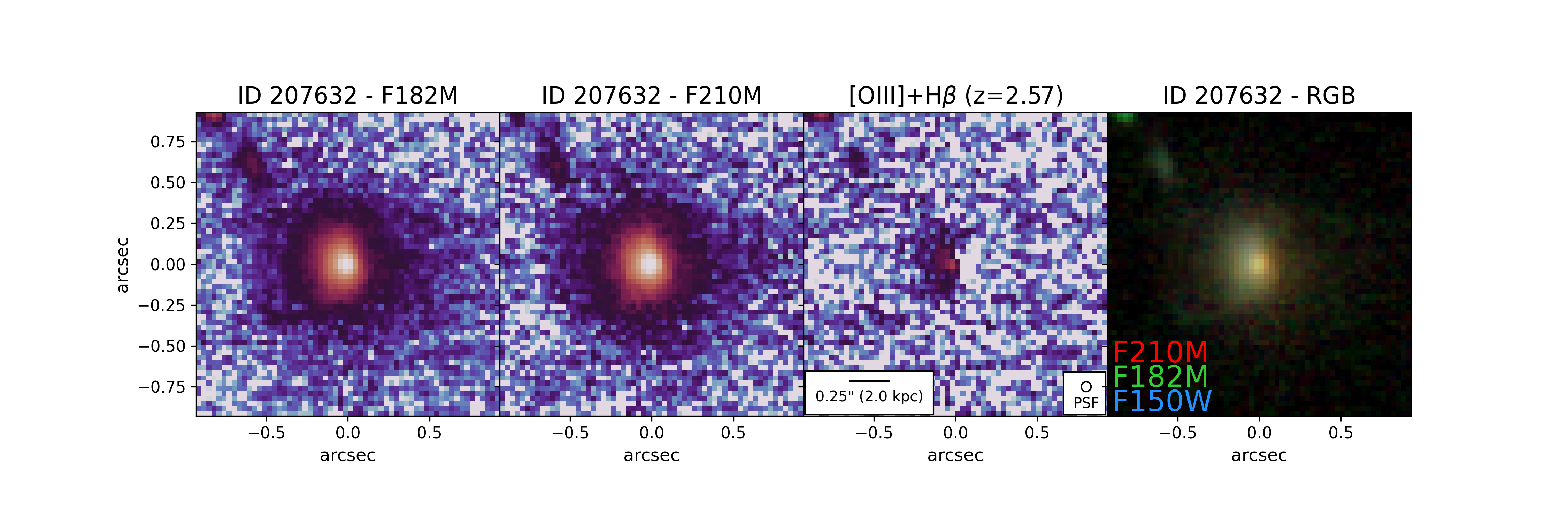"}
        \includegraphics[width=160mm,height=38mm,trim={0 1.5cm 0 2cm},clip]{"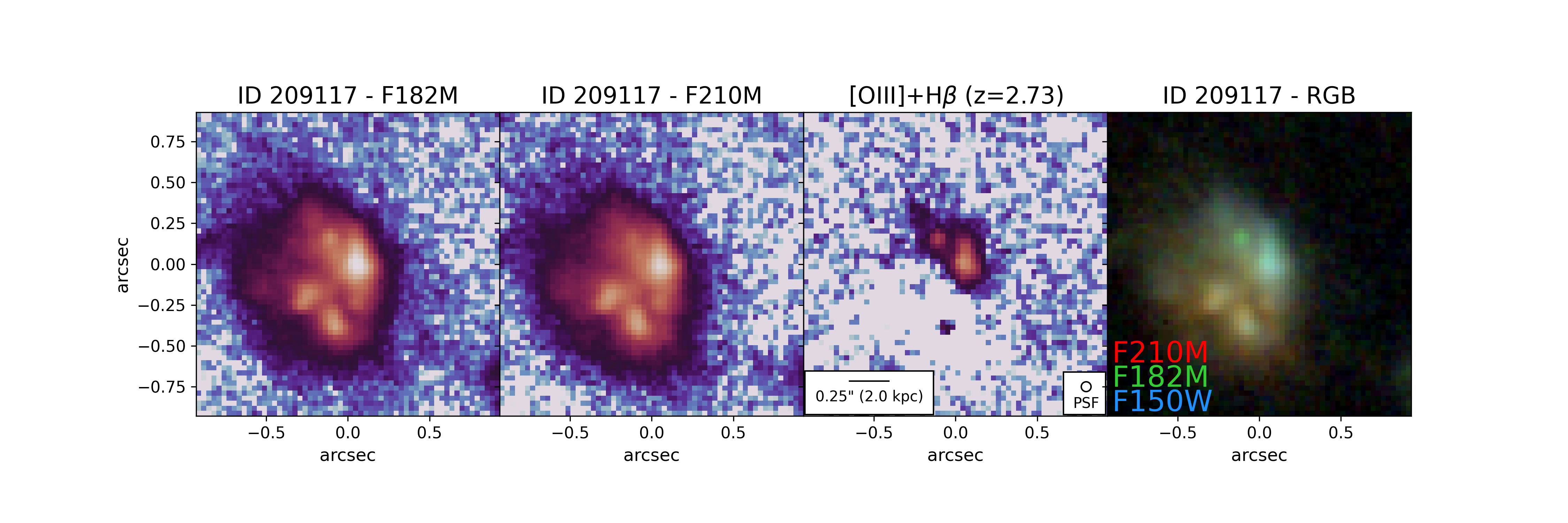"}
        \includegraphics[width=160mm,height=38mm,trim={0 1.5cm 0 2cm},clip]{"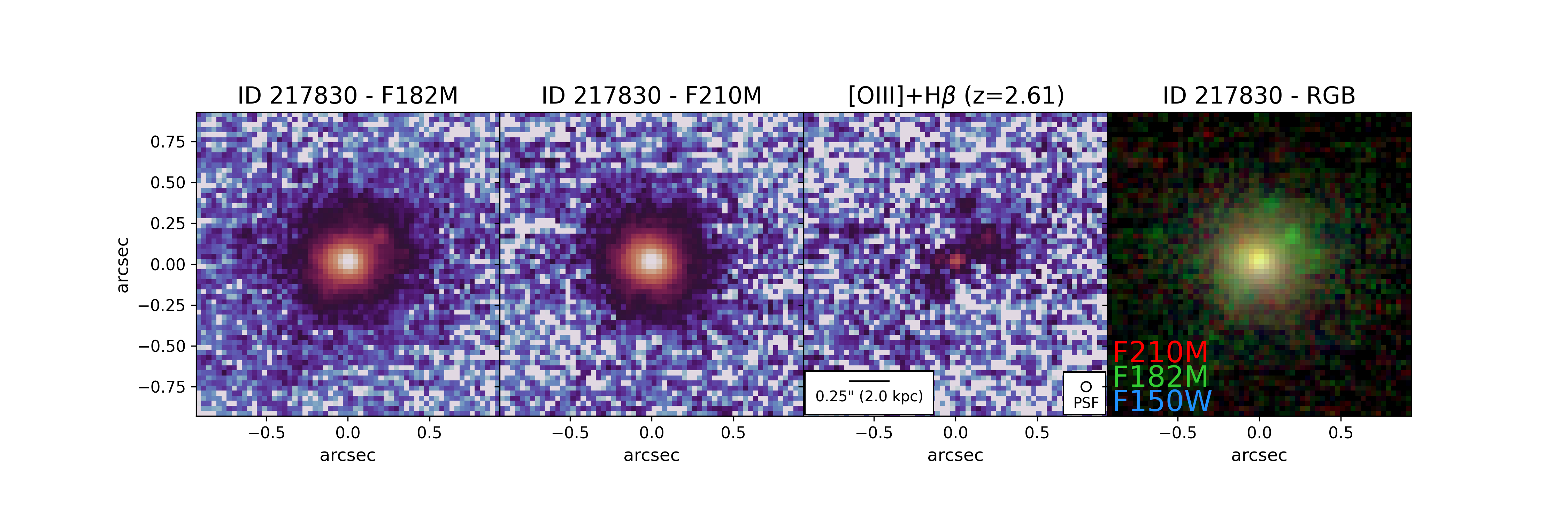"}
        \caption{Montage of NIRCam objects with AGN NLRs that show possible cone or bicone morphology.} \label{fig:Bicone}
\end{figure*}

We categorize our objects by their [\ion{O}{3}]$+\mathrm{H}\beta$ morphologies as either extended, compact, or no detected [\ion{O}{3}]$+\mathrm{H}\beta$ emission. We classify these objects as extended if their [\ion{O}{3}]$+\mathrm{H}\beta$ emission visibly appears elongated across multiple pixels, exhibiting features such as rings, spiral arms, or conical morphologies ($17/27$ are extended). Within this sample, objects with conical [\ion{O}{3}]$+\mathrm{H}\beta$ morphology or asymmetric, extended [\ion{O}{3}]$+\mathrm{H}\beta$ emission inconsistent with a spiral arm or ring form a subsample we refer to as our "cone" sample ($6$ objects). These objects are shown in Figure \ref{fig:Bicone}. Objects with extended [\ion{O}{3}]$+\mathrm{H}\beta$ morphology consistent with star-forming rings or spiral arms ($11$ objects) are shown in Figures \ref{fig:Extended_a} and \ref{fig:Extended_b} in Appendix \ref{subsubsection:extended}. 

\begin{figure*}
        \centering
        \includegraphics[width=160mm,height=38mm,trim={0 1.5cm 0 2cm},clip]{"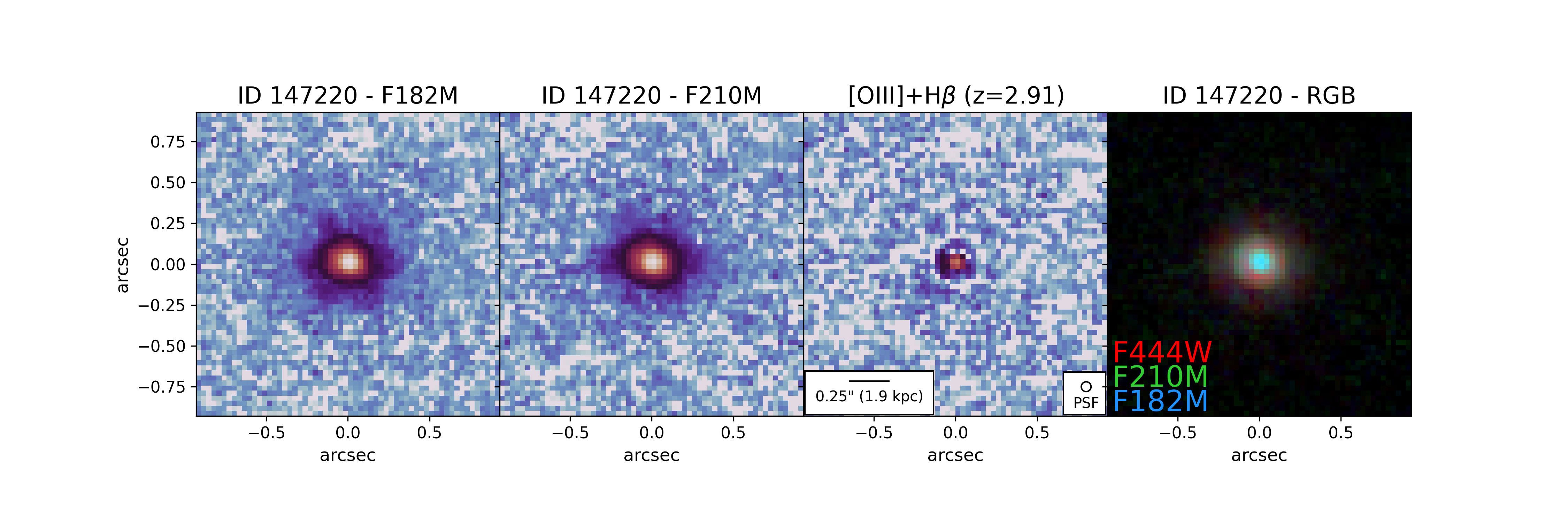"}  
        \includegraphics[width=160mm,height=38mm,trim={0 1.5cm 0 2cm},clip]{"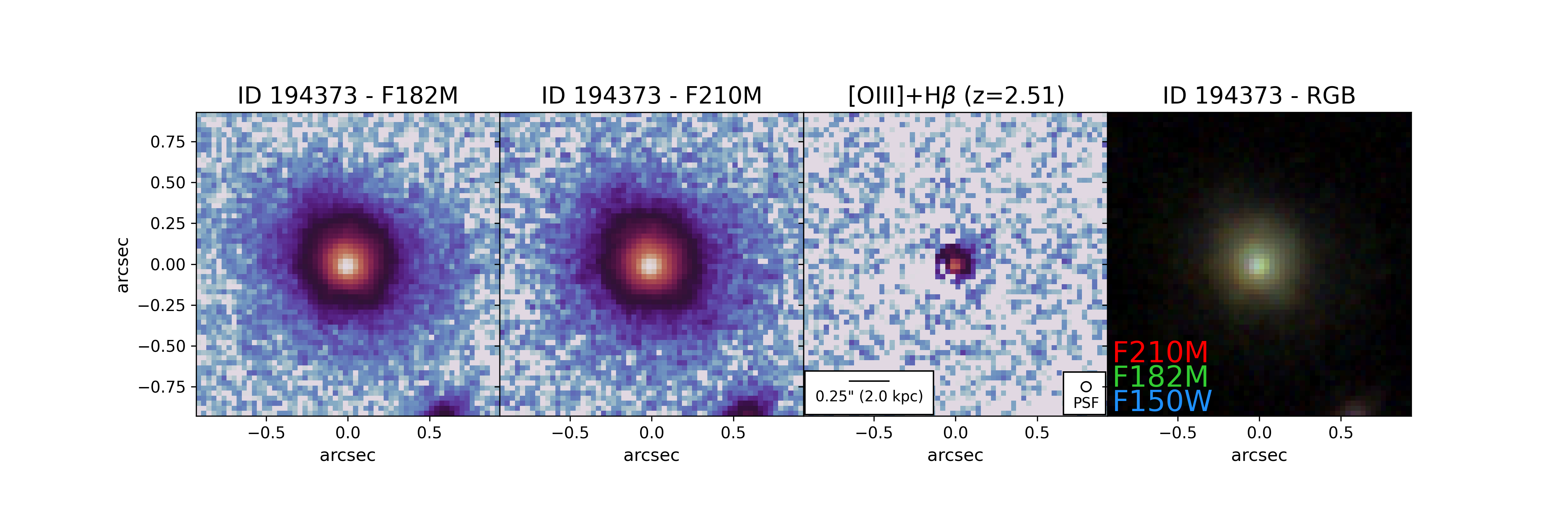"}  
        \includegraphics[width=160mm,height=38mm,trim={0 1.5cm 0 2cm},clip]{"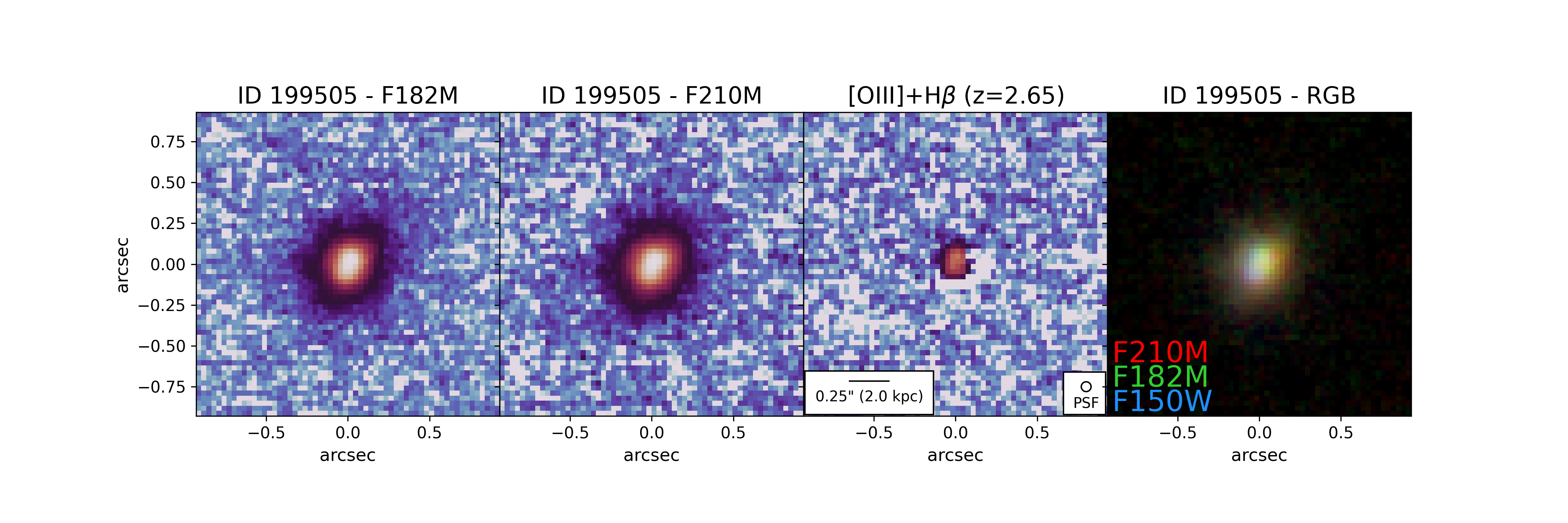"}
        \includegraphics[width=160mm,height=38mm,trim={0 1.5cm 0 2cm},clip]{"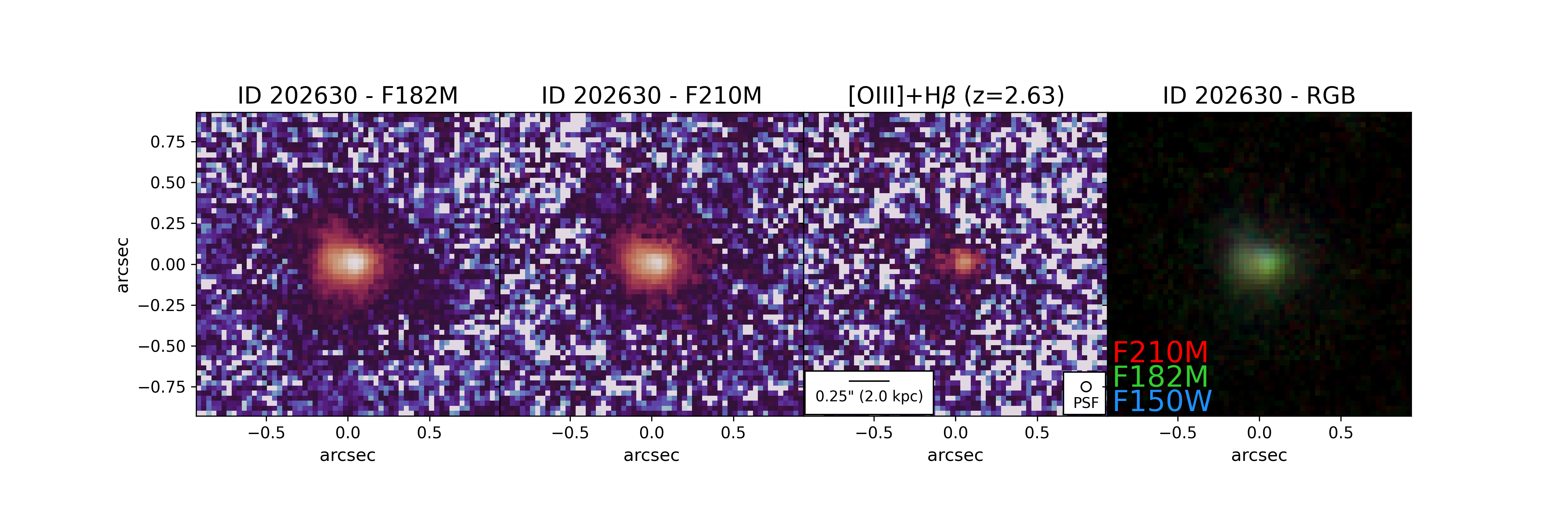"}
        \caption{Montage of NIRCam objects with centrally concentrated [\ion{O}{3}]$+\mathrm{H}\beta$ (likely compact NLRs).} \label{fig:Compact}
\end{figure*}

We classify objects as having compact [\ion{O}{3}]$+\mathrm{H}\beta$ morphology if their [\ion{O}{3}]$+\mathrm{H}\beta$ emission visually appears to be symmetrically concentrated around a central nucleus ($7/27$ are compact). We note that this classification does not consider whether the source is resolved ($>2\times$ PSF). Figure \ref{fig:Compact} shows objects with centrally concentrated [\ion{O}{3}]$+\mathrm{H}\beta$ emission ($4$ objects). Figure \ref{fig:Companion} in Appendix \ref{subsubsection:compact} shows compact objects with potential companion galaxies or other ionized features ($3$ objects). We also find three AGN in our sample that show no detectable [\ion{O}{3}]$+\mathrm{H}\beta$ emission (see Figure \ref{fig:No OIII} in Appendix \ref{subsubsection:no_OIII}). By visually inspecting the [\ion{O}{3}]$+\mathrm{H}\beta$ morphologies of our AGN sample, we find six AGN with [\ion{O}{3}]$+\mathrm{H}\beta$ emission consistent with extended NLRs and/or ionization cones. We classify the four AGN with symmetric, centrally concentrated [\ion{O}{3}]$+\mathrm{H}\beta$ emission as candidate compact NLRs, bringing our observed NLR sample size to ten AGN. However, we note that our classification method may exclude some AGN in the ``extended'' and ``companion'' subcategories that could have an NLR that is obscured by dust or not clearly visible due to the geometry of the system. As a result, our observed NLR sample includes only AGN that we consider to have strong evidence of an NLR, constituting approximately one-third of our total observed sample. We note that the majority of these sources are spatially resolved compared to the NIRCam PSF, with only two cases being unresolved ($<2\times$ PSF). For reference purposes, we add the NIRCam PSF size to each figure. We investigate the NLR sizes of these objects as a function of AGN luminosity in the following section. 

\subsection{Exploring the Observed NLR Size-AGN Luminosity Relation at Cosmic Noon} \label{subsection:NIRCam_size_lum}
We measure the sizes of the NLRs for objects with conical [\ion{O}{3}]$+\mathrm{H}\beta$ morphology and compact [\ion{O}{3}]$+\mathrm{H}\beta$ morphology. We do not include objects with extended emission from star-forming regions or compact objects with potential companions since the NLR radii of these objects cannot be accurately measured. Therefore, our "cone" sample includes six AGN, and our "compact" sample includes four AGN, bringing our sample of measured NLR sizes to ten. The NLR radii and AGN luminosities for the observed NLR sample are detailed in Table \ref{table:NLR_sizes}. 

\begin{figure*}
        \centering        
        \includegraphics[width=140mm,height=75mm,trim={0 0 0 0},clip]{"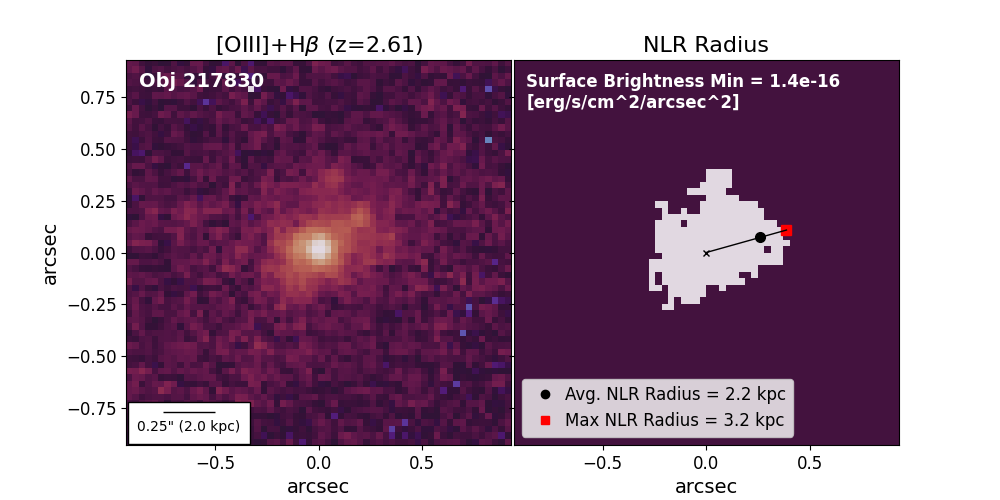"}
        \caption{Example of approximating the NLR radius in Object 217830 by measuring the extent of the [\ion{O}{3}]$+\mathrm{H}\beta$ emission above a cutoff surface brightness level of $1.4 \times 10^{-16}$ erg s$^{-1}$ cm$^{-2}$ arcsec$^{-2}$. Left: [\ion{O}{3}]$+\mathrm{H}\beta$ map ($F182M-F210M$). Right: The [\ion{O}{3}]$+\mathrm{H}\beta$ emission above the limiting surface brightness is shown by the gray region. The average NLR radius is shown by the black circle and the maximum NLR radius is shown by the red square.} \label{fig:NLR_3sig}
\end{figure*}

Since our simulated noise images were constructed to have the same noise properties as our observed NLR sample, we measure the NLR radii for our observed sample using the same limiting surface brightness as our simulated noise sample (SB = $1.4 \times 10^{-16}$ erg s$^{-1}$ cm$^{-2}$ arcsec$^{-2}$). We show an example of our measurements of the NLR radius for Object 217830 (cone subsample) in Figure \ref{fig:NLR_3sig}. For this object, we report the average NLR radius (black circle) and maximum NLR radius (red square). The average NLR radii is calculated as discussed in Section \ref{subsection:MUSE_NLR_size}. The maximum NLR radius is calculated as the distance from the centroid to the furthest edge pixel reaching the limiting surface brightness. We consider the maximum NLR radius as the ionization cone often extends along one axis, which could result in a lower estimate of the average NLR radius for objects with an asymmetric NLR. 

\begin{deluxetable}{l l c c c}
\tabletypesize{\footnotesize}
\tablecolumns{5}
\tablewidth{0pt}
\tablecaption{NIRCam NLR radii for observed NLR sample \label{table:NLR_sizes}}
\tablehead{JADES ID & [\ion{O}{3}]$+\mathrm{H}\beta$ & Log[$\mathrm{L}_{\mathrm{bol}}$] & Avg. NLR & Max. NLR \\
\colhead{} & Morphology & [erg s$^{-1}$] & radius (kpc) & radius (kpc)}
\startdata
    144830 & Cone & 44.06 & 1.1 & 1.9 \\
    147220 & Compact & 44.97 & 1.5 & 2.2 \\ 
    182930 & Cone & 45.24 & 1.4 & 2.3 \\
    188852 & Cone & 43.46 & 2.2 & 3.5 \\ 
    194373 & Compact & 43.09 & 1.5 & 1.9 \\ 
    199505 & Compact & 45.02 & 1.8 & 2.5 \\ 
    202630 & Compact & 43.37 & 1.4 & 2.0 \\ 
    207632 & Cone & 44.96 & 1.5 & 2.4 \\ 
    209117 & Cone & 43.70 & 2.5 & 3.6 \\ 
    217830 & Cone & 45.13 & 2.2 & 3.2   
\enddata
\tablecomments{See Figure \ref{fig:NLR} for plot of NLR radii as a function of AGN luminosity for these objects.}
\end{deluxetable}

\begin{figure}
        \centering       
        \includegraphics[width=90mm,trim={0.5cm 0 0 1cm},clip]{"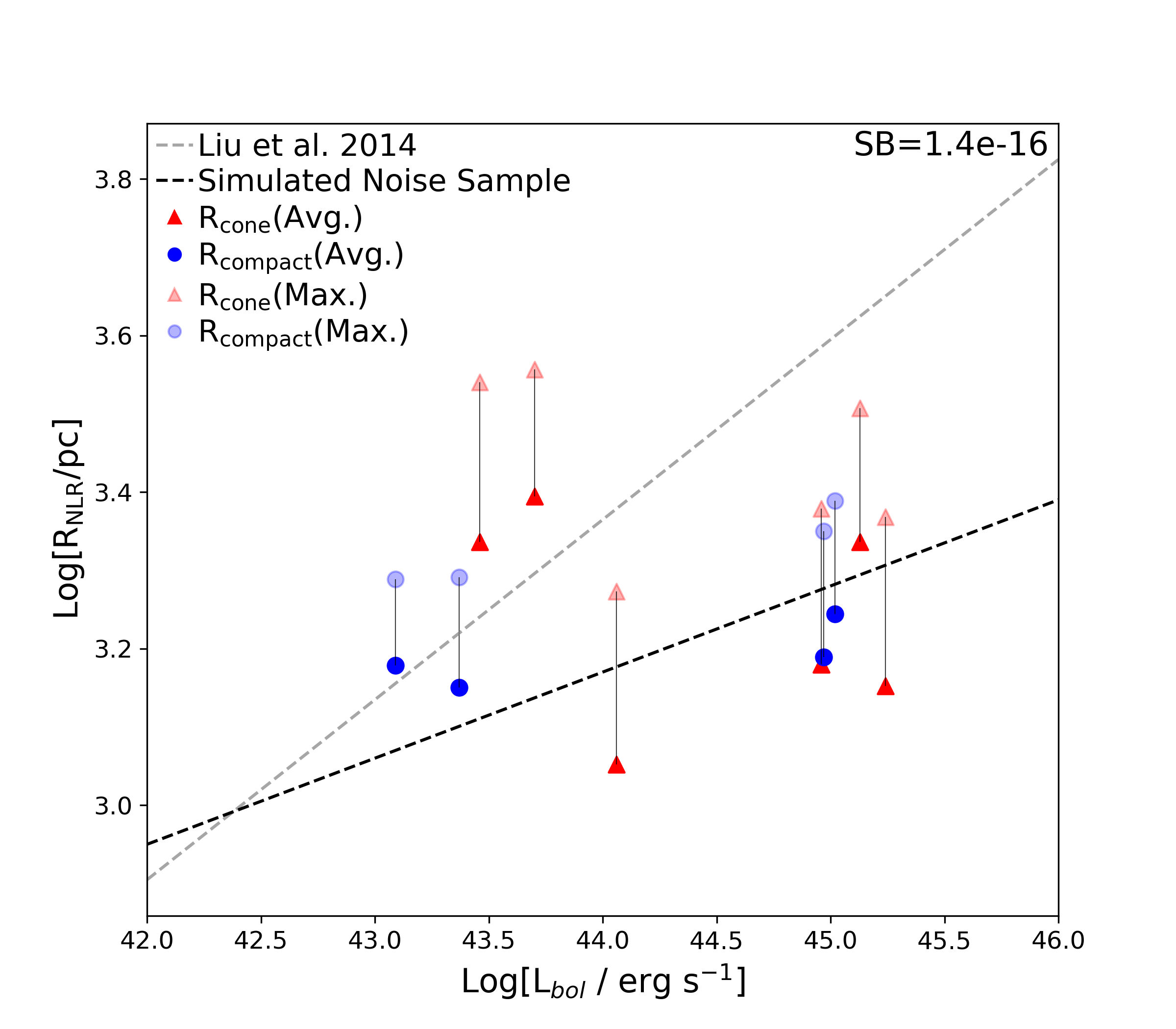"}
        \caption{Average and maximum NLR radii (kpc) versus AGN bolometric luminosity (erg s$^{-1}$) for the observed sample. NLR radii are calculated down to a fixed surface brightness level of $1.4 \times 10^{-16}$ erg s$^{-1}$ cm$^{-2}$ arcsec$^{-2}$. The cone sample is represented by red triangles and the compact sample is represented by blue circles. The opaque points show the measured average NLR radii and the faint points show the measured maximum NLR radii. The gray dashed line is the low-redshift best-fit line from \cite{Liu2014} (\ref{eq:R_NLR}). The black dashed line represents the linear fit to the $z=2.7$ simulated noise sample (black points in Figure \ref{fig:MUSE_NLR_plot}).} \label{fig:NLR}
\end{figure}

The average and maximum NLR radii (kpc) versus the AGN bolometric luminosity (measured in erg s$^{-1}$) for our cone and compact subsamples are plotted in Figure \ref{fig:NLR}. The cone sample is represented by red triangles and the compact sample is represented by blue circles. The opaque points show the average NLR radii and the faint points show the maximum NLR radii (representing an upper limit). For comparison with the observed low-redshift trend and our simulated trend, we overlay the low-redshift best-fit line from \cite{Liu2014} (gray dashed line) and the linear fit to the $z=2.7$ simulated noise sample (black dashed line) presented in Figure \ref{fig:MUSE_NLR_plot}. The median average NLR radii is $1.5$ kpc with a standard deviation of $0.4$ kpc. The median average NLR radii for the cone subsample is $1.8$ kpc with a standard deviation of $0.5$ kpc and the median average NLR radii for the compact subsample is $1.5$ kpc with a standard deviation of $0.1$ kpc. The median maximum NLR radii is $2.4$ kpc with a standard deviation of $0.6$ kpc. The median maximum NLR radii for the cone subsample is $2.8$ kpc with a standard deviation of $0.6$ kpc and the median maximum NLR radii for the compact subsample is $2.1$ kpc with a standard deviation of $0.2$ kpc. 

We find a general trend of increasing NLR radius with AGN bolometric luminosity, but a tight correlation cannot be discerned due to the significant scatter introduced by our small sample size and observational biases (i.e. cosmological dimming, instrumental noise, and medium-band imaging). The observed average NLR radii appear to be in closer agreement with the linear fit to the simulated noise sample (black dashed line) than the observed low-redshift line from \cite{Liu2014} (gray dashed line), while the maximum NLR radii are in closer agreement with the low-redshift, best-fit line. We elaborate further on how instrument noise, dimming, and contamination with H$\beta$ and continuum emission in the medium-band may influence the observed NLR size-AGN luminosity trend at cosmic noon in Section \ref{subsection:NLR_sizes}.

\section{Discussion} \label{sec:discussion}
In Section \ref{sec:simulation_results}, we demonstrated through our simulated images that NIRCam is able to resolve [\ion{O}{3}] ionization cones in cosmic noon galaxies ($z=2-3$) with narrow- and medium-band imaging. In Section \ref{sec:observation_results}, we presented NIRCam medium-band observations of the [\ion{O}{3}]$+\mathrm{H}\beta$ emission in 27 AGN host galaxies at $2.4<z<3.4$. We detected [\ion{O}{3}]$+\mathrm{H}\beta$ emission in 24 out of 27 of our objects. We classified six AGN as hosting conical [\ion{O}{3}]$+\mathrm{H}\beta$ morphology consistent with an AGN ionization cone and an additional four AGN with candidate compact NLRs. In this section, we discuss the significance and implications of our simulated and observed images. In Section \ref{subsection:NLR_sizes}, we discuss how observational biases influence observed NLR sizes as a function of AGN luminosity at cosmic noon. In Section \ref{subsection:galaxy_evolution}, we consider the implications of our study on the AGN-host galaxy relationship at cosmic noon. In Section \ref{subsection:limitations}, we discuss the limitations of using imaging to study AGN ionization cones in cosmic noon galaxies.

\subsection{Constraining the NLR Size-AGN Luminosity Trend at Cosmic Noon} \label{subsection:NLR_sizes}
The physical size of the NLR serves as an important parameter for evaluating the influence of the AGN on its host's environment. Literature studies at low redshift ($z<0.6$) indicate that the typical sizes of the extended NLR range from $2$ to $20$ kpc, and are steeply correlated with AGN mid-IR luminosity at lower luminosities and flatten at approximately $10$ kpc for high AGN luminosities. ($L_{\mathrm{bol}}>10^{46} $ erg s$^{-1}$) \citep{Liu2013a, Liu2014, Hainline2013, Bennert2006, Haineline2014, Sun2017}. Although the NLR size-AGN luminosity relationship has been studied extensively at low redshifts, it has remained largely unexplored at higher redshifts, particularly in the range $z\sim2.5-3$, where atmospheric absorption significantly hinders ground-based observations of [\ion{O}{3}]. As a result, studies of AGN ionization cones at cosmic noon have predominantly probed AGN at $z<2.5$ \citep{Harrison2016b, Circosta2018, Kakkad2020, Tozzi2024} or the brightest quasars (log[$L_{\mathrm{bol}}$/ erg s$^{-1}$] $\sim 47$) \citep{Wylezalek2022, Vayner2023, Vayner2024}. Given the range of redshifts ($2.5<z<3.3$) and AGN luminosities ((log[$L_{\mathrm{bol}}$/ erg s$^{-1}$] $\sim 43-45$) of our observed NLR sample, our observations offer the opportunity to extend the NLR size-AGN luminosity relation to a previously unexplored parameter space. Moreover, our simulated images allow us to interpret the results of our observations and assess how observational biases may influence and potentially skew the observed NLR size-AGN luminosity trend at high redshift. We discuss the biasing effects of instrumental noise and cosmological dimming in \ref{subsubsection:noise_bias} and of contamination in the medium-band in \ref{subsubsection:bandwidth_bias}.

\subsubsection{Biasing Effects of Noise and Cosmological Dimming}
\label{subsubsection:noise_bias}
We discuss how observational biases affect the observed NLR size-AGN luminosity trend at cosmic noon, using the simulated trend (with and without noise) to explore the effects of noise and dimming on the observed trend. The simulated $z=2.7$ NLR size-AGN luminosity relation for our MUSE AGN sample (with and without noise), was presented in Figure \ref{fig:MUSE_NLR_plot} and the observed NLR size-AGN luminosity relation was presented in Figure \ref{fig:NLR}. We find that the range of NLR radii for our simulated sample without instrument noise (SB limit $=2.4\times10^{-17}$ erg s$^{-1}$ cm$^{-2}$ arcsec$^{-2}$) is $2.2-5.6$ kpc, consistent with typical NLR sizes reported in the literature \citep{Bennert2006, Sun2017}. With the addition of instrument noise (SB limit $=1.4\times10^{-16}$ erg s$^{-1}$ cm$^{-2}$), this range is reduced by $\sim0.5$ dex, with characteristic NLR sizes ($0.9-2.8$ kpc) lying below typical literature estimates. These simulated NLR sizes are comparable to those measured in our observed sample (SB limit $=1.4\times10^{-16}$ erg s$^{-1}$ cm$^{-2}$), which possesses characteristic NLR sizes of $1.1-2.5$ kpc (using an average measured radius) and $1.9-3.6$ kpc (using a maximum measured radius) (see Section \ref{subsection:NIRCam_size_lum} for an explanation of these radii). The measured NLR radii as a function of AGN luminosity for the observed sample show significantly more scatter compared to the simulated trend, likely arising from the small size of our observed NLR sample. Nevertheless, most of the NLR radii for the observed sample fall well-below the derived low-redshift trend from \citet{Liu2014} (gray dashed line), aligning more closely with the linear fit to the simulated noise sample (black dashed line).

\begin{figure}
        \centering
        \includegraphics[width=90mm,trim={0.5cm 0.5cm 0 1.5cm},clip]{"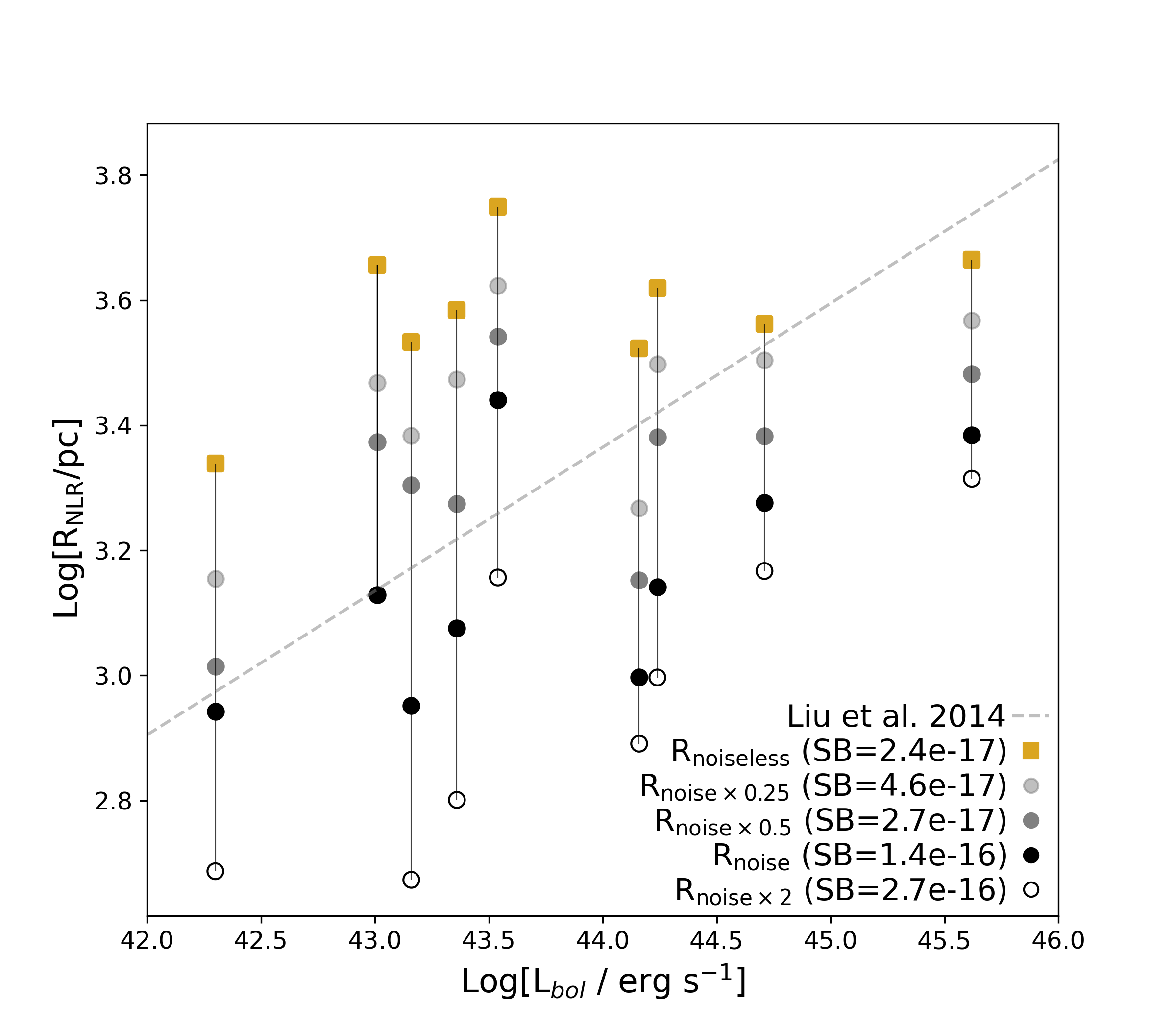"}
        \caption{Plot of average NLR radii versus their AGN bolometric luminosities measured from medium-bands for our full MUSE sample at $z=2.7$ with varying levels of simulated instrument noise. The yellow squares represent the NLR radii measured from the [\ion{O}{3}]$+\mathrm{H}\beta$ images without noise (SB limit $= 2.4\times10^{-17}$ erg s$^{-1}$ cm$^{-2}$). The circular points represent the NLR radii measured from the [\ion{O}{3}]$+\mathrm{H}\beta$ images with varying levels of simulated noise added. Faint gray points: simulated noise level equal to a quarter of the observed noise level (SB limit $= 4.6\times10^{-17}$ erg s$^{-1}$ cm$^{-2}$). Gray points: simulated noise level equal to a half of the observed noise level (SB limit $= 2.7\times10^{-17}$ erg s$^{-1}$ cm$^{-2}$). Black points: simulated noise level equivalent to the measured noise level of observations (SB limit $= 1.4\times10^{-16}$ erg s$^{-1}$ cm$^{-2}$). Open points: simulated noise level equal to twice the observed noise level (SB limit $= 2.7\times10^{-16}$ erg s$^{-1}$ cm$^{-2}$). The appropriate surface brightness limits for each sample are determined using the same method described in Section \ref{subsection:MUSE_NLR_size}, in which the limiting surface brightness level is equivalent to the $3 \sigma$ mean noise level for each sample. The gray dashed line shows the observed low-redshift best-fit line from \cite{Liu2014}.}
        \label{fig:NLR_plot_noises}
\end{figure}

The small relative measured NLR sizes in the simulated noise and observed samples at $z\sim2-3$, compared to the typical low-redshift literature values, likely reflects an observational bias rather than an inherent trend. As we showed in Section \ref{subsection:MUSE_NLR_size}, NLR size measurements at $z\sim2-3$ are highly sensitive to the choice of limiting surface brightness, primarily due to cosmological surface brightness dimming and instrumental noise, which reduce the signal-to-noise ratio and make it more challenging to detect extended [\ion{O}{3}] emission above the instrument's noise threshold. To quantify this effect, we explore the dependency of NLR size on the instrumental noise by measuring the NLR sizes of the simulated $z=2.7$ sample at multiple noise levels. In Figure \ref{fig:NLR_plot_noises}, we show the NLR sizes measured from simulated [\ion{O}{3}]$+\mathrm{H}\beta$ images under different noise conditions: no simulated noise (yellow squares), simulated noise at one-quarter of the observed noise level (faint gray points), half the observed noise level (gray points), equivalent to the observed noise level (black points), and twice the observed noise level (open points). We find that doubling the simulated noise level decreases the average NLR radius by a factor of 1.7 (0.3 dex). Conversely, reducing the noise level by a factor of $2$ increases the average NLR radius by a factor of 1.8 (0.2 dex), while reducing the noise level by a factor of 4 increases the average NLR radius by a factor of 2.2 (0.3 dex). 

Interestingly, reducing the simulated noise level by a factor of $4$ nearly restores the NLR radii measured from the noiseless [\ion{O}{3}]$+\mathrm{H}\beta$ images, with a median difference of $0.1$ dex between the two samples. However, quadrupling the signal-to-noise ratio requires a factor of sixteen increase in exposure time, making it challenging to recover the faint, extended [\ion{O}{3}]$+\mathrm{H}\beta$ emission through medium-band observations \citep[for reference, our NIRCam AGN sample was observed for $7.8$ hours in F210M and $9.9$ hours in F182M; ][]{Oesch2023, Williams2023}. As a result, observational biases from instrumental noise and cosmological dimming hinder our ability to accurately measure the intrinsic sizes of NLRs at higher redshifts. These biases systematically lower the observed NLR size–AGN luminosity relation at cosmic noon compared to low redshift and may also lead to a high incidence of NLRs appearing with artificially compact morphologies. In fact, this effect is observed in our simulated sample for IC 5063, ESO 428-14, NGC 5728, and IC 1481, where the visible extent of the [\ion{O}{3}] cones is reduced in the noise images, leading to a more compact [\ion{O}{3}] morphology (see Figures \ref{fig:Montage_a} and \ref{fig:Montage_b} in Appendix \ref{section:Appendix_A}). Indeed, this could explain the high number of AGN in our observed sample showing centrally concentrated [\ion{O}{3}]$+\mathrm{H}\beta$ emission. 

Additionally, objects with centrally concentrated [\ion{O}{3}]$+\mathrm{H}\beta$ morphologies may host small [\ion{O}{3}] cones or compact NLRs that NIRCam is unable to resolve. Other factors, such as the viewing angle of the NLR (i.e. face-on versus edge-on orientation) and the availability of gas for the AGN to ionize, may contribute to the smaller relative sizes of the NLR \citep{Joh2021, Haineline2014, He2018}. Nevertheless, our results from the noise simulations motivate longer exposure times to effectively recover the faint, extended [\ion{O}{3}] emission at the outskirts of the host galaxy. In our observed NLR sample, we find that the [\ion{O}{3}] cone of the faintest object in our "cone" sample (Object 144830) is detected at a SNR of $3$ (calculated by taking the ratio of the maximum measured surface brightness in a $5x5$ pixel box centered on the [\ion{O}{3}] cone to the mean noise level of the [\ion{O}{3}]$+\mathrm{H}\beta$ map). Applying this method to each object in our "cone" sample, we find that the faintest extended [\ion{O}{3}]$+\mathrm{H}\beta$ emission for each object is detected at a SNR of $\sim 3$. 

We also find that both of the simulated samples (noise/noiseless) show a positive relationship between NLR radius and AGN luminosity, although the slopes are shallower compared to the trend observed at low redshift. Instrumental noise may contribute to the shallower slope of the simulated noise sample linear fit (0.11) compared to the slope of the best-fit line (0.23) from \cite{Liu2014}. However, the simulated noiseless sample also exhibits a shallow slope (0.053). In fact, eight out of nine objects in our sample without simulated noise (yellow squares) lie above the low-redshift relation. These measurements may reflect overestimated NLR sizes for our MUSE AGN sample resulting from our choice to measure the NLR sizes from simulated medium-band images. We explore the impact of contamination in the medium-band in the following section.

\subsubsection{Medium-band Contamination}\label{subsubsection:bandwidth_bias}
As discussed in Section \ref{subsection:filter_choice}, extragalactic surveys are more likely to utilize medium- or wide-band imaging due to the broader redshift range over which [\ion{O}{3}] can be targeted and lower exposure times required to reach the same signal-to-noise as in a narrow-band. We have validated the use of NIRCam medium-band imaging for detecting the [\ion{O}{3}]$+\mathrm{H}\beta$ emission associated with AGN ionization cones and revealing their morphologies. Additionally, other studies have demonstrated that NIRCam broadband imaging can be used to identify photometric excesses due to [\ion{O}{3}]$+\mathrm{H}\beta$ \citep{Rinaldi2023, Solimano2025}, motivating the use of large broadband extragalactic datasets to search for AGN with candidate NLRs. While broadband imaging offers advantages in redshift coverage and dataset size, narrow-band imaging has the potential to more clearly detect the [\ion{O}{3}] cone by minimizing contamination from the galaxy’s stellar continuum and nearby emission or absorption lines, such as H$\beta$. This contamination can lead to overestimated NLR radii from medium-band imaging, especially for objects that are bright in H$\beta$. Although we cannot simulate narrow-band images with NIRCam noise due to the lack of NIRCam narrow-band AGN observations at a comparable redshift (to measure the noise properties), we can simulate narrow-band images at $z=2.7$ without NIRCam noise, using the narrow-band filter F187N to target [\ion{O}{3}]. These narrow-band images can provide more accurate measurements of NLR sizes for our MUSE AGN sample and aid in our assessment of the impact of H$\beta$ contamination in the medium-band band.

\begin{figure*}
        \centering
        \includegraphics[width=170mm,trim={0 0 0 1.5cm},clip]{"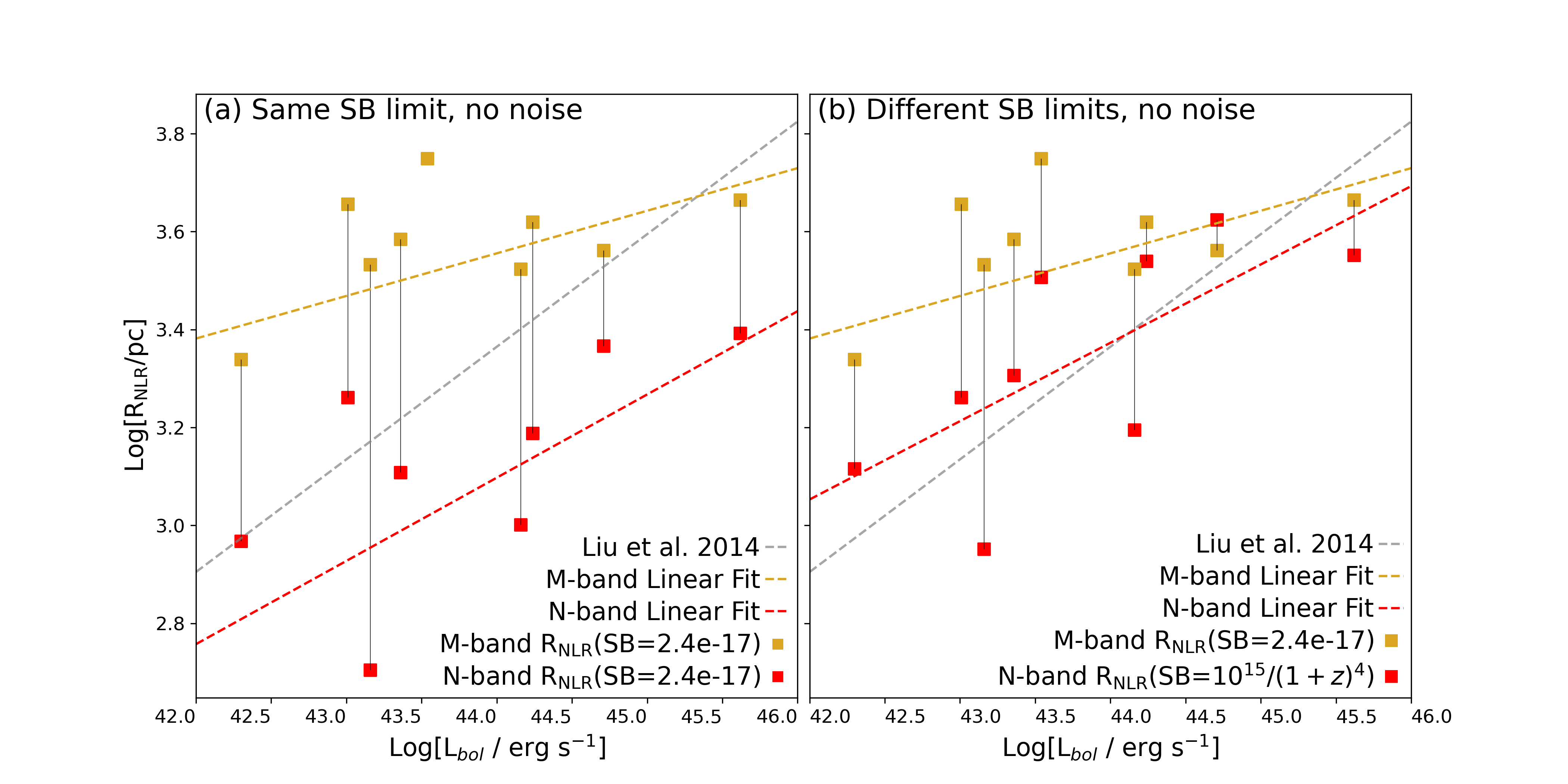"}
        \caption{Plot of average NLR radii versus their AGN bolometric luminosities measured from narrow and medium-bands for our full MUSE sample without instrument noise at $z=2.7$. The NLR radii measured from the medium-band images are represented by the yellow squares and are calculated with a limiting surface brightness of $2.4\times10^{-17}$ erg s$^{-1}$ cm$^{-2}$. The NLR radii measured from the narrow-band images are represented by the red squares and are calculated with a limiting surface brightness of $2.4\times10^{-17}$ erg s$^{-1}$ cm$^{-2}$ in panel (a) and $10^{-15}/(1+z)^{4}$ erg s$^{-1}$ cm$^{-2}$ arcsec$^{-2}$ in panel (b).} The gray dashed line shows the observed low-redshift best-fit line from \cite{Liu2014}. The yellow dashed line represents the linear fit to the simulated medium-band sample without noise (yellow squares). The red dashed line represents the linear fit to the simulated narrow-band sample without noise (red squares).
        \label{fig:Nband_size_plot}
\end{figure*}

In Figure \ref{fig:Nband_size_plot}, we plot the NLR sizes of our MUSE AGN sample (without noise) as a function of AGN luminosity for our simulated, narrow-band sample (red squares) and our simulated, medium-band sample (yellow squares) at $z=2.7$, measured using the same surface brightness limit (panel a) and different surface brightness limits (panel b). The dashed red and yellow lines represent the linear fits to these samples, respectively, and the low-redshift relation from \cite{Liu2014} (gray dashed line) is again overlaid for reference. When measuring the NLR radii from the simulated narrow-band images, the challenge once again arises of determining the suitable surface brightness limit for this sample. Conventionally, NLR sizes are typically measured using a limiting surface brightness level of $10^{-15}/(1+z)^{4}$ erg s$^{-1}$ cm$^{-2}$ arcsec$^{-2}$, where the $(1+z)^{4}$ term corrects for cosmological surface brightness dimming. As discussed in Section \ref{subsection:MUSE_NLR_size}, we are unable to measure the NLR sizes in our $z=2.7$, medium-band images using this surface brightness limit, as it is similar to the uncertainty in the residual background. Instead, we choose the more suitable limiting surface brightness of $2.4\times10^{-17}$ erg s$^{-1}$ cm$^{-2}$. Yet, by avoiding contamination with $H\beta$ in the narrow-band images, we now find that the conventional limiting surface brightness of $10^{-15}/(1+z)^{4}$ erg s$^{-1}$ cm$^{-2}$ arcsec$^{-2}$ is above the residual background level of the narrow-band [\ion{O}{3}] map, making it an appropriate isophotal radius for our narrow-band sample. To illustrate the impact of the surface brightness limit on the narrow-band NLR sizes, we plot the narrow-band NLR radii in panel (a) using the same surface brightness limit as the medium-band sample ($2.4\times10^{-17}$ erg s$^{-1}$ cm$^{-2}$), and in panel (b) using a more suitable limit of $10^{-15}/(1+z)^{4}$ erg s$^{-1}$ cm$^{-2}$ arcsec$^{-2}$.

We find that when measured using the same limiting surface brightness as the medium-band sample, the median NLR radius for the narrow-band sample is $1.3$ kpc with a standard deviation of $0.8$ kpc, approximately $0.4$ dex smaller than the medium-band sample. When measured down to the limiting surface brightness of $10^{-15}/(1+z)^{4}$ erg s$^{-1}$ cm$^{-2}$ arcsec$^{-2}$, the median NLR radius is $2.0$ kpc with a standard deviation of $1.1$ kpc, approximately $0.2$ dex smaller than the medium-band sample. The linear fit (red dashed line) to the narrow-band sample in panel (a) is given by:

\begin{equation}
\begin{aligned}
&\log(\frac{R_{\mathrm{NLR_{N-band}}}(a)}{pc}) = 0.17 \times \log(\frac{L_{[\mathrm{OIII}]}}{10^{42} \mathrm{erg s}^{-1}}) + 3.36
\label{eq:R_NLR_Nband_a}
\end{aligned}
\end{equation} 

The linear fit (red dashed line) to the narrow-band sample in panel (b) is given by:

\begin{equation}
\begin{aligned}
&\log(\frac{R_{\mathrm{NLR_{N-band}}}(b)}{pc}) = 0.16 \times \log(\frac{L_{[\mathrm{OIII}]}}{10^{42} \mathrm{erg s}^{-1}}) + 3.62
\label{eq:R_NLR_Nband_b}
\end{aligned}
\end{equation} 

Both linear fits to the narrow-band sample show a slope closer to the slope ($0.23$) of the low-redshift best-fit line from \cite{Liu2014} (Equation \ref{eq:R_NLR}) than to the slope of the linear fit to the medium-band sample. However, the linear fit in panel (b) has an intercept of $3.63$, which is much closer to the best-fit line intercept of $3.72$, likely due to the use of the same limiting surface brightness as in \cite{Liu2014}. These findings highlight the advantages of narrow-band imaging for measuring the sizes of NLRs at cosmic noon by 1) providing more accurate measurements of the NLR size through reducing contamination from H$\beta$ and stellar continuum, and 2) enabling measurements of the [\ion{O}{3}] extent at higher redshifts using the same limiting surface brightness level adopted in low-redshift studies \citep{Liu2013a, Liu2013b, Liu2014, Hainline2013, Haineline2014, Sun2017}. The second point is particularly important because it ensures uniformity in methodology, facilitating direct comparisons of the NLR size-AGN luminosity relationship across different redshifts. Accurate measurements of the [\ion{O}{3}] extent could be further improved with IFU spectroscopy, which allows for the creation of [\ion{O}{3}] maps by isolating the line at each spectral pixel through spectral slicing. We elaborate on the advantages of IFU spectroscopy for investigating the properties of AGN ionization cones in Section \ref{subsection:limitations}. 

As the number of objects in our MUSE AGN sample and NIRCam observation samples with NLRs are both small, it is challenging to compare how NLR sizes change from cosmic noon to low redshift. Our simulated images with instrument noise show that NIRCam may have difficulty recovering the intrinsic size of the NLR below a signal-to-noise ratio of $3$. The biasing effects of instrumental noise and cosmological dimming may result in smaller, more compact observed NLRs at cosmic noon. Although, given the correlation between AGN luminosity and NLR size, it is possible that the lower, "non-quasar" AGN luminosities (L$_{\mathrm{bol}} < 10^{46}$ erg s$^{-1}$) of our observed AGN sample could contribute towards the smaller relative NLR sizes. Ultimately, constraining the NLR size-AGN luminosity trend at cosmic noon must involve sampling AGN at a range of luminosities, including quasars. Although the quasar number density peaks at cosmic noon, low-to-moderate luminosity AGN are still more common than quasars at both low and high redshifts. As such, larger survey areas or pointed observations of pre-selected quasar candidates will be needed to select for a greater number of high-luminosity AGN.

\subsection{Implications for AGN Feedback at Cosmic Noon}
\label{subsection:galaxy_evolution}
Evidence of the close link between supermassive black holes (SMBHs) and their host galaxies has been accumulating over the last few decades \citep{Soltan1982, Kormendy1995, Magorrian1998}. \citet{Ferrarese2000} and \citet{Gebhardt2000} identified a strong correlation between the SMBH and the properties of the host galaxy, particularly the notable M$_{\mathrm{BH}}$-$\sigma$ relation between the mass of the SMBH and stellar velocity dispersion in the galactic bulge. Subsequent simulations revealed that active SMBHs can serve as a central engine, capable of regulating the galactic environment and driving the evolution of their host galaxies through AGN feedback \citep{DiMatteo2005, Springel2005, Bower2006, Croton2006}. In fact, many of the observed galaxy properties (i.e. stellar mass, star formation rates, black hole growth, and AGN accretion activity) appear to peak at cosmic noon and decline from $z=2$ to $z=0$ \citep{Madau2014}, an observation that is commonly attributed to AGN feedback. In this section, we discuss the importance of identifying and investigating NLRs at cosmic noon, and their role in probing AGN feedback during this pivotal evolutionary era.

Extended NLRs highlight the substantial potential of AGN to shape the galactic environment of their hosts, offering a valuable setting for studying the relationship between AGN and their host galaxies. NLRs have frequently been linked to "radiative-mode" AGN feedback through ionized outflows, propelled by radiation pressure from the accretion of matter onto the AGN \citep{Veilleux2005, Heckman&Best2014, Zakamska2014}. Identifying NLR outflows is especially important for understanding the coevolution of AGN and their host galaxies as these outflows have the capacity to regulate star formation in the galactic bulge and the surrounding local interstellar medium \citep{Cano-diaz2012, Liu2013a, Liu2013b, Revalski2019, Revalski2021, Ventur&Marconi2021}. In our NIRCam observations, we detect NLRs in $37\%$ of our objects and extended NLRs in $22\%$, suggesting that extended NLRs may be widespread at cosmic noon. Our detection fraction is in agreement with a low-redshift study ($0.01<z<0.15$) of $170$ AGN from the Mapping Nearby Galaxies at APO (MaNGA) survey with comparable bolometric luminosities (log[$L_{\mathrm{bol}}$/ erg s$^{-1}$] $\sim 42-45$) to our observed AGN sample. This study reported spatially resolved NLRs in $32\%$ of their sample \citep{Machado2022}. Additionally, our detection ratio may represent a lower limit, as we excluded AGN in the ``companion'' subcategory and those with extended [\ion{O}{3}]$+\mathrm{H}\beta$ emission resembling star-forming regions, both of which could host an undetected NLR obscured by dust or system geometry \citep{Hickox&Alexander2018}. Nevertheless, objects with a detected NLR provide a valuable opportunity to investigate AGN feedback during the peak epoch of galaxy formation by identifying outflows within their NLRs.  

Yet, an important theoretical question we must consider is whether NLR outflows are powerful enough to exert feedback on their host system. Many studies assess the effectiveness of feedback using a common threshold based on the energy outflow rate. For outflows to significantly impact the galactic environment, they must reach a peak energy outflow rate of approximately $\sim 0.05-5\%$ of the AGN bolometric luminosity \citep{DiMatteo2005, Hopkins&Elvis2010, Revalski2021}. Although observations of ionized outflows in the NLRs of low-redshift galaxies are common \citep{Gonzalez-Martin, Durre&Mould2018, Lopez-coba, Juneau, Ventur&Marconi2021, Revalski2021}, evidence of effective feedback induced by these NLR outflows remains largely inconclusive. In one study of six nearby galaxies ($z<0.6$) with detected ionized outflows, four out of the six met the criteria for effective feedback with typical outflow sizes of $0.1-3$ kpc, maximum mass outflow rates of $\dot{M}_{out} \approx 3-12 M_{\odot}$ yr$^{-1}$, and kinetic energies of $E \approx 10^{54}-10^{56}$ er\citep{Revalski2019, Revalski2021}. The ionized gas masses, outflow rates, and radial extents were all found to be positively correlated with the AGN bolometric luminosities across this sample. Yet, in another study of $18$ luminous AGN (L$_{\mathrm{bol}} \gtrsim 10^{45}$ erg s$^{-1}$) at $0.3<z<0.6$, no strong evidence for effective feedback or the large-scale ($\sim \mathrm{several}-15$ kpc) ionized outflows commonly reported for quasars \citep{Liu2013b}, was discovered \citep{Villar2016}. 

Models suggest that radiatively driven ionized outflows may be more prevalent and impactful at cosmic noon due to the high incidence of quasars and intense AGN accretion activity during this period \citep{Veilleux2001, Fischer2010, Heckman&Best2014}. In a study of $21$ Type-1 AGN at $z \sim 2$, ionized outflows were found to be ubiquitous in this sample, exhibiting typical radial extents up to $6$ kpc and mass outflow rates of $0.1-1000 M_{\odot}$ yr$^{-1}$ \citep{Kakkad2020}. Further evidence of effective feedback driven by NLR outflows were found in two bright quasars (log[$L_{\mathrm{bol}}$/ erg s$^{-1}$] $\sim 46-47$) at $z=1.6$ \citep{Cresci2023} and $z=2.9$ \citep{Wylezalek2022, Vayner2023, Vayner2024}. These studies underscore the potential influence of ionized outflows in shaping the galactic environment at cosmic noon. However, understanding the full impact AGN can have on their host galaxies through NLR outflows requires probing AGN across a range of luminosities, not just the brightest quasars. Given the moderate-to-high AGN luminosities of our observed AGN sample with extended NLRs, follow-up studies should investigate the prevalence and impact of ionized outflows in this sample, as these objects are strong candidates for exhibiting AGN feedback. A kinematic analyses of the [\ion{O}{3}] emission utilizing integral field unit (IFU) spectroscopy would aid in revealing AGN feedback in our NIRCam observations. We consider this in more depth in the following section and discuss the limitations of using imaging to study NLRs at cosmic noon. 

\subsection{Limitations of Imaging Studies} \label{subsection:limitations}
Unveiling the connection between AGN and their host galaxies through their extended NLRs at cosmic noon remains a significant challenge in extragalactic astronomy. While our simulated and observed images show that NIRCam can resolve the extended NLRs of AGN at cosmic noon, observational biases such as instrument noise and cosmological dimming will impact our ability to accurately recover the full extent of the [\ion{O}{3}] emission at this epoch. These factors will also make it challenging to visually determine NLR morphologies, especially if the fainter, extended [\ion{O}{3}] emission falls below the noise threshold, potentially biasing observations towards more compact NLRs. Additionally, the NLR morphologies and measured NLR sizes are subject to some uncertainty due to the scaling factor, $f$, used in the continuum subtraction to account for the continuum slope (see Section \ref{subsection:observation_methods} for details about the scaling factor). Throughout this paper, we determine $f$ using the ratio of the average continuum fluxes through F182M and F210M, assuming it remains spatially constant across the image. However, this assumption may lead to an imperfectly subtracted continuum, as the continuum slope may not be uniform across the image — particularly if the galaxy continuum is bluer or redder on its outskirts relative to the averaged galaxy continuum. To assess the impact of this assumption, we tested different $f$-values and found that the resulting uncertainty in the measured NLR sizes is small. A $10\%$ decrease in $f$ leads to a $\sim 5\%$ ($0.03$ dex) increase in the median average NLR radius of our observed NLR sample, while a $10\%$ increase in $f$ leads to only a $\sim 2\%$ ($0.02$ dex) decrease. Although the uncertainty introduced by $f$ is minimal, it can be further reduced using spectroscopy, which would enable the creation of an $f$-factor map that accounts for spatial variations in the continuum slope across the galaxy and may provide tighter constraints on the NLR morphologies (i.e. conical versus compact).

Furthermore, accurately determining NLR sizes through imaging is heavily influenced by the filter width used in observations, which can introduce contamination from H$\beta$ and stellar continuum in the medium-band. While narrow-band imaging offers the potential to more accurately measure NLR sizes, imaging studies do not provide the means to investigate the gas kinematics of the NLR, a necessary element for identifying outflows. By providing spectral information at each pixel of the spectral image, IFU spectroscopy enables the creation of spatial, velocity, and dispersion maps, which can aid in uncovering the gas kinematics in the NLR. Ionized outflows can then be identified by the broadening of the wings in the [\ion{O}{3}] line profile. Moreover, spectroscopic observations have the advantage to more accurately constrain NLR sizes compared to imaging. This is achieved by constructing [\ion{O}{3}] maps that isolate line emission through spectral slicing, effectively avoiding contamination from continua or other emission lines.

Prior to JWST, spectroscopic observations at $z>1$ were primarily obtained by ground-based NIR spectrographs, such as the Spectrograph for INtegral Field Observations in the Near Infrared (SINFONI) and the K-band Multi Object Spectrograph (KMOS) \citep{Nesvadba2008, Kakkad2020}, which limited the wavelength ranges that could be targeted in the infrared due to atmospheric transmission windows. Promisingly, recent JWST studies have shown that the IFU spectrograph, NIRSpec, is able to reveal ionized outflows at cosmic noon with unprecedented resolution and sensitivity. One such study revealed a quasar-driven [\ion{O}{3}] outflow with a reported radial extent of $\sim13$ kpc \citep{Cresci2023}. Another study discovered biconical outflows of [\ion{O}{3}] gas in a powerful, obscured quasar at $z=2.94$ exhibiting strong evidence of AGN feedback \citep{Wylezalek2022, Vayner2023, Vayner2024}. Follow-up IFU spectroscopic observations with NIRSpec can provide the kinematic information necessary to search for outflows in our observed NLR sample, complementing the imaging methods that we have validated for NIRCam with this work.

\section{Conclusions} \label{sec:conclusion}
In this paper, we presented simulated and observed NIRCam images of the [\ion{O}{3}] and/or [\ion{O}{3}]$+\mathrm{H}\beta$ emission in cosmic noon galaxies. We summarize our findings as follows:

\begin{enumerate}
    \item Using simulations of low-redshift ($z<0.05$) AGN with confirmed extended NLRs, we demonstrate that NIRCam, with narrow- and medium-band imaging, can resolve [\ion{O}{3}] cones in AGN host galaxies at cosmic noon ($z\sim2-3$).
    \item We use NIRCam medium-band observations targeting [\ion{O}{3}]$+\mathrm{H}\beta$ emission to search for extended NLRs in 27 previously identified AGN within the GOODS-S field at $2.4<z<3.4$. Among these, we identify six AGN with conical [\ion{O}{3}]$+\mathrm{H}\beta$ morphologies consistent with extended NLRs and four AGN with compact NLRs, resulting in a total observed NLR sample of ten.
    \item We measure the characteristic NLR sizes for our $z=2.7$ simulated sample (with and without NIRCam noise)  and observed NLR sample. In the simulated sample without noise, the median NLR radius is $3.8 \pm 0.9$ kpc. With the addition of simulated NIRCam noise, the median NLR radius is reduced to $1.3 \pm 0.6$ kpc. Our observed NLR sample possesses a median NLR radius of $1.5 \pm 0.4$ kpc, consistent with the predictions from the noise-added simulated images. 
    \item We find a positive relationship between NLR radius and AGN luminosity in both our simulated (noiseless and with noise) and observed samples. However, the correlation between NLR sizes and AGN luminosity appears shallower than the low-redshift trend reported in the literature.
    \item Using our simulated images, we investigate how observational biases—such as instrumental noise, cosmological surface brightness dimming, and medium-band imaging—impact NLR size measurements in our NIRCam observations. We find that these factors constrain the limiting surface brightness in which NLR sizes can be measured down to, leading to a bias that lowers the NLR size-AGN luminosity relation observed at low redshift by a factor of $\sim 2$ at cosmic noon.
    
\end{enumerate}

Our study demonstrates the power of JWST imaging in probing the relationship between supermassive black holes and their host galaxies through their extended NLRs. The detection of prominent extended NLRs in approximately $20\%$ of our AGN sample highlights the significant and potentially widespread impact of AGN on shaping the galactic environment of their hosts during this early epoch in cosmic history. Future spectroscopic observations focusing on the [\ion{O}{3}] emission in our observed NLR sample will provide critical insights into the role of ionized outflows in the early regulation of star formation, providing a much-needed link between the early, growing galaxies observed at cosmic noon and the quiescent ellipticals seen in the local universe.

\section{Acknowledgments} 
We thank the referee for their careful reading and constructive feedback, which has substantially improved this manuscript. NIRCam was built by a team at the University of Arizona (UofA) and Lockheed Martin's Advanced Technology Center, led by Prof. Marcia Rieke at UoA. JADES data taken under the JWST/NIRCam contract to the University of Arizona NAS5-02015. Based on observations collected at the European Southern Observatory Science Archive Facility under ESO programme(s) TP.C-094.B-0321(A), 60.A-9339(A), 095.B-0934(A), 097.D-0408(A), 097.B-0640(A), 106.21C7.001, 097.D-0408(A), 106.21C7.002, 095.B-0015(A). This research uses services or data provided by the Astro Data Lab, which is part of the Community Science and Data Center (CSDC) Program of NSF NOIRLab. NOIRLab is operated by the Association of Universities for Research in Astronomy (AURA), Inc. under a cooperative agreement with the U.S. National Science Foundation. This work is supported [in part] through the Arizona NASA Space Grant Consortium, cooperative agreement 80NSSC20M0041.

This work is based [in part] on observations made with the NASA/ESA/CSA James Webb Space Telescope. The data were obtained from the Mikulski Archive for Space Telescopes at the Space Telescope Science Institute, which is operated by the Association of Universities for Research in Astronomy, Inc., under NASA contract NAS 5-03127 for JWST. These observations are associated with programs: JADES DOI: http://dx.doi.org/10.17909/8tdj-8n28 \citep{JADES_doi}, JADES DR2 DOI: http://dx.doi.org/10.17909/z2gw-mk31 \citep{JADES_DR2_doi}, JEMS DOI: https://dx.doi.org/10.17909/fsc4-dt61 \citep{JEMS_doi}, and FRESCO DOI: http://dx.doi.org/10.17909/gdyc-7g80 \citep{FRESCO_doi}. The work of SJ and CCW is supported by NOIRLab, which is managed by the Association of Universities for Research in Astronomy (AURA) under a cooperative agreement with the National Science Foundation. CCW acknowledges funding from program JWST-GO-1963 that was provided by NASA through a grant from the Space Telescope Science Institute, which is operated by the Association of Universities for Research in Astronomy, Inc., under NASA contract NAS 5-03127.

We respectfully acknowledge the University of Arizona is on the land and territories of Indigenous peoples. Today, Arizona is home to 22 federally recognized tribes, with Tucson being home to the O’odham and the Yaqui. Committed to diversity and inclusion, the University strives to build sustainable relationships with sovereign Native Nations and Indigenous communities through education offerings, partnerships, and community service.  

\vspace{3mm}
\facilities{JWST (NIRCam), VLT (MUSE), Astro Data Lab}
\software{\texttt{numpy} \citep{harris2020}, \texttt{matplotlib} \citep{Hunter2007}, \texttt{astropy} \citep{astropy2013, astropy2018, astropy2022}, \texttt{WebbPSF} \citep{Perrin2014}, \texttt{astro-datalab} \citep{Fitzpatrick2014,Nikutta2020, Juneau2021_jupyter}}

\appendix
\section{Notes on Simulated Sample} \label{section:Appendix_A}
\begin{figure*}
        \includegraphics[width=200mm,trim={4.5cm 0 0 0},clip]{"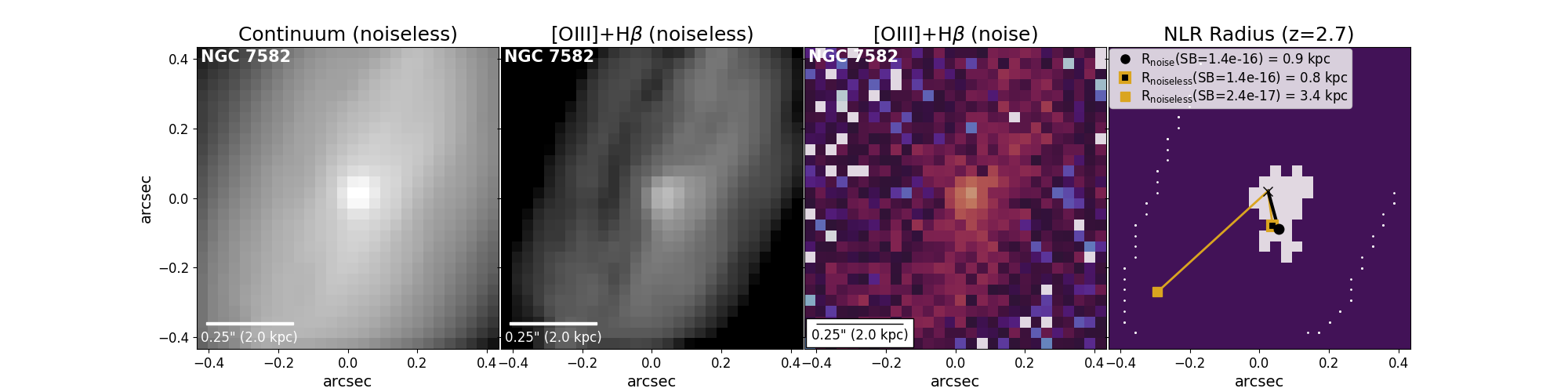"}  
        \includegraphics[width=200mm,trim={4.4cm 0 0 0},clip]{"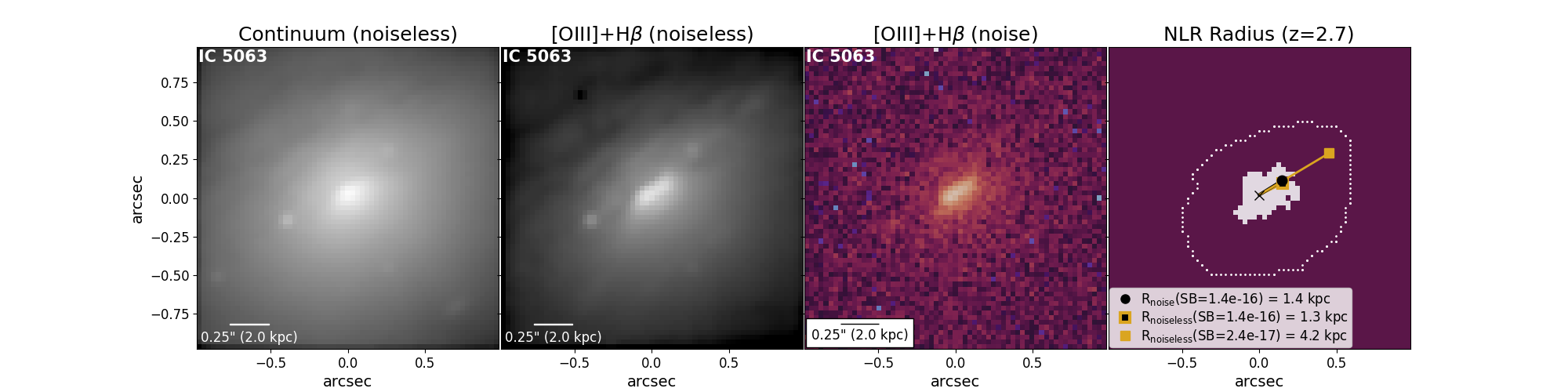"}  
        \includegraphics[width=200mm,trim={4.5cm 0 0 0},clip]{"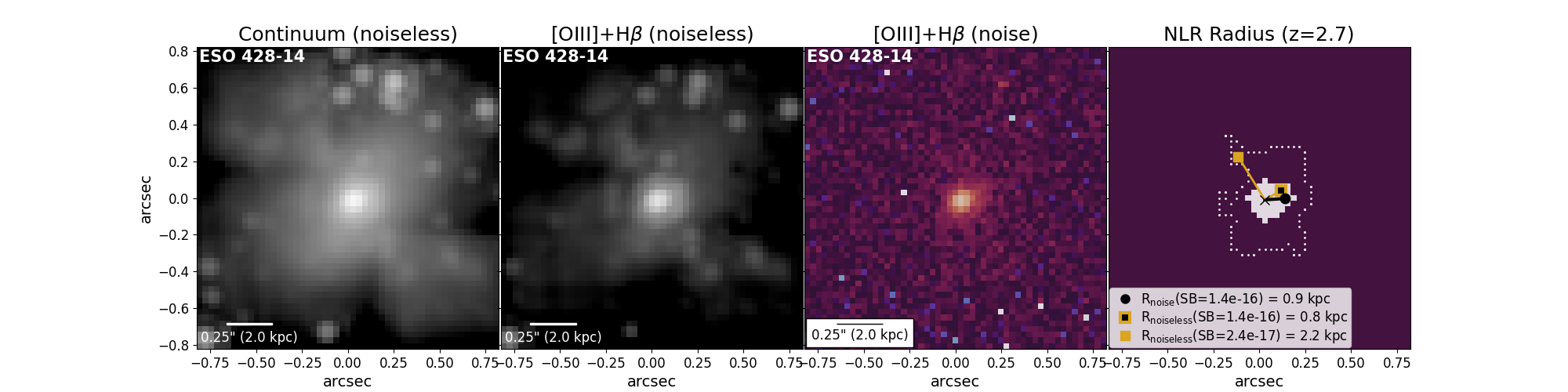"}
        \includegraphics[width=200mm,trim={4.5cm 0 0 0},clip]{"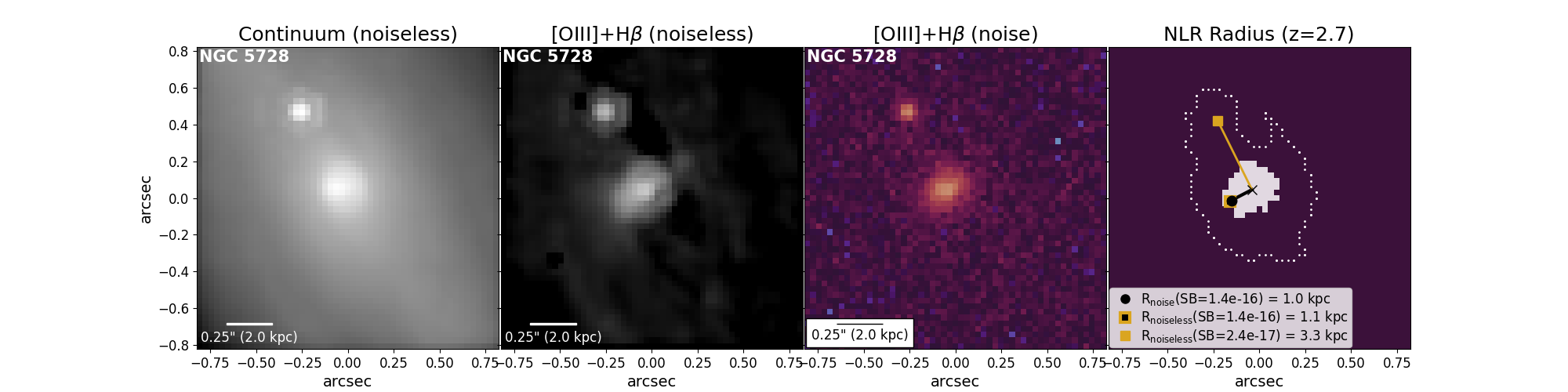"}  
        \caption{Montage of z=2.7 NIRCam simulated images of our MUSE sample. First column: continuum emission (F210M). Second column: [\ion{O}{3}]$+\mathrm{H}\beta$ map ($F182M - F210M$). Third column: [\ion{O}{3}]$+\mathrm{H}\beta$ map with simulated NIRCam noise (third column). Fourth column: Plot of the NLRs in the noise (gray area) and noiseless (white outline) [\ion{O}{3}]$+\mathrm{H}\beta$ images. The circular NLR radius represented with the dashed blue circle and the average NLR radius is shown by the black circle for the noise image. The average NLR radius for the noiseless image is shown by the yellow square.}
        \label{fig:Montage_a}
\end{figure*}

\begin{figure*}
        \includegraphics[width=200mm,trim={4.5cm 0 0 0},clip]{"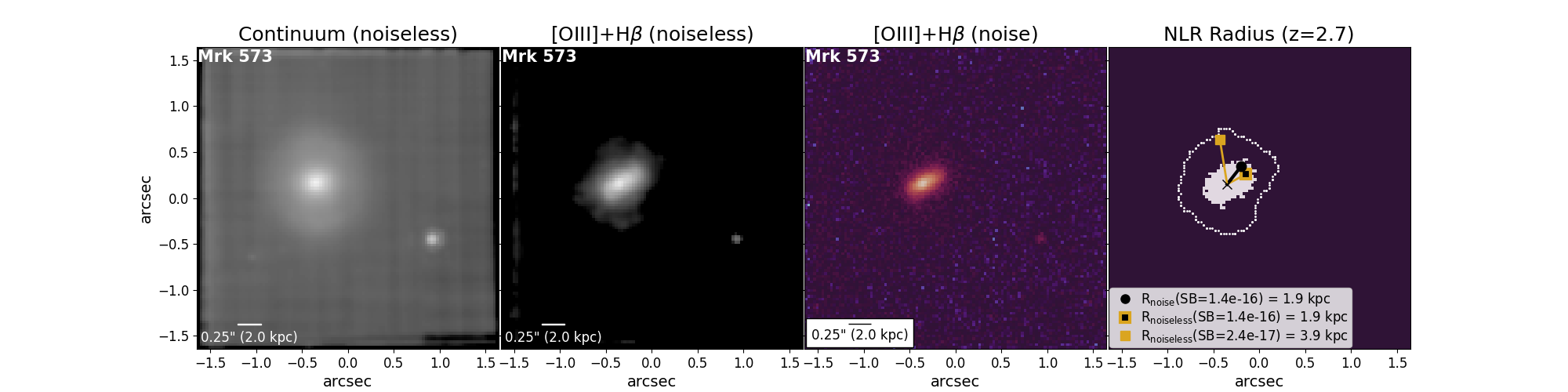"}  
        \includegraphics[width=200mm,trim={4.5cm 0 0 0},clip]{"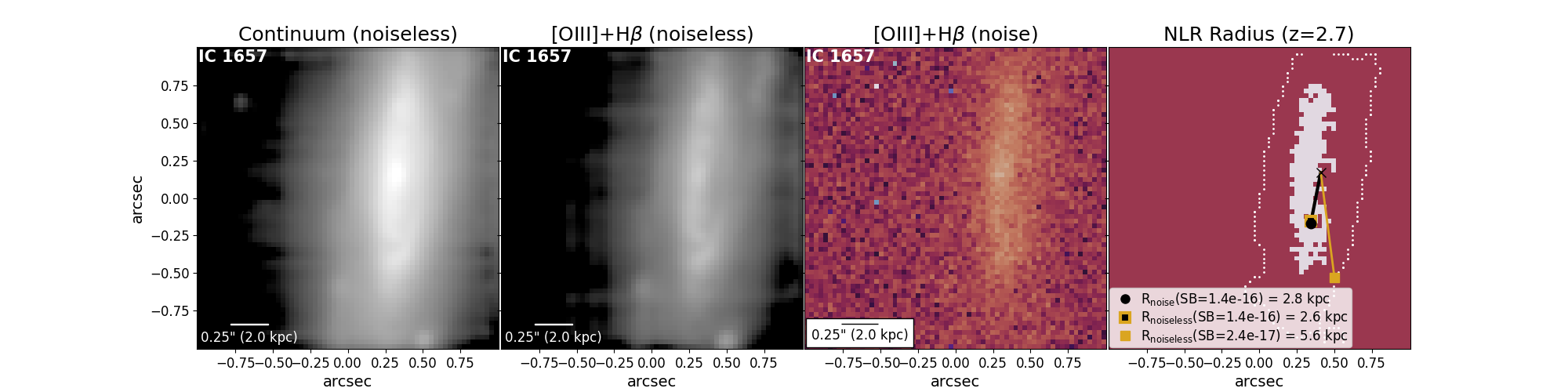"}  
        \includegraphics[width=200mm,trim={4.5cm 0 0 0},clip]{"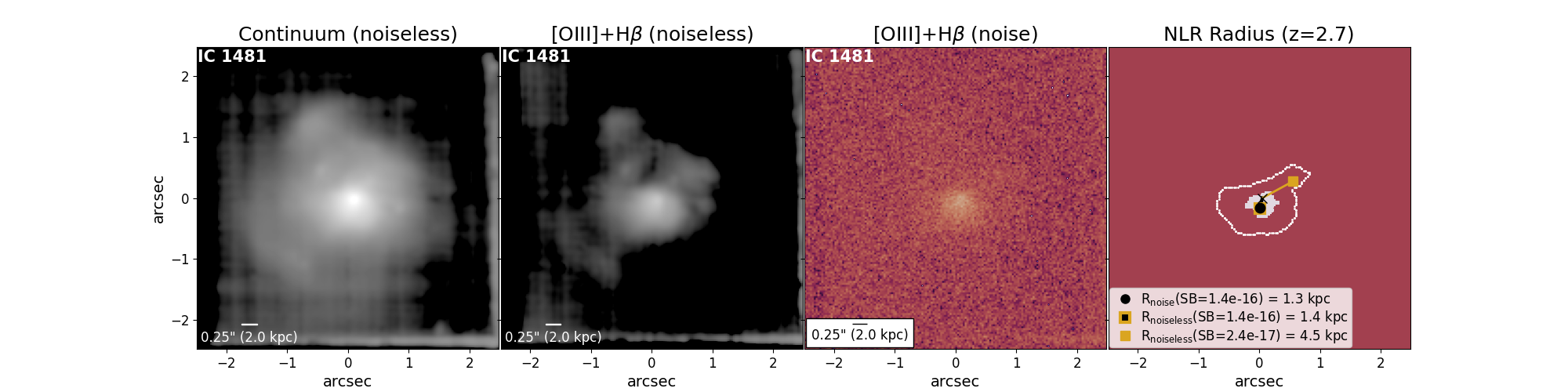"}  
        \includegraphics[width=200mm,trim={4.5cm 0 0 0},clip]{"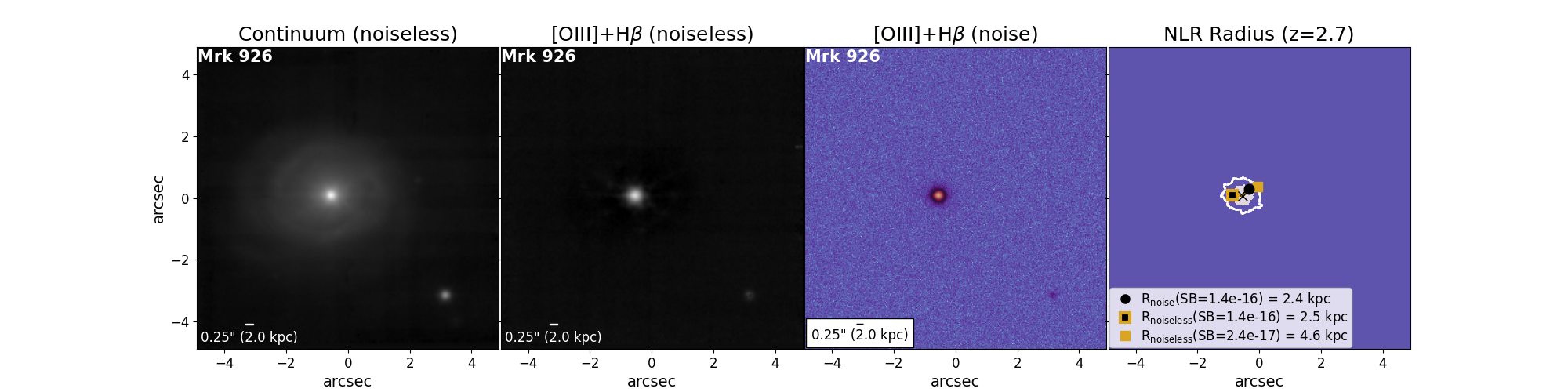"}  
        \caption{Montage of z=2.7 NIRCam simulated images of our MUSE sample. First column: continuum emission (F210M). Second column: [\ion{O}{3}]$+\mathrm{H}\beta$ map ($F182M - F210M$). Third column: [\ion{O}{3}]$+\mathrm{H}\beta$ map with simulated NIRCam noise (third column). Fourth column: Plot of the NLRs in the noise (gray area) and noiseless (white outline) [\ion{O}{3}]$+\mathrm{H}\beta$ images. The circular NLR radius represented with the dashed blue circle and the average NLR radius is shown by the black circle for the noise image. The average NLR radius for the noiseless image is shown by the yellow square.} \label{fig:Montage_b}
\end{figure*}

We report notes on individual galaxies for our simulation sample. Figures \ref{fig:Montage_a} and \ref{fig:Montage_b} shows a compilation of nearby AGN host galaxies simulated at $z=2.7$.  Each row shows the continuum emission in F210M (first column), the [\ion{O}{3}]$+\mathrm{H}\beta$ map without simulated noise (second column), the [\ion{O}{3}]$+\mathrm{H}\beta$ map with simulated instrument noise (third column), and our measurements of the NLR in the [\ion{O}{3}]$+\mathrm{H}\beta$ images with noise (gray region) and without noise (dashed white outline), (fourth column). The average NLR radii measured from the noisy [\ion{O}{3}]$+\mathrm{H}\beta$ maps are shown by the black circles (SB limit $=1.4x10^{-16}$ erg s$^{-1}$ cm$^{-2}$ arcsec$^{-2}$). The average NLR radii measured from the noiseless [\ion{O}{3}]$+\mathrm{H}\beta$ maps are shown by the yellow squares (SB limit $=2.4x10^{-17}$ erg s$^{-1}$ cm$^{-2}$ arcsec$^{-2}$). We also plot the NLR radii measured from the noiseless [\ion{O}{3}]$+\mathrm{H}\beta$ map using a limiting surface brightness level of $1.4x10^{-16}$ erg s$^{-1}$ cm$^{-2}$ arcsec$^{-2}$ (shown by the black squares with yellow outlines).

\textbf{NGC 7582} ($z=0.0053$) is a nearby spiral galaxy classified as a Type 2 Seyfert, with an obscured, Compton-thick AGN and bolometric AGN luminosity of 43.16 erg s$^{-1}$ \citep{Riffel2009}. NGC 7582 is reported to host a diffuse radio structure, ionization bicone, and conical [\ion{O}{3}] outflows \citep{Juneau}. The $z=2.7$ simulated image reveals one ionization cone, but not the back cone reported in \cite{Juneau}. The [\ion{O}{3}]$+\mathrm{H}\beta$ emission appears to dominate in the cone. The cone extends from the center with an average radius of $3.4$ kpc in the noiseless [\ion{O}{3}]$+\mathrm{H}\beta$ image (second column). This radius is reduced to $0.9$ kpc in the [\ion{O}{3}]$+\mathrm{H}\beta$ noise image (third column). 

\textbf{IC 5063} ($z=0.011$) is a massive, early-type Seyfert 2 with a bright bolometric AGN luminosity of 44.24 erg s$^{-1}$ \citep{Kakkad2022}. IC 5063 is also a radio-loud galaxy hosting a powerful radio jet that propagates over several hundred parsecs \citep{Morganti1998}. The $z=2.7$ simulated image shows both [\ion{O}{3}] cones, with an average radius of $4.1$ kpc in the noiseless image (second column). This radius is reduced in the noise image (third column) with an average NLR size of $1.4$ kpc. The nucleus is bright and dominated by [\ion{O}{3}] while the [\ion{O}{3}] cones appear fainter. 

\textbf{ESO 428-14} ($z=0.006$) is a Compton-thick Seyfert II galaxy with a bolometric AGN luminosity of 42.3 erg s$^{-1}$ \citep{May2018}. ESO 428-14 host a double-sided radio jet aligned in the direction of the  ionization cone reported in \cite{Falcke1998, Lopez-coba}. The $z=2.7$ NIRCam simulation shows a compact region of [\ion{O}{3}]$+\mathrm{H}\beta$ gas concentrated in the central region of the galaxy, however the ionized gas structure shown in \cite{Lopez-coba} is not resolved. The spiral pattern is an artifact from the convolution with WebbPSF. The average NLR radius of the noiseless [\ion{O}{3}]$+\mathrm{H}\beta$ image (second column) is $2.2$ kpc and the average NLR radius of the noise [\ion{O}{3}]$+\mathrm{H}\beta$ image (third column) is $0.9$ kpc.
 
\textbf{NGC 5728} ($z=0.009$) is a Seyfert 2 with an obscured AGN and a bolometric AGN luminosity of 44.16 erg s$^{-1}$ \citep{Durre&Mould2019}. NGC 5729 is reported to host an ionization bicone and radio jet extending in the NW-SE direction \cite{Durre&Mould2018}. The [\ion{O}{3}] bicone is visible in the $z=2.7$ simulated image. The front cone is larger and brighter than the back cone with an average NLR radius of $3.3$ kpc in the noiseless image (second column). However, this radius is likely overstated due to contamination by the foreground star in the NLR approximation. The object shows a more compact morphology in the noise image (third column) with an average NLR radius of $1.0$ kpc. The simulated image of the [\ion{O}{3}] cones is consistent with the MUSE image of [\ion{O}{3}] shown in \cite{Lopez-coba}.

\textbf{Mrk 573} ($z=0.017$) is a nearby Compton-thick Seyfert galaxy with a radio source and bolometric AGN luminosity of 44.71 erg s$^{-1}$ \citep{Falcke1998, Gonzalez-Martin}. Observations of Mrk 573 have revealed a [\ion{O}{3}] bicone \cite{Gonzalez-Martin} The $z=2.7$ simulated image (second column) shows both the front and back cones extending SE and NW with an average NLR radius of $3.6$ kpc. The [\ion{O}{3}]$+\mathrm{H}\beta$ emission appears more compact in the noise image (third column) with an average NLR radius of $1.9$ kpc. The NW cone appears to be slightly brighter than the SE cone and may be the front cone. The SE and NW cones are in agreement with the MUSE-imaged [\ion{O}{3}] bicone in \cite{Gonzalez-Martin}.

\textbf{IC 1657} ($z=0.012$) is a highly obscured, Compton-thick, Seyfert 2 with a bolometric AGN luminosity of 43.54 erg s$^{-1}$ \citep{Kakkad2022}. IC 1657 is reported to possess a [\ion{O}{3}] cone extending east from the plane of the disk \citep{Lopez-coba}. The $z=2.7$ simulated [\ion{O}{3}]$+\mathrm{H}\beta$ image (second column) does not reveal the [\ion{O}{3}] cone extending east from the central nucleus of the galaxy described in \cite{Lopez-coba}. The low AGN luminosity may be connected to the lack of a strong detection of the AGN ionization cones. [\ion{O}{3}]$+\mathrm{H}\beta$ emission is also concentrated in the spiral arms of the galaxy indicating low metallicity, star forming regions. The average NLR radius is $5.6$ kpc in the noiseless image (second column), however this measurement is likely dominated by [\ion{O}{3}]$+\mathrm{H}\beta$ emission originating from the spiral arms. The average NLR radius is $2.8$ kpc in the noise image (third column). 

\textbf{IC 1481} ($z=0.02$) is a spiral galaxy with a bolometric AGN luminosity of at least 43.01 erg s$^{-1}$. The bolometric AGN luminosity is calculated from the [\ion{O}{3}] luminosity reported in \cite{Keel2021} using the relation $L_{\mathrm{bol}}/L_{[\mathrm{OIII}]}$ $\simeq 3500$ for Type I AGN \citep{Lamastra2009}. This relation does not account for dust extinction in the narrow line region, making the calculated bolometric AGN luminosity a lower estimate. \cite{Keel2021} reports low-ionization nuclear emission-line region (LINER) in the nucleus and theorizes that IC 1481 may be in a post-merger state. The $z=2.7$ simulated image shows extended [\ion{O}{3}]$+\mathrm{H}\beta$ emission projecting east and west of the central nucleus, forming the east and west ionization cones discussed in \cite{Lopez-coba} and \cite{Keel2021}. The average NLR radius is $4.5$ kpc in the noiseless image (second column) and appears more compact in the noise image (third column) with an average radius of $1.3$ kpc. 

\textbf{Mrk 926} ($z=0.047$) is a highly variable AGN hosting a flat spectrum radio source with a bolometric AGN luminosity of 45.62 erg s$^{-1}$ \citep{Kollatschny2022, Kakkad2022}.  \cite{Lopez-coba} reports a small [\ion{O}{3}] cone located NE of the central nucleus. The simulated [\ion{O}{3}]$+\mathrm{H}\beta$ map of Mrk 926 reveals a central point source, but the reported ionization cone is not detected in the simulated image. This cone is likely too small to be resolved by NIRCam at $z=2.7$. The average NLR radius for the noiseless image (second column) is $4.6$ kpc. The average NLR radius for the noise image (third column) is $2.4$ kpc.

\section{Extension to Higher Redshifts} \label{subsection:Appendix_B}
\begin{figure*}
        \centering
        \includegraphics[width=180mm,trim={1cm 1cm 1cm 2cm},clip]{"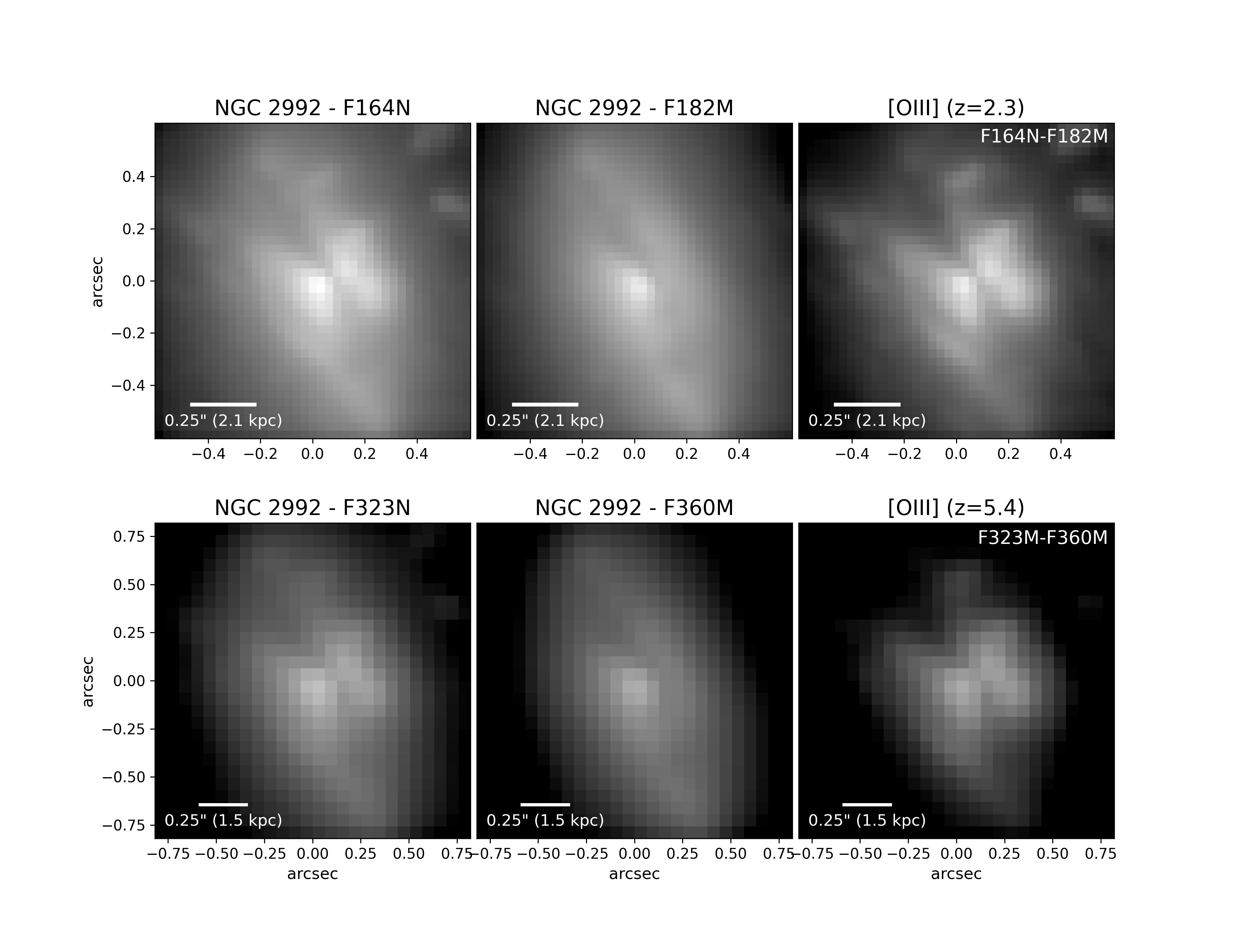"}        
        \caption{NIRCam simulated images of NGC 2992 at $z=2.3$ (top row) and $z=5.4$ (bottom row). Top row. Left: F164N ([\ion{O}{3}]), Middle: F182M (continuum), and Right: $F164N - F182M$ ([\ion{O}{3}] map). Bottom Row. Left: F323N ([\ion{O}{3}]), Middle: F360M (continuum), and Right: $F323N - F360M$ ([\ion{O}{3}] map).} \label{fig:Redshift}
\end{figure*}

Identifying AGN ionization cones at higher redshifts may provide insight into the AGN-host galaxy connection during earlier evolutionary periods in the cosmic history of a galaxy. However, cosmological surface brightness dimming, which takes the form of $(1+z)^{-4}$, poses challenges for resolving the NLR in high redshift objects as distant objects will appear fainter due to the expansion of space and curvature \citep{Calvi2014}. In this section, we extend our simulation to $z=5.4$ to test the ability of NIRCam to detect AGN ionization cones at higher redshifts. In the second row of Figure \ref{fig:Redshift}, we show simulated images of NGC 2992 at $z=5.4$ using F323N to target the [\ion{O}{3}] emission (left) and F360M to capture the continuum emission (middle). The right image shows the continuum-subtracted [\ion{O}{3}] map. In the first row, we show simulated images of NGC 2992 at $z=2.3$ for reference purposes. 

This figure shows that the [\ion{O}{3}] bicone is not as well resolved by NIRCam at $z=5.4$ compared to the simulated $z=2.3$ image, with the [\ion{O}{3}] emission appearing much less extended at $z=5.4$. The spatial resolution of the the object also decreases as the redshift increases. At $z=2.3$, 0.25$^{\prime\prime}=2.1$ kpc while at $z=5.4$, 0.25$^{\prime\prime}=1.5$ kpc. While the simulated $z=5.4$ image appears fainter and more blurred compared to its $z=2.3$ counterpart, the [\ion{O}{3}] bicone is still detected at this redshift, indicating that NIRCam could potentially resolve AGN ionization cones at redshifts above that of our NIRCam observations ($z>3.4$). However, we note that this simulated image is an optimistic image as it does not include simulated NIRCam noise. The reduction in image detail and the apparent size of the bicone at $z=5.4$ aligns with expectations from cosmological dimming and is further exacerbated by the smaller visual angle subtended by distant objects on a telescope's field of view at higher redshifts. In the next section, we discuss the importance of searching for extended NLRs at high redshift, particularly at cosmic noon.

\section{Notes on Observed Sample} \label{section:Appendix_C}
\begin{figure*}
        \centering
        \includegraphics[width=160mm,height=38mm,trim={0 1.5cm 0 2cm},clip]{"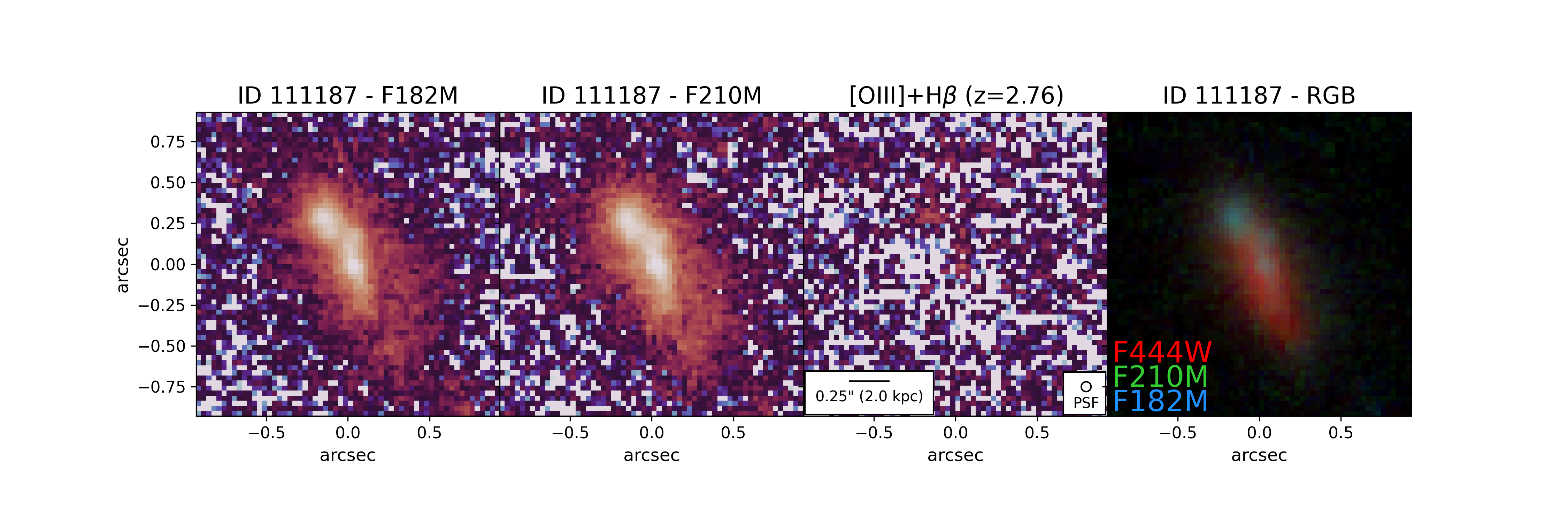"}      
        \includegraphics[width=160mm,height=38mm,trim={0 1.5cm 0 2cm},clip]{"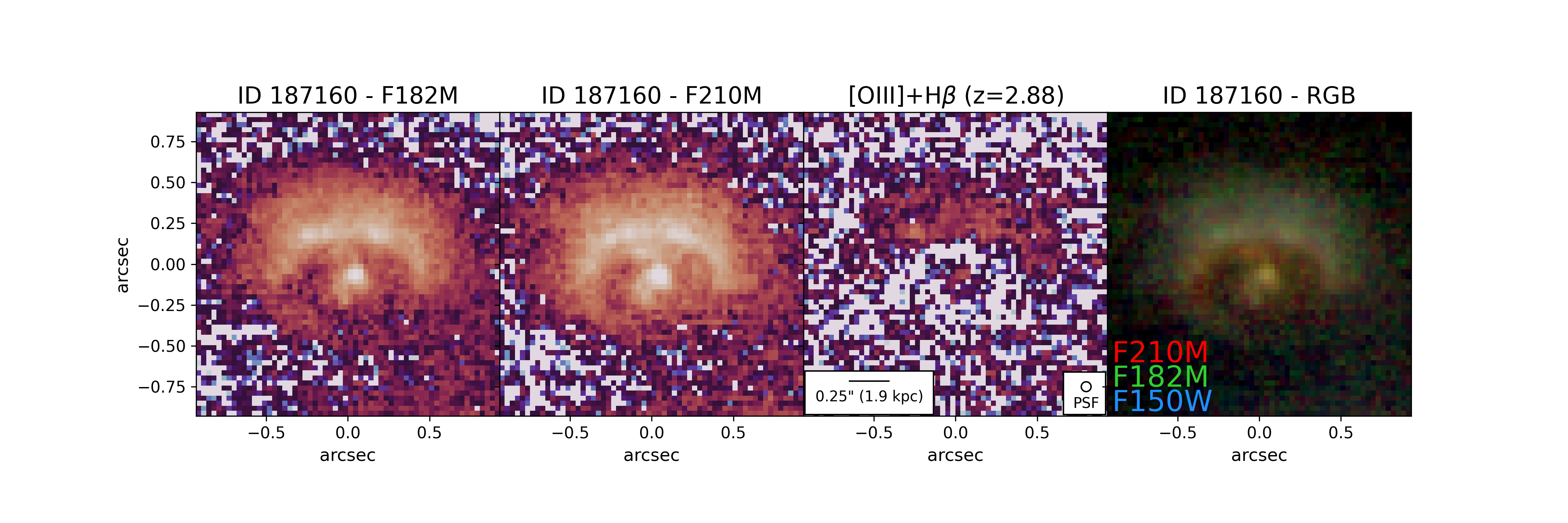"}     
        \includegraphics[width=160mm,height=38mm,trim={0 1.5cm 0 2cm},clip]{"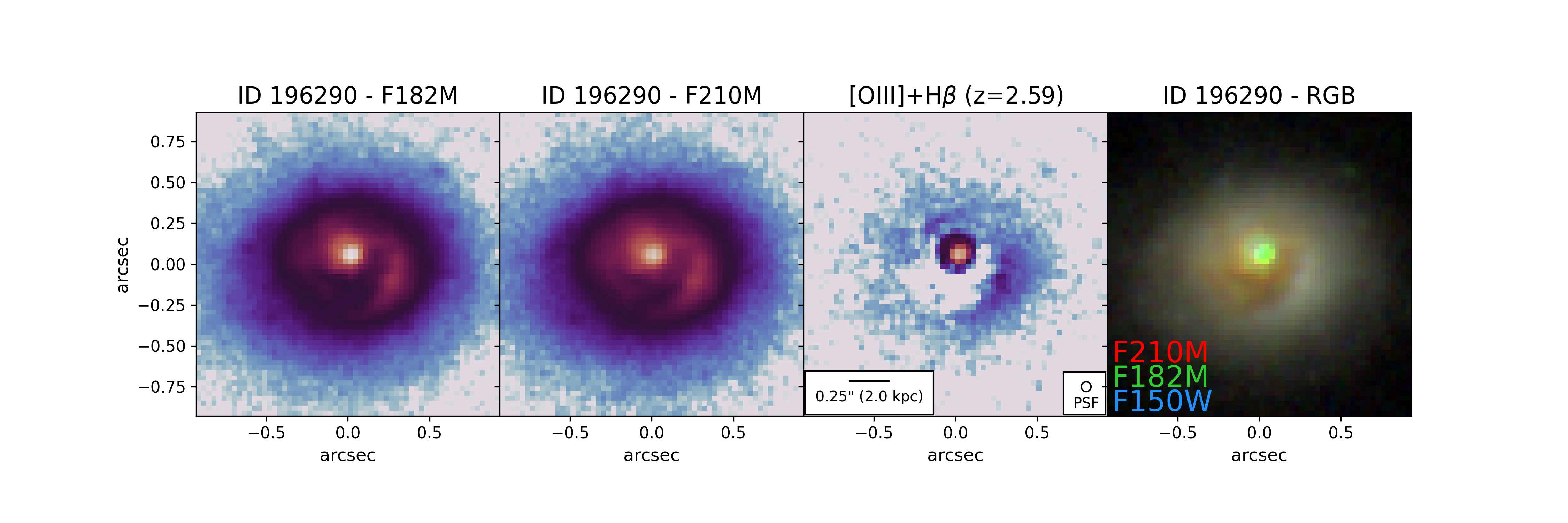"} 
        \includegraphics[width=160mm,height=38mm,trim={0 1.5cm 0 2cm},clip]{"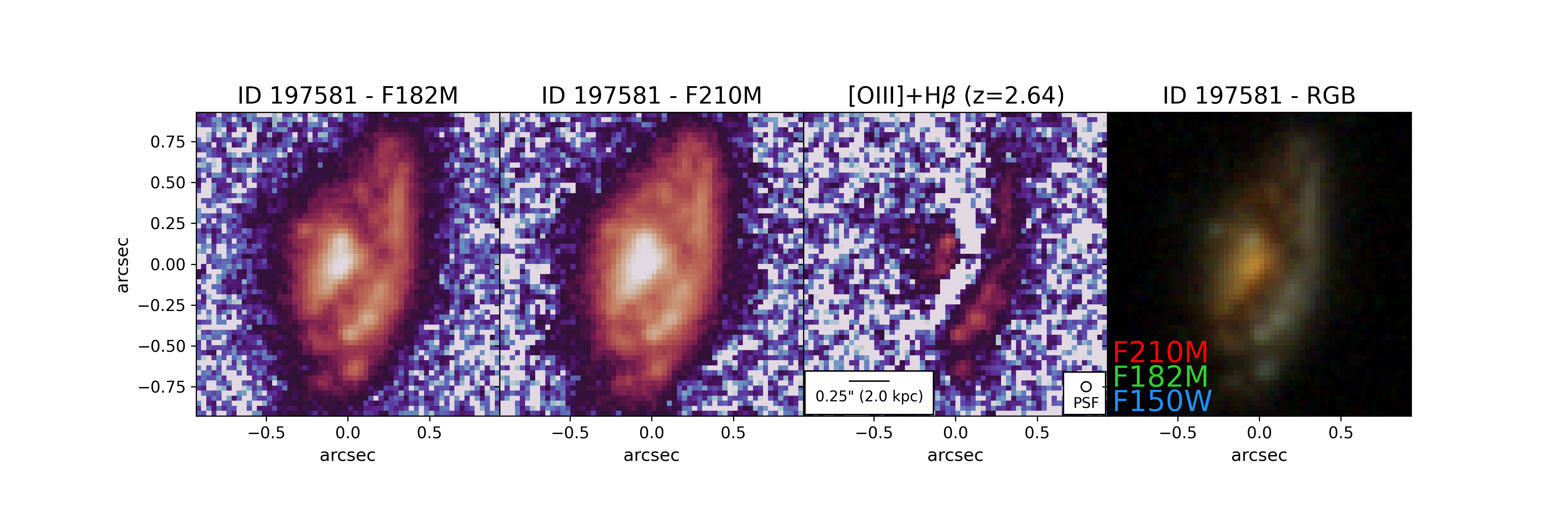"}
        \includegraphics[width=160mm,height=38mm,trim={0 1.5cm 0 2cm},clip]{"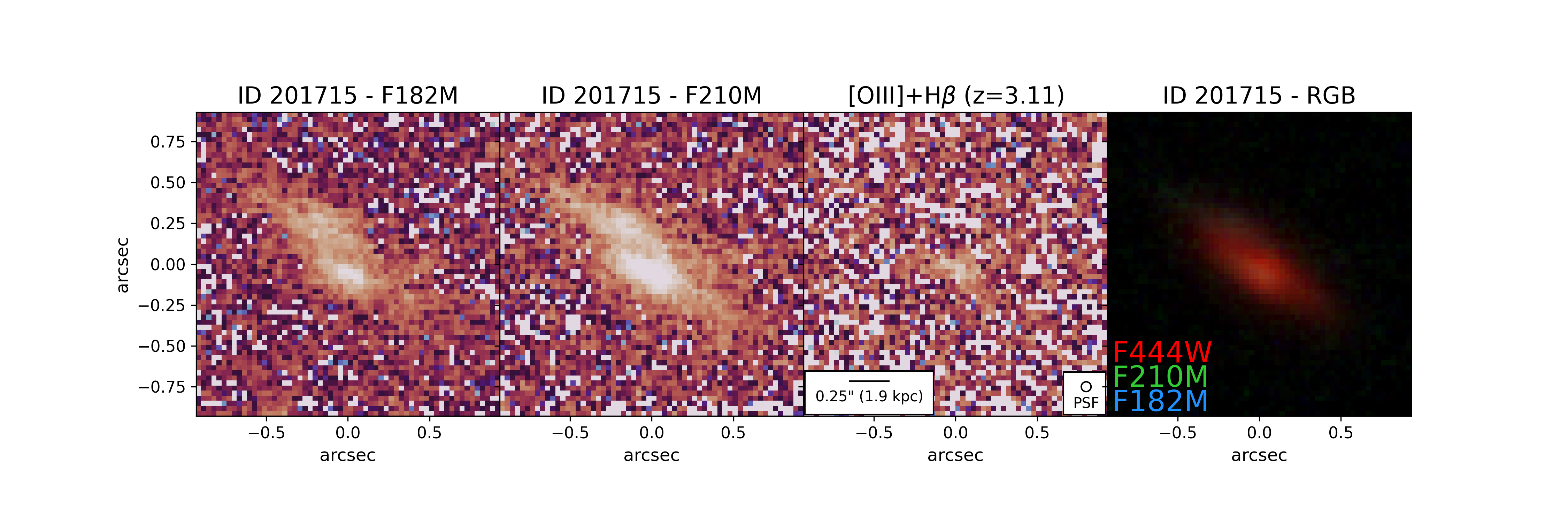"}
        \includegraphics[width=160mm,height=38mm,trim={0 1.5cm 0 2cm},clip]{"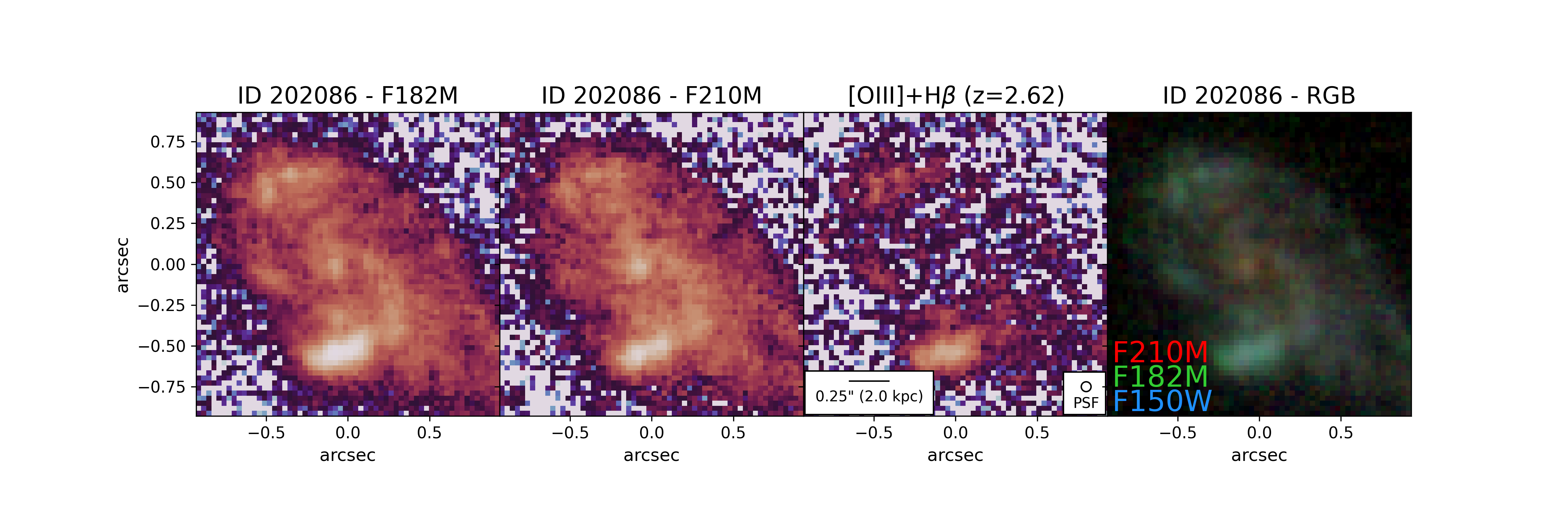"}
        \caption{Montage of NIRCam objects with [\ion{O}{3}]$+\mathrm{H}\beta$ from SF regions such as rings or spiral arms.} \label{fig:Extended_a}
\end{figure*}

\begin{figure*}   
        \centering
        \includegraphics[width=160mm,height=38mm,trim={0 1.5cm 0 2cm},clip]{"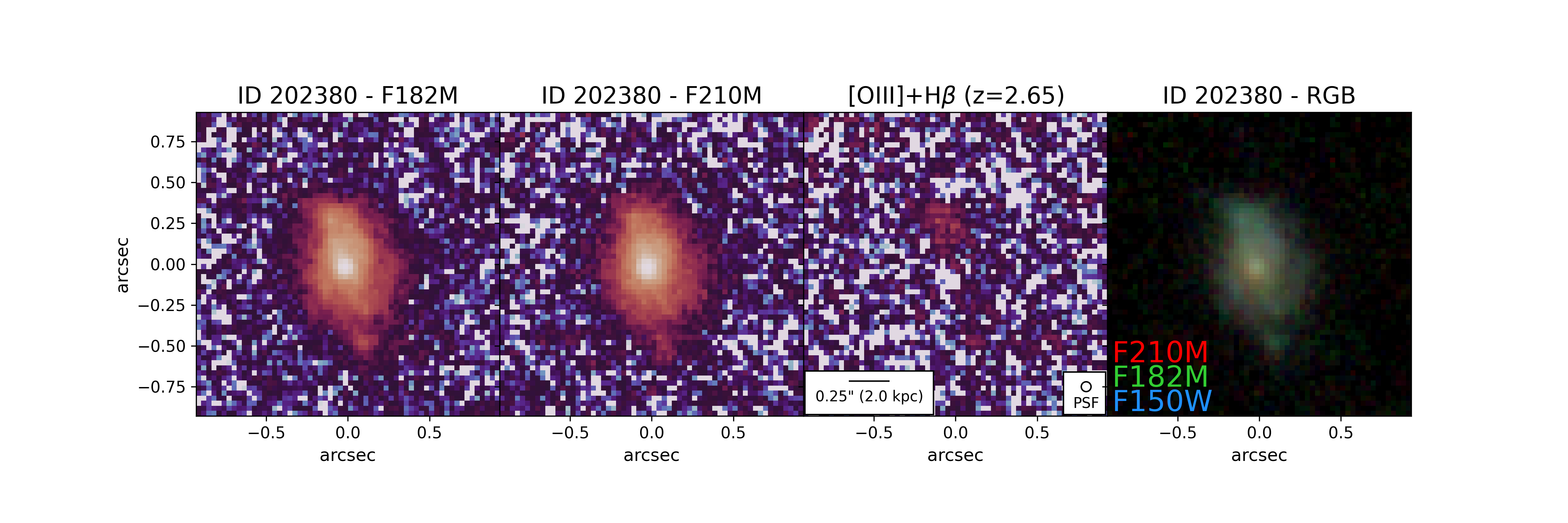"}      
        \includegraphics[width=160mm,height=38mm,trim={0 1.5cm 0 2cm},clip]{"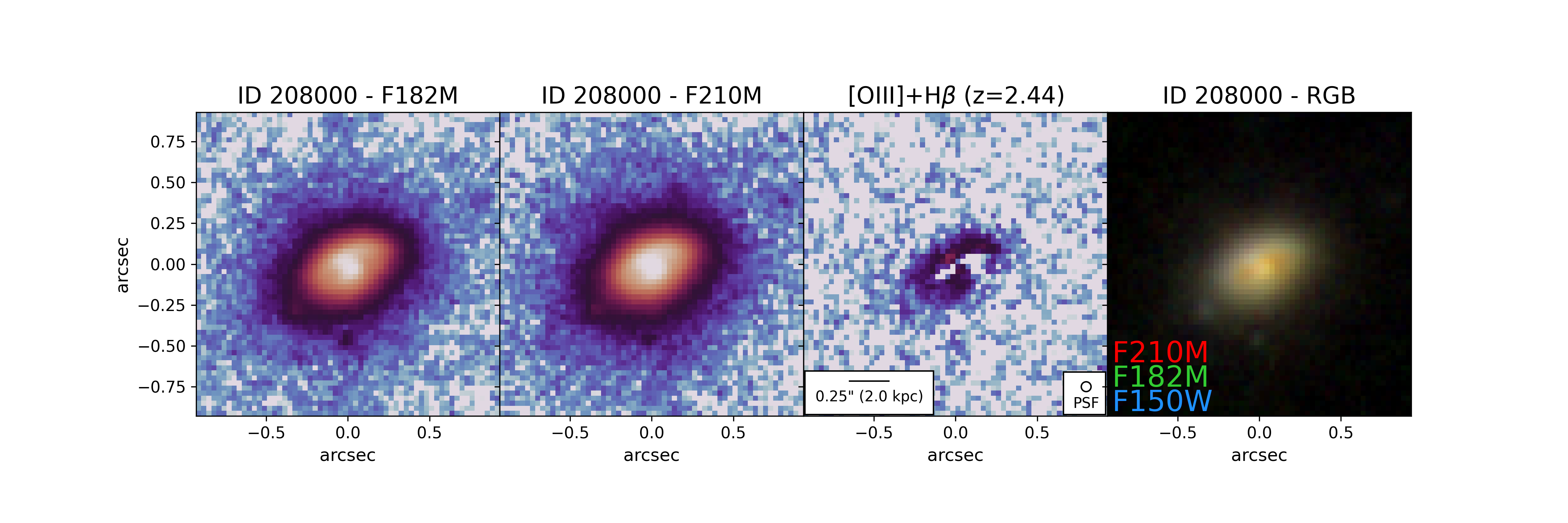"}
        \includegraphics[width=160mm,height=38mm,trim={0 1.5cm 0 2cm},clip]{"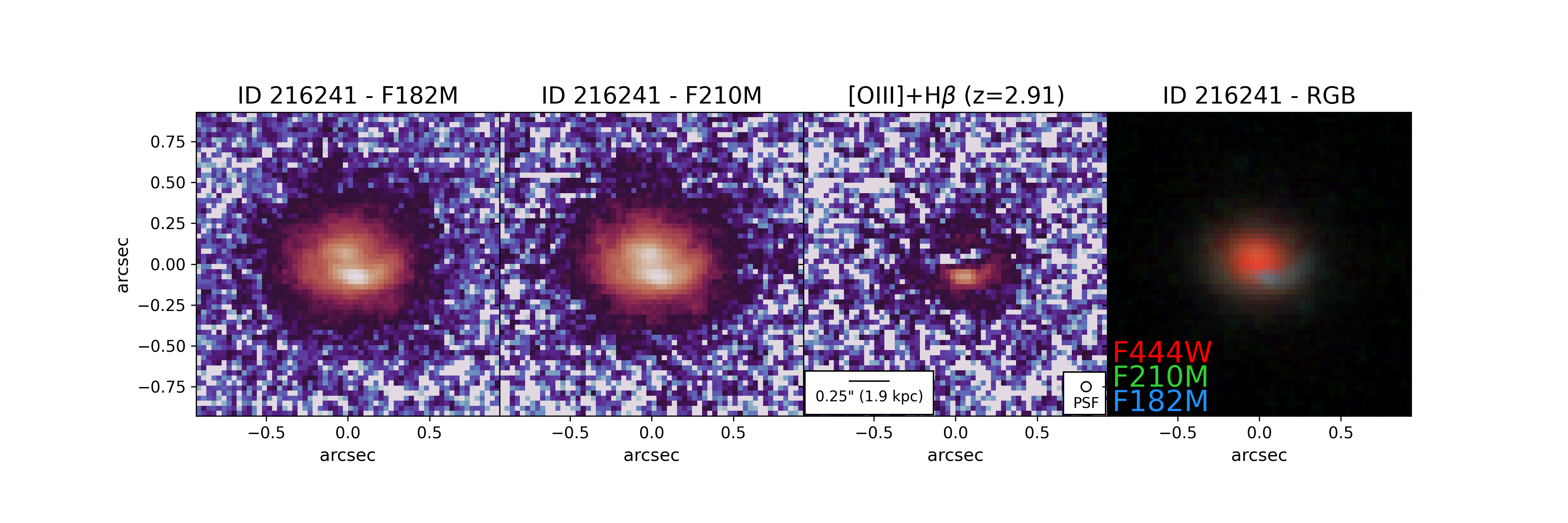"}
        \includegraphics[width=160mm,height=38mm,trim={0 1.5cm 0 2cm},clip]{"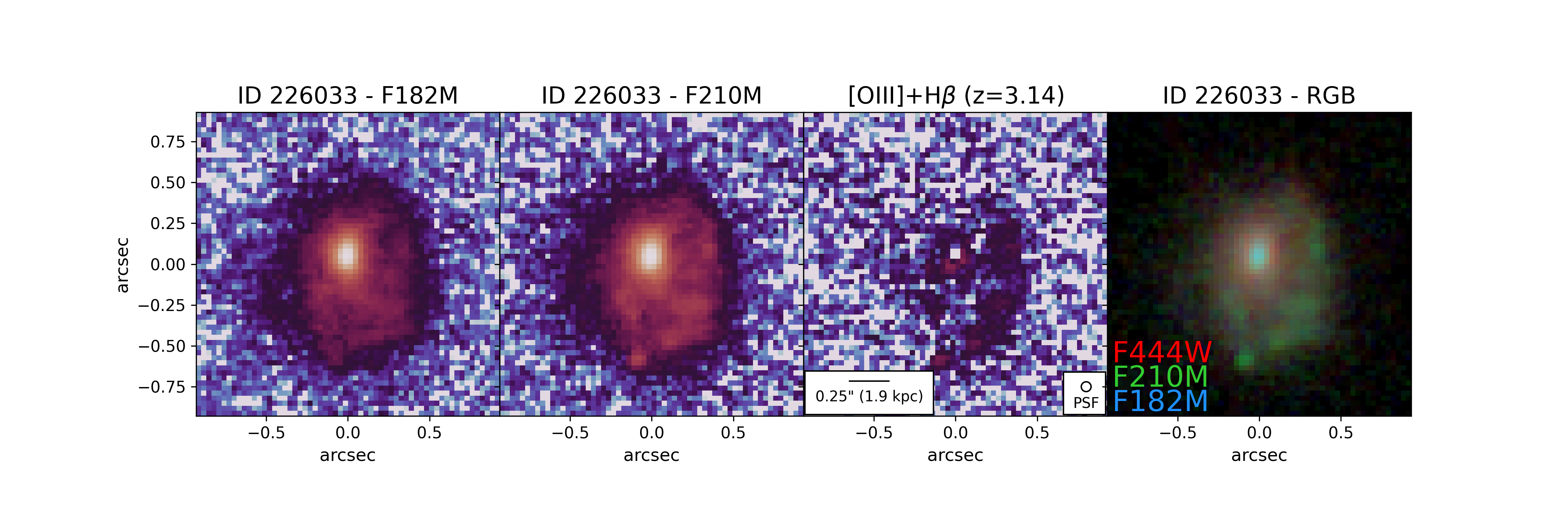"}     
        \includegraphics[width=160mm,height=38mm,trim={0 1.5cm 0 2cm},clip]{"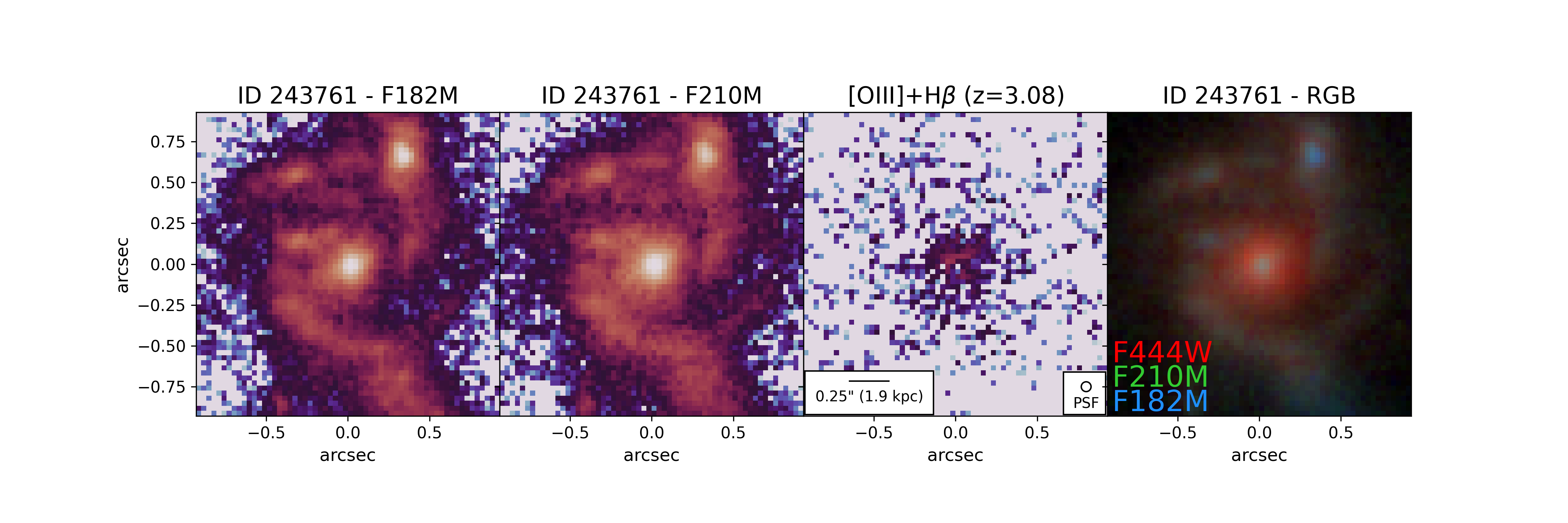"}
        \caption{Montage of NIRCam objects with [\ion{O}{3}]$+\mathrm{H}\beta$ from SF regions such as rings or spiral arms.} \label{fig:Extended_b}
\end{figure*}

\begin{figure*}[p]
        \centering
        \includegraphics[width=160mm,height=38mm,trim={0 1.5cm 0 2cm},clip]{"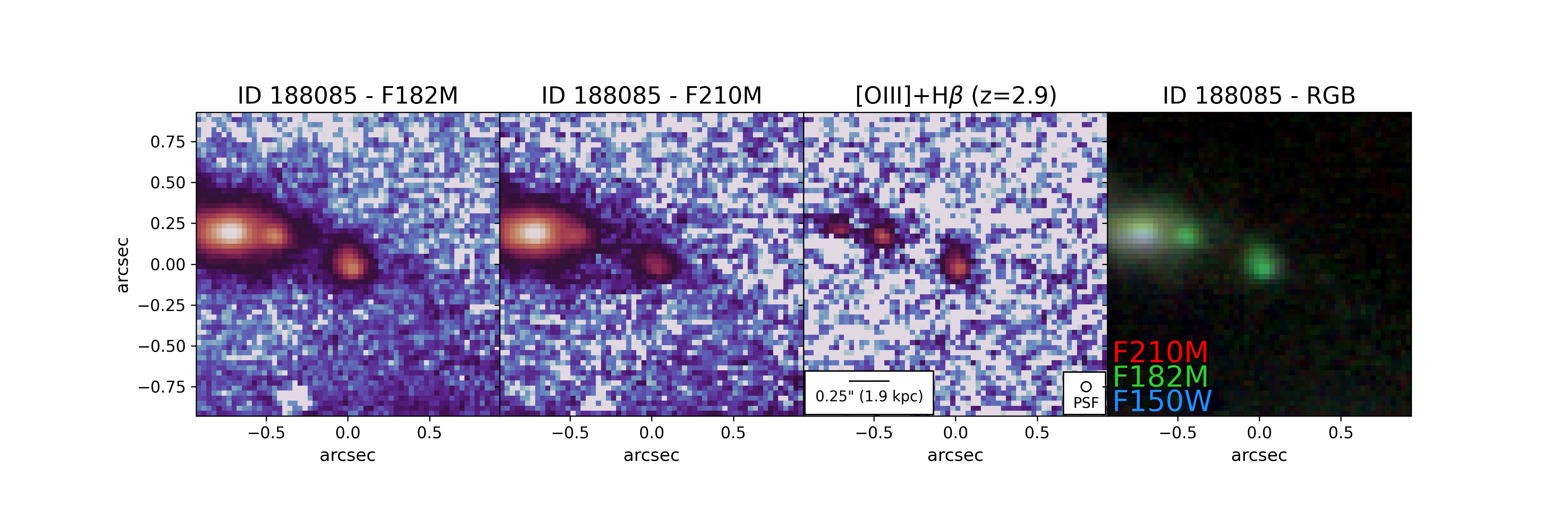"}  
        \includegraphics[width=160mm,height=38mm,trim={0 1.5cm 0 2cm},clip]{"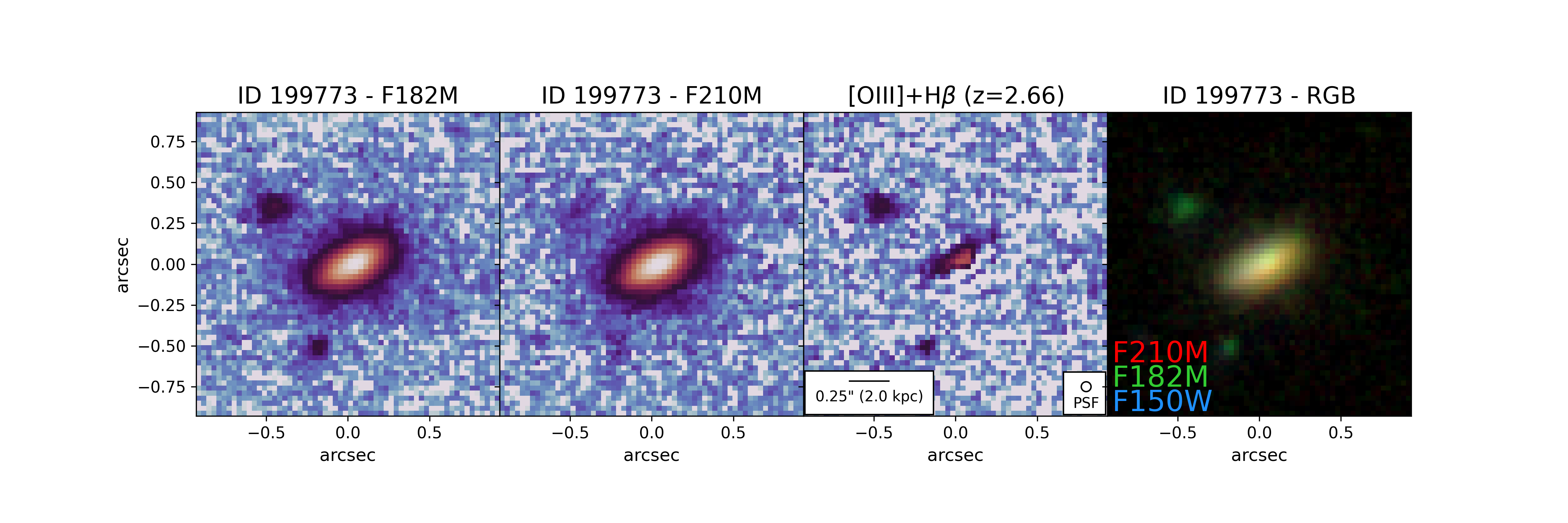"}
        \includegraphics[width=160mm,height=38mm,trim={0 1.5cm 0 2cm},clip]{"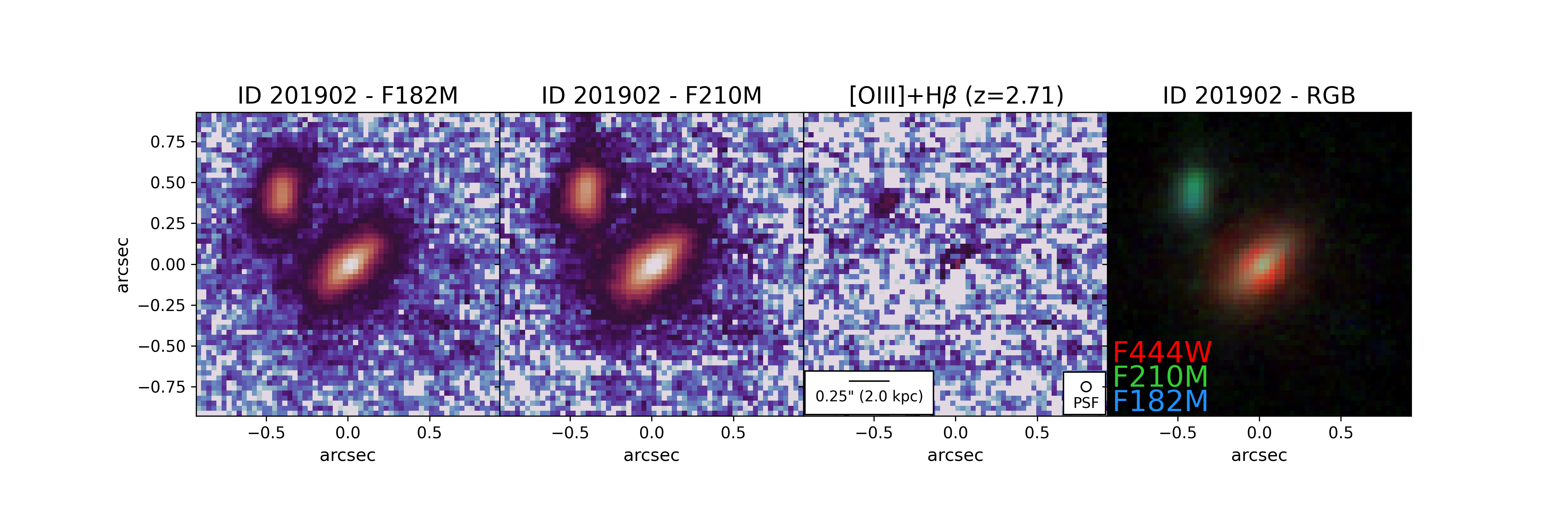"}
        \caption{Montage of NIRCam objects with possible companions at similar redshifts.} \label{fig:Companion}
        \includegraphics[width=160mm,height=38mm,trim={0 1.5cm 0 2cm},clip]{"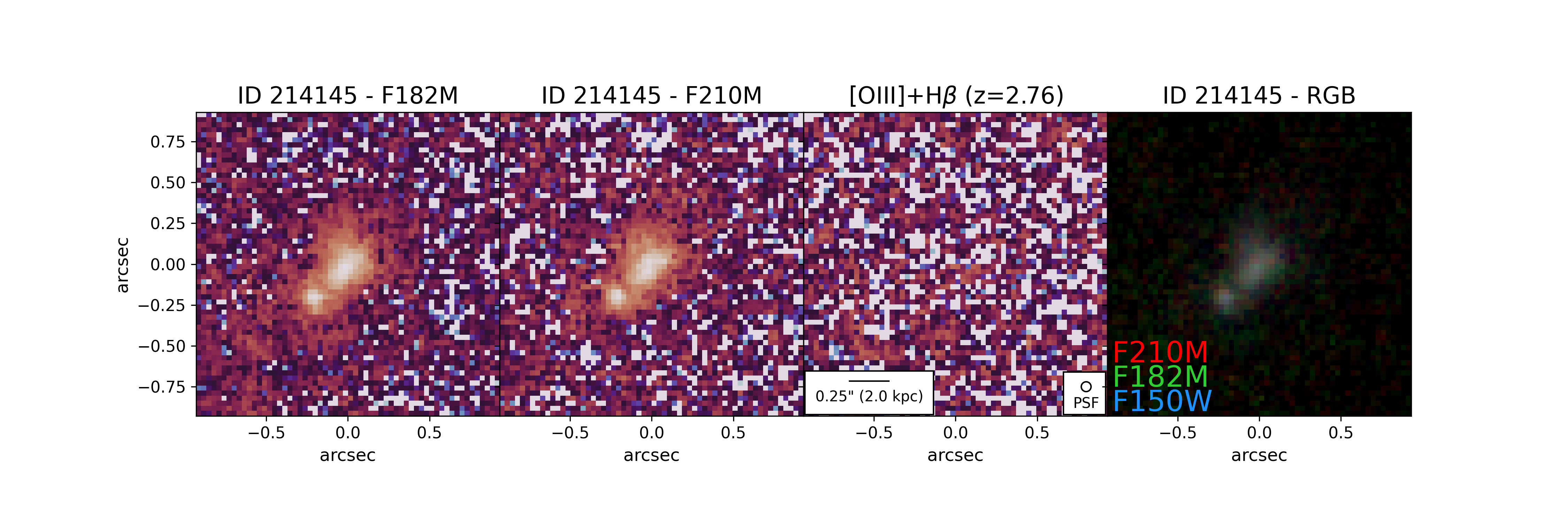"}  
        \includegraphics[width=160mm,height=38mm,trim={0 1.5cm 0 2cm},clip]{"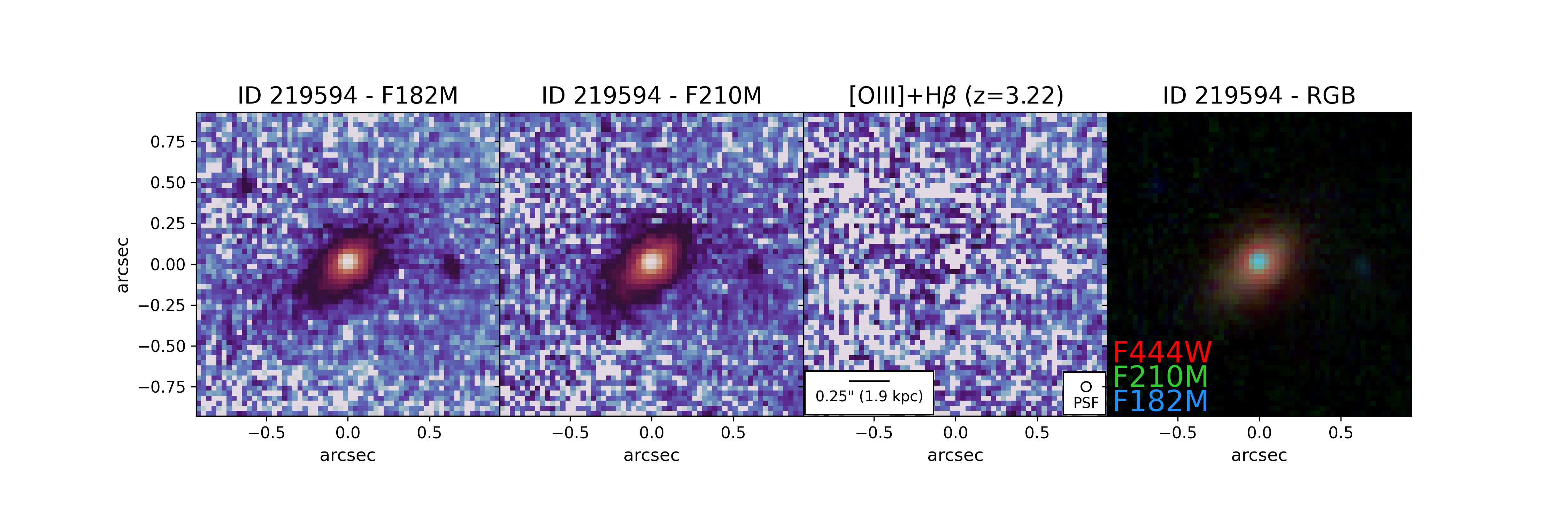"} 
        \includegraphics[width=160mm,height=38mm,trim={0 1.5cm 0 2cm},clip]{"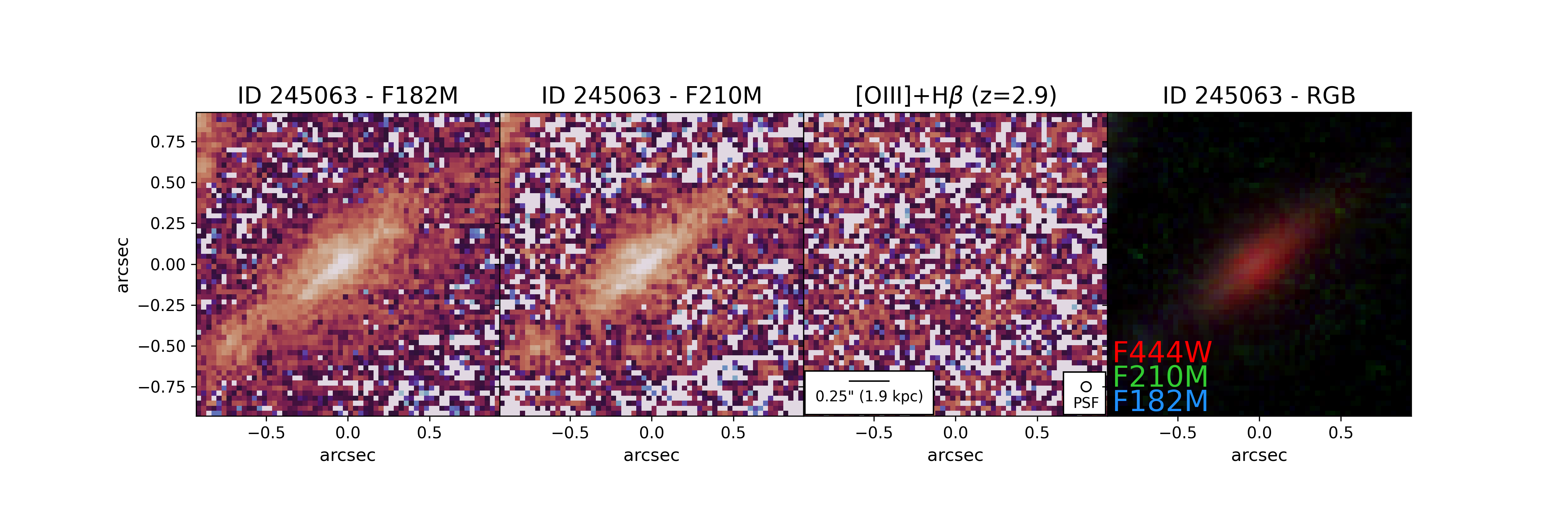"}
        \caption{Montage of NIRCam objects with no detectable [\ion{O}{3}].} \label{fig:No OIII}
\end{figure*}

We report notes on individual objects for our observed sample. We categorize our objects by their [\ion{O}{3}]$+\mathrm{H}\beta$ morphology as extended (Section \ref{subsubsection:extended}), compact (Section \ref{subsubsection:compact}), or no [\ion{O}{3}]$+\mathrm{H}\beta$ detection (Section \ref{subsubsection:no_OIII}). Figures \ref{fig:Extended_a} and \ref{fig:Extended_b} show NIRCam objects with extended [\ion{O}{3}]$+\mathrm{H}\beta$ emission consistent with star-forming regions or spiral arms. Figure \ref{fig:Companion} shows objects with potential companion galaxies or other ionized features. Figure \ref{fig:No OIII} shows three examples of NIRCam objects with no detectable [\ion{O}{3}]$+\mathrm{H}\beta$. 

\subsubsection{Extended [\ion{O}{3}]$+\mathrm{H}\beta$ Emission}\label{subsubsection:extended}

\textbf{Object 144830} (R.A.:53.108582, Dec:-27.757321, $z_{\mathrm{phot}}=2.9$) is characterized as an X-ray source with a highly obscured, Compton-thick AGN \citep{Luo2017, Li2019, Straatman2016}. The calculated average NLR radius is $1.1$ kpc and max NLR radius is $1.9$ kpc. The [\ion{O}{3}]$+\mathrm{H}\beta$ image (3rd column) shows [\ion{O}{3}]$+\mathrm{H}\beta$ emission extending south from the central region which may indicate a southern [\ion{O}{3}] cone.  

\textbf{Object 182930} (R.A.:53.161524, Dec:-27.85607, $z_{\mathrm{phot}}=3.04$ , $z_{spec}=3.03$ \citep{McLure2017}) is classified as a X-ray source, quasar, and obscured AGN hosting a Ly-$\alpha$ halo and a population of young stars at the same photometric redshift as the quasar \citep{Hsu2014, Rauch2013}. This young stellar population in the vicinity of the AGN may point towards positive AGN feedback occuring in this system \citep{Rauch2013}. The [\ion{O}{3}]$+\mathrm{H}\beta$ image (3rd column) shows biconical morphology with [\ion{O}{3}]$+\mathrm{H}\beta$ emission extending $\sim1$ kpc NE and $\sim2$ kpc SW of the nucleus. The calculated average NLR radius is $1.4$ kpc and max NLR radius is $2.3$ kpc. 

\textbf{Object 188852} (R.A.:53.173974, Dec:-27.843388, $z_{\mathrm{phot}}=2.73$) is classified as an X-ray source and AGN \citep{Luo2017}. The [\ion{O}{3}]$+\mathrm{H}\beta$ image (3rd column) shows possible bicone morphology extending approximately $1$ kpc SE and $2$ kpc NW from the central AGN. The measured average NLR radius is $2.2$ kpc and max NLR radius is $3.5$ kpc. 

\textbf{Object 207632} (R.A.:53.134037, Dec:-27.781039, $z_{\mathrm{phot}}=2.57$) is classified as an X-ray source with a highly obscured, Compton-thick AGN \citep{Hsu2014, Luo2017, Li2019}. The [\ion{O}{3}]$+\mathrm{H}\beta$ image (3rd column) shows [\ion{O}{3}]$+\mathrm{H}\beta$ emission extending approximately $2$ kpc north and $1$ kpc south of the central AGN indicating possible biconical morphology. The average NLR radius is $1.5$ kpc and the max NLR radius is $2.4$ kpc.  

\textbf{Object 209117} (R.A.:53.160609, Dec:-27.776249, $z_{\mathrm{phot}}=2.73$) is classified as an X-ray source, AGN, and faint radio source \citep{Hsu2014, Xue2011, Alberts2020} The [\ion{O}{3}]$+\mathrm{H}\beta$ image (3rd column) shows a bright clump of [\ion{O}{3}]$+\mathrm{H}\beta$ emission with some [\ion{O}{3}]$+\mathrm{H}\beta$ extending $\sim2$ kpc in the NW direction. This extended [\ion{O}{3}]$+\mathrm{H}\beta$ emission corresponds to the green in the RGB image (right) which may indicate a singular cone. The average NLR radius is $2.5$ kpc and the max NLR radius is $3.6$ kpc.  

\textbf{Object 217830} (R.A.:53.162737, Dec:-27.744344, $z_{\mathrm{phot}}=2.61$) is classified as an X-ray source with an obscured, Compton-thick AGN \citep{Hsu2014, Li2019, Xue2011}. The [\ion{O}{3}]$+\mathrm{H}\beta$ image (3rd column) and RGB image (right) show obvious bicone morphology extending $\sim 2$ kpc left and right of the center. The average NLR radius is $2.2$ kpc and the max NLR radius is $3.2$ kpc. 

\textbf{Object 111187} (R.A.: 53.092843, Dec: -27.80131, $z_{\mathrm{phot}}=2.76$, $z_{spec}=3.85$) \citep{Franco2018} is a massive star-forming galaxy and faint radio source \citep{Franco2020, Gim2019}. The [\ion{O}{3}]$+\mathrm{H}\beta$ image (3rd column) shows an small cluster of [\ion{O}{3}]$+\mathrm{H}\beta$ emission extending $4$ kpc NE to SW, aligned along the NW edge of the galaxy. The arm may trace a region of star formation. 

\textbf{Object 187160} (R.A.:53.063882, Dec:-27.843767, $z_{\mathrm{phot}}=2.88$, $z_{spec}=2.92$) \citep{Franco2018} is an X-ray source and dusty star-forming galaxy \citep{Gomez-Guijarro2022, Franco2020}. The [\ion{O}{3}]$+\mathrm{H}\beta$ image (3rd column) shows a possible star-forming spiral arm extending approximately $8$ kpc west to east, north of the bright spiral seen in the continuum image (2nd column).  

\textbf{Object 196290} (R.A.:53.148855, Dec:-27.821178, $z_{\mathrm{phot}}=2.59$, $z_{spec}=2.58$) \citep{Brusa2009} is classified as a X-ray source, faint radio source, and Compton-thick AGN \citep{Hsu2014, Alberts2020, Veron-Cetty2010, Guo2021}. The [\ion{O}{3}]$+\mathrm{H}\beta$ image (3rd column) shows a ring of [\ion{O}{3}]$+\mathrm{H}\beta$ around the central bright spot that appears to trace the spiral structure seen in the continuum image (2nd column). The [\ion{O}{3}]$+\mathrm{H}\beta$ emission in this spiral structure is likely dominated by star formation. 

\textbf{Object 197581} (R.A.:53.141682, Dec:-27.816629, $z_{\mathrm{phot}}=2.64$) is a dusty, star-forming galaxy, X-ray source, and AGN \citep{Fujimoto2018, Hsu2014, Xue2011}. The [\ion{O}{3}]$+\mathrm{H}\beta$ image (2nd column) shows the central nucleus and a possible star-forming spiral arms aligned to the right of the bright center of the galaxy in the continuum image (2nd column). The star-forming spiral arm extends approximately 8 kpc from north to south.  

\textbf{Object 201715} (R.A.:53.104026, Dec:-27.802457, $z_{\mathrm{phot}}=3.11$) is classified as a galaxy \citep{Straatman2016}. The [\ion{O}{3}]$+\mathrm{H}\beta$ image (3rd column) shows a small tail of [\ion{O}{3}]$+\mathrm{H}\beta$ emission which may indicated a region of star formation. 

\textbf{Object 202086} (R.A.:53.110846, Dec:-27.800666, $z_{\mathrm{phot}}=2.62$, $z_{spec}=2.61$) \citep{Momcheva2016} is a X-ray source \citep{Luo2017, Guo2020}. The [\ion{O}{3}]$+\mathrm{H}\beta$ image (3rd column) shows spiral-shaped clusters of [\ion{O}{3}]$+\mathrm{H}\beta$ that likely trace star-forming spiral arms. However, the patch of [\ion{O}{3}]$+\mathrm{H}\beta$ emission in the NE region of the [\ion{O}{3}]$+\mathrm{H}\beta$ image has a photometric redshift of $z=4.3$, likely originating from a background galaxy. There is also a bright region of [\ion{O}{3}]$+\mathrm{H}\beta$ approximately $2$ kpc south of the central region, possibly indicating a patch of heavy star formation. 

\textbf{Object 202380} (R.A.:53.129801, Dec:-27.799204, $z_{\mathrm{phot}}=2.65$) is classified as a X-ray source and AGN \citep{Luo2017, Guo2020}. The [\ion{O}{3}]$+\mathrm{H}\beta$ image (3rd column) shows a tail of [\ion{O}{3}]$+\mathrm{H}\beta$ extending approximately $2$ kpc north of the central region. 

\textbf{Object 208000} (R.A.:53.14616, Dec:-27.779939, $z_{\mathrm{phot}}=2.44$, $z_{spec}=2.50$) \citep{Dunlop2017} is a dusty, star-forming galaxy, X-ray source, faint radio source, and AGN \citep{Pantoni2021, Hsu2014, Alberts2020, Xue2011}. The [\ion{O}{3}]$+\mathrm{H}\beta$ image (3rd column) shows a ring of [\ion{O}{3}]$+\mathrm{H}\beta$ appearing to wrap around the galaxy's center with a diameter of $4$ kpc. 

\textbf{Object 216241} (R.A.:53.121861, Dec:-27.752775, $z_{\mathrm{phot}}=2.91$) is classified as a X-ray source, faint radio source, and AGN \citep{Hsu2014, Alberts2020, Xue2011}. \cite{Fujimoto2018} also reports Object 216241 as a dusty, star-forming galaxy. The [\ion{O}{3}]$+\mathrm{H}\beta$ image (3rd column) shows [\ion{O}{3}]$+\mathrm{H}\beta$ emission in the nucleus and along the southern edge of the galaxy extending approximately $4$ kpc. 

\textbf{Object 226033} (R.A.:53.153855, Dec:-27.840255, $z_{\mathrm{phot}}=3.14$, $z_{spec}=2.30$) \citep{Tasca2017} is classified as an X-ray source with a highly obscured, Compton-thick AGN \citep{Luo2017, Li2019}. The [\ion{O}{3}]$+\mathrm{H}\beta$ image (3rd column) shows a cluster of [\ion{O}{3}]$+\mathrm{H}\beta$ emission in the central region of the galaxy and extended in a semicircle approximately $8$ kpc in length. The extended [\ion{O}{3}]$+\mathrm{H}\beta$ emission on the edge of the galaxy may indicate a star-forming arm. 

\textbf{Object 243761} (R.A.:53.178092, Dec:-27.861805, $z_{\mathrm{phot}}=3.08$, $z_{spec}=2.52$) \citep{Straatman2016} is a X-ray source and AGN \citep{Guo2020, Luo2017}. In \cite{Ciesla2018}, Object 243761 is reported as having experienced a recent drop off in star formation with a SFR below the main sequence. The [\ion{O}{3}]$+\mathrm{H}\beta$ image (3rd column) shows scattered [\ion{O}{3}]$+\mathrm{H}\beta$ emission. There is also a second bright spot in the NW region of the galaxy viewable in both the continuum (1st column) and [\ion{O}{3}]$+\mathrm{H}\beta$ image (3rd column). This bright spot has a photometric redshift of $z=2.98$. It's location in the spiral arm indicates this bright spot may be a star-forming knot.

\subsubsection{Compact [\ion{O}{3}]} \label{subsubsection:compact}
Figure \ref{fig:Compact} and \ref{fig:Companion} shows NIRCam objects with compact [\ion{O}{3}]. Objects are classified as having compact [\ion{O}{3}$+\mathrm{H}\beta$] if their [\ion{O}{3}]$+\mathrm{H}\beta$ emission visually appears to be concentrated around a central nucleus or if the [\ion{O}{3}]$+\mathrm{H}\beta$ appears to be a point source. 

\textbf{Object 147220} (R.A.:53.108139, Dec:-27.754065, $z_{\mathrm{phot}}=2.91$, $z_{\mathrm{spec}}=2.73$) \citep{Barro2014} is a massive, compact star-forming galaxy with an X-ray luminous AGN \citep{Barro2014, Luo2017}. The [\ion{O}{3}]$+\mathrm{H}\beta$ image (3rd column) shows a point source of [\ion{O}{3}]. There is a small [\ion{O}{3}]$+\mathrm{H}\beta$ ring on the east side of the point source. The measured average NLR radius is $1.5$ kpc and the max NLR radius is $2.2$ kpc. 

\textbf{Object 194373} (R.A.:53.140062, Dec:-27.826514, $z_{\mathrm{phot}}=2.51$) is classified as a X-ray source and AGN \citep{Luo2017}. The [\ion{O}{3}]$+\mathrm{H}\beta$ image (3rd column) shows a centrally concentrated [\ion{O}{3}]$+\mathrm{H}\beta$ ring. The measured average NLR radius is $1.5$ kpc and the max NLR radius is $1.9$ kpc. 

\textbf{Object 199505} (R.A.:53.185844, Dec:-27.810029, $z_{\mathrm{phot}}=2.65$, $z_{spec}=2.58$) \citep{Balestra2010} is classified as a X-ray source and faint radio source with an obscured AGN \citep{Hsu2014, Alberts2020, Szokoly2004}. In \cite{Szokoly2004}, Object 199505 is listed as a X-ray Type 1 quasar based on its extreme X-ray luminosity. The [\ion{O}{3}]$+\mathrm{H}\beta$ image (3rd column) shows a point source of [\ion{O}{3}]. The average NLR radius is $1.8$ kpc and the max NLR radius is $2.5$.

\textbf{Object 202630} (R.A.:53.145598, Dec:-27.798975, $z_{\mathrm{phot}}=2.63$) has a J-H color consistent with a faint red galaxy and is dominated primarily by an old stellar population with a small amount of ongoing star formation \citep{Stutz2008}. The [\ion{O}{3}]$+\mathrm{H}\beta$ image (3rd column) shows scattered [\ion{O}{3}]$+\mathrm{H}\beta$ emission concentrated in the center with some [\ion{O}{3}]$+\mathrm{H}\beta$ extending $\sim 1$ kpc north and south of the nucleus. The average NLR radius is $1.4$ kpc and the max NLR radius is $2.0$. This scattered [\ion{O}{3}]$+\mathrm{H}\beta$ emission may point towards the ongoing star formation in this galaxy discussed in \cite{Stutz2008}. 

\textbf{Object 188085} (right bright spot) (R.A.:53.137716, Dec:-27.845074, $z_{\mathrm{phot}}=2.9$, $z_{\mathrm{spec}}=2.79$) and Object 188086 (left bright spot) (R.A.:53.137943, Dec:-27.845026, $z_{\mathrm{phot}}=2.86$, $z_{spec}=2.79$), \citep{Momcheva2016}, are separated by $2$ kpc in the continuum (2nd column) and [\ion{O}{3}]$+\mathrm{H}\beta$ (3rd column) images . These objects are not distinguished in the literature and are listed as one galaxy \citep{Straatman2016}. The right bright spot (Object 188085) is smaller than the left (Object 188086) and could be a separate small galaxy or an ejected gas cloud resulting from an AGN echo from Object 188086. 

\textbf{Object 199773} (R.A.:53.163241, Dec:-27.809057, $z_{\mathrm{phot}}=2.66$) is classified as a X-ray source and obscured, Compton-thick AGN \citep{Hsu2014, Li2019, Luo2017}. In \cite{Szomoru2012}, Object 199773 is listed as a quiescent galaxy based off its $U-V$ and $V-J$ colors and specific star formation rate. The [\ion{O}{3}]$+\mathrm{H}\beta$ image (3rd column) shows two [\ion{O}{3}]$+\mathrm{H}\beta$ bright spots that may form possible companions $2$ kpc NE and $2$ kpc south of the bright nucleus. These [\ion{O}{3}]$+\mathrm{H}\beta$ knots may be an artifact produced by ionizing radiation from the AGN. 

\textbf{Object 201902} (right object) (R.A.:53.093815, Dec:-27.801388, $z_{\mathrm{phot}}=2.71$) is classified as a star-forming galaxy, X-ray source, faint radio source, and AGN \citep{Luo2017, Gim2019, Xue2011}. Object 201904 (left object) (R.A.:53.093938, Dec:-27.801273, $z_{\mathrm{phot}}=2.33$) is a X-ray detected galaxy \citep{Hsu2014}. The [\ion{O}{3}]$+\mathrm{H}\beta$ image (3rd column) shows two [\ion{O}{3}]$+\mathrm{H}\beta$ spots. The NE spot is brighter and corresponds to Object 201904 and the fainter SW spot corresponds to Object 201902. These spots are separated by approximately $4$ kpc. The RGB image (right) also shows that Object 201904 is very bright in [\ion{O}{3}]$+\mathrm{H}\beta$ indicating significant AGN ionization. 

\subsubsection{Undetected [\ion{O}{3}]} \label{subsubsection:no_OIII}

\textbf{Object 214145} (R.A.:53.181181, Dec:-27.760363, $z_{\mathrm{phot}}=2.76$) is classified as a galaxy \citep{Straatman2016}. The [\ion{O}{3}]$+\mathrm{H}\beta$ image (3rd column) shows no detection. 

\textbf{Object 219594} (R.A.:53.17447, Dec:-27.733386, $z_{\mathrm{phot}}=3.22$) is a massive, compact star-forming galaxy with an X-ray luminous AGN \citep{Barro2014, Hsu2014, Sarajedini2011}. The [\ion{O}{3}]$+\mathrm{H}\beta$ image (3rd column) shows no [\ion{O}{3}]$+\mathrm{H}\beta$ emission. 

\textbf{Object 245063} (R.A.:53.082562, Dec:-27.755328, $z_{\mathrm{phot}}=2.9$, $z_{spec}=2.93$) \citep{Garilli2021} is a X-ray source and AGN \citep{Fiore2012, Xue2011}. The [\ion{O}{3}]$+\mathrm{H}\beta$ image (3rd column) shows no [\ion{O}{3}]$+\mathrm{H}\beta$ detection. 

\clearpage

\bibliography{ms}{}

\begin{thebibliography}{}
\expandafter\ifx\csname natexlab\endcsname\relax\def\natexlab#1{#1}\fi
\providecommand{\url}[1]{\href{#1}{#1}}
\providecommand{\dodoi}[1]{doi:~\href{http://doi.org/#1}{\nolinkurl{#1}}}
\providecommand{\doeprint}[1]{\href{http://ascl.net/#1}{\nolinkurl{http://ascl.net/#1}}}
\providecommand{\doarXiv}[1]{\href{https://arxiv.org/abs/#1}{\nolinkurl{https://arxiv.org/abs/#1}}}

\bibitem[{{Alberto} {et~al.}(2019){Alberto}, {Magda}, {Nausicaa}, {Vincenzo},
  {Nathalie}, {Olivier}, {Uwe}, {Mubashir}, {Laura}, {Joerg}, {Martino},
  {Devendra}, {Chiara}, {Malgorzata}, {Felix}, {Ignacio}, \&
  {Stefano}}]{Alberto2019}
{Alberto}, M., {Magda}, A., {Nausicaa}, D., {et~al.} 2019, in Astronomical
  Society of the Pacific Conference Series, Vol. 523, Astronomical Data
  Analysis Software and Systems XXVII, ed. P.~J. {Teuben}, M.~W. {Pound}, B.~A.
  {Thomas}, \& E.~M. {Warner}, 433

\bibitem[{{Alberts} {et~al.}(2020){Alberts}, {Rujopakarn}, {Rieke},
  {Jagannathan}, \& {Nyland}}]{Alberts2020}
{Alberts}, S., {Rujopakarn}, W., {Rieke}, G.~H., {Jagannathan}, P., \&
  {Nyland}, K. 2020, \apj, 901, 168, \dodoi{10.3847/1538-4357/abb1a0}

\bibitem[{{Antonucci}(1993)}]{Antonucci1993}
{Antonucci}, R. 1993, \araa, 31, 473,
  \dodoi{10.1146/annurev.aa.31.090193.002353}

\bibitem[{{Astropy Collaboration} {et~al.}(2013{\natexlab{a}}){Astropy
  Collaboration}, {Robitaille}, {Tollerud}, {Greenfield}, {Droettboom}, {Bray},
  {Aldcroft}, {Davis}, {Ginsburg}, {Price-Whelan}, {Kerzendorf}, {Conley},
  {Crighton}, {Barbary}, {Muna}, {Ferguson}, {Grollier}, {Parikh}, {Nair},
  {Unther}, {Deil}, {Woillez}, {Conseil}, {Kramer}, {Turner}, {Singer}, {Fox},
  {Weaver}, {Zabalza}, {Edwards}, {Azalee Bostroem}, {Burke}, {Casey},
  {Crawford}, {Dencheva}, {Ely}, {Jenness}, {Labrie}, {Lim}, {Pierfederici},
  {Pontzen}, {Ptak}, {Refsdal}, {Servillat}, \& {Streicher}}]{astropy:2013}
{Astropy Collaboration}, {Robitaille}, T.~P., {Tollerud}, E.~J., {et~al.}
  2013{\natexlab{a}}, \aap, 558, A33, \dodoi{10.1051/0004-6361/201322068}

\bibitem[{{Astropy Collaboration} {et~al.}(2013{\natexlab{b}}){Astropy
  Collaboration}, {Robitaille}, {Tollerud}, {Greenfield}, {Droettboom}, {Bray},
  {Aldcroft}, {Davis}, {Ginsburg}, {Price-Whelan}, {Kerzendorf}, {Conley},
  {Crighton}, {Barbary}, {Muna}, {Ferguson}, {Grollier}, {Parikh}, {Nair},
  {Unther}, {Deil}, {Woillez}, {Conseil}, {Kramer}, {Turner}, {Singer}, {Fox},
  {Weaver}, {Zabalza}, {Edwards}, {Azalee Bostroem}, {Burke}, {Casey},
  {Crawford}, {Dencheva}, {Ely}, {Jenness}, {Labrie}, {Lim}, {Pierfederici},
  {Pontzen}, {Ptak}, {Refsdal}, {Servillat}, \& {Streicher}}]{astropy2013}
---. 2013{\natexlab{b}}, \aap, 558, A33, \dodoi{10.1051/0004-6361/201322068}

\bibitem[{{Astropy Collaboration} {et~al.}(2018{\natexlab{a}}){Astropy
  Collaboration}, {Price-Whelan}, {Sip{\H{o}}cz}, {G{\"u}nther}, {Lim},
  {Crawford}, {Conseil}, {Shupe}, {Craig}, {Dencheva}, {Ginsburg}, {Vand
  erPlas}, {Bradley}, {P{\'e}rez-Su{\'a}rez}, {de Val-Borro}, {Aldcroft},
  {Cruz}, {Robitaille}, {Tollerud}, {Ardelean}, {Babej}, {Bach}, {Bachetti},
  {Bakanov}, {Bamford}, {Barentsen}, {Barmby}, {Baumbach}, {Berry}, {Biscani},
  {Boquien}, {Bostroem}, {Bouma}, {Brammer}, {Bray}, {Breytenbach},
  {Buddelmeijer}, {Burke}, {Calderone}, {Cano Rodr{\'\i}guez}, {Cara},
  {Cardoso}, {Cheedella}, {Copin}, {Corrales}, {Crichton}, {D'Avella}, {Deil},
  {Depagne}, {Dietrich}, {Donath}, {Droettboom}, {Earl}, {Erben}, {Fabbro},
  {Ferreira}, {Finethy}, {Fox}, {Garrison}, {Gibbons}, {Goldstein}, {Gommers},
  {Greco}, {Greenfield}, {Groener}, {Grollier}, {Hagen}, {Hirst}, {Homeier},
  {Horton}, {Hosseinzadeh}, {Hu}, {Hunkeler}, {Ivezi{\'c}}, {Jain}, {Jenness},
  {Kanarek}, {Kendrew}, {Kern}, {Kerzendorf}, {Khvalko}, {King}, {Kirkby},
  {Kulkarni}, {Kumar}, {Lee}, {Lenz}, {Littlefair}, {Ma}, {Macleod},
  {Mastropietro}, {McCully}, {Montagnac}, {Morris}, {Mueller}, {Mumford},
  {Muna}, {Murphy}, {Nelson}, {Nguyen}, {Ninan}, {N{\"o}the}, {Ogaz}, {Oh},
  {Parejko}, {Parley}, {Pascual}, {Patil}, {Patil}, {Plunkett}, {Prochaska},
  {Rastogi}, {Reddy Janga}, {Sabater}, {Sakurikar}, {Seifert}, {Sherbert},
  {Sherwood-Taylor}, {Shih}, {Sick}, {Silbiger}, {Singanamalla}, {Singer},
  {Sladen}, {Sooley}, {Sornarajah}, {Streicher}, {Teuben}, {Thomas},
  {Tremblay}, {Turner}, {Terr{\'o}n}, {van Kerkwijk}, {de la Vega}, {Watkins},
  {Weaver}, {Whitmore}, {Woillez}, {Zabalza}, \& {Astropy
  Contributors}}]{astropy:2018}
{Astropy Collaboration}, {Price-Whelan}, A.~M., {Sip{\H{o}}cz}, B.~M., {et~al.}
  2018{\natexlab{a}}, \aj, 156, 123, \dodoi{10.3847/1538-3881/aabc4f}

\bibitem[{{Astropy Collaboration} {et~al.}(2018{\natexlab{b}}){Astropy
  Collaboration}, {Price-Whelan}, {Sip{\H{o}}cz}, {G{\"u}nther}, {Lim},
  {Crawford}, {Conseil}, {Shupe}, {Craig}, {Dencheva}, {Ginsburg},
  {VanderPlas}, {Bradley}, {P{\'e}rez-Su{\'a}rez}, {de Val-Borro}, {Aldcroft},
  {Cruz}, {Robitaille}, {Tollerud}, {Ardelean}, {Babej}, {Bach}, {Bachetti},
  {Bakanov}, {Bamford}, {Barentsen}, {Barmby}, {Baumbach}, {Berry}, {Biscani},
  {Boquien}, {Bostroem}, {Bouma}, {Brammer}, {Bray}, {Breytenbach},
  {Buddelmeijer}, {Burke}, {Calderone}, {Cano Rodr{\'\i}guez}, {Cara},
  {Cardoso}, {Cheedella}, {Copin}, {Corrales}, {Crichton}, {D'Avella}, {Deil},
  {Depagne}, {Dietrich}, {Donath}, {Droettboom}, {Earl}, {Erben}, {Fabbro},
  {Ferreira}, {Finethy}, {Fox}, {Garrison}, {Gibbons}, {Goldstein}, {Gommers},
  {Greco}, {Greenfield}, {Groener}, {Grollier}, {Hagen}, {Hirst}, {Homeier},
  {Horton}, {Hosseinzadeh}, {Hu}, {Hunkeler}, {Ivezi{\'c}}, {Jain}, {Jenness},
  {Kanarek}, {Kendrew}, {Kern}, {Kerzendorf}, {Khvalko}, {King}, {Kirkby},
  {Kulkarni}, {Kumar}, {Lee}, {Lenz}, {Littlefair}, {Ma}, {Macleod},
  {Mastropietro}, {McCully}, {Montagnac}, {Morris}, {Mueller}, {Mumford},
  {Muna}, {Murphy}, {Nelson}, {Nguyen}, {Ninan}, {N{\"o}the}, {Ogaz}, {Oh},
  {Parejko}, {Parley}, {Pascual}, {Patil}, {Patil}, {Plunkett}, {Prochaska},
  {Rastogi}, {Reddy Janga}, {Sabater}, {Sakurikar}, {Seifert}, {Sherbert},
  {Sherwood-Taylor}, {Shih}, {Sick}, {Silbiger}, {Singanamalla}, {Singer},
  {Sladen}, {Sooley}, {Sornarajah}, {Streicher}, {Teuben}, {Thomas},
  {Tremblay}, {Turner}, {Terr{\'o}n}, {van Kerkwijk}, {de la Vega}, {Watkins},
  {Weaver}, {Whitmore}, {Woillez}, {Zabalza}, \& {Astropy
  Contributors}}]{astropy2018}
---. 2018{\natexlab{b}}, \aj, 156, 123, \dodoi{10.3847/1538-3881/aabc4f}

\bibitem[{{Astropy Collaboration} {et~al.}(2022{\natexlab{a}}){Astropy
  Collaboration}, {Price-Whelan}, {Lim}, {Earl}, {Starkman}, {Bradley},
  {Shupe}, {Patil}, {Corrales}, {Brasseur}, {N{"o}the}, {Donath}, {Tollerud},
  {Morris}, {Ginsburg}, {Vaher}, {Weaver}, {Tocknell}, {Jamieson}, {van
  Kerkwijk}, {Robitaille}, {Merry}, {Bachetti}, {G{"u}nther}, {Aldcroft},
  {Alvarado-Montes}, {Archibald}, {B{'o}di}, {Bapat}, {Barentsen}, {Baz{'a}n},
  {Biswas}, {Boquien}, {Burke}, {Cara}, {Cara}, {Conroy}, {Conseil}, {Craig},
  {Cross}, {Cruz}, {D'Eugenio}, {Dencheva}, {Devillepoix}, {Dietrich},
  {Eigenbrot}, {Erben}, {Ferreira}, {Foreman-Mackey}, {Fox}, {Freij}, {Garg},
  {Geda}, {Glattly}, {Gondhalekar}, {Gordon}, {Grant}, {Greenfield}, {Groener},
  {Guest}, {Gurovich}, {Handberg}, {Hart}, {Hatfield-Dodds}, {Homeier},
  {Hosseinzadeh}, {Jenness}, {Jones}, {Joseph}, {Kalmbach}, {Karamehmetoglu},
  {Ka{l}uszy{'n}ski}, {Kelley}, {Kern}, {Kerzendorf}, {Koch}, {Kulumani},
  {Lee}, {Ly}, {Ma}, {MacBride}, {Maljaars}, {Muna}, {Murphy}, {Norman},
  {O'Steen}, {Oman}, {Pacifici}, {Pascual}, {Pascual-Granado}, {Patil},
  {Perren}, {Pickering}, {Rastogi}, {Roulston}, {Ryan}, {Rykoff}, {Sabater},
  {Sakurikar}, {Salgado}, {Sanghi}, {Saunders}, {Savchenko}, {Schwardt},
  {Seifert-Eckert}, {Shih}, {Jain}, {Shukla}, {Sick}, {Simpson},
  {Singanamalla}, {Singer}, {Singhal}, {Sinha}, {Sip{H{o}}cz}, {Spitler},
  {Stansby}, {Streicher}, {{{S}}umak}, {Swinbank}, {Taranu}, {Tewary},
  {Tremblay}, {Val-Borro}, {Van Kooten}, {Vasovi{'c}}, {Verma}, {de Miranda
  Cardoso}, {Williams}, {Wilson}, {Winkel}, {Wood-Vasey}, {Xue}, {Yoachim},
  {Zhang}, {Zonca}, \& {Astropy Project Contributors}}]{astropy:2022}
{Astropy Collaboration}, {Price-Whelan}, A.~M., {Lim}, P.~L., {et~al.}
  2022{\natexlab{a}}, \apj, 935, 167, \dodoi{10.3847/1538-4357/ac7c74}

\bibitem[{{Astropy Collaboration} {et~al.}(2022{\natexlab{b}}){Astropy
  Collaboration}, {Price-Whelan}, {Lim}, {Earl}, {Starkman}, {Bradley},
  {Shupe}, {Patil}, {Corrales}, {Brasseur}, {N{\"o}the}, {Donath}, {Tollerud},
  {Morris}, {Ginsburg}, {Vaher}, {Weaver}, {Tocknell}, {Jamieson}, {van
  Kerkwijk}, {Robitaille}, {Merry}, {Bachetti}, {G{\"u}nther}, {Aldcroft},
  {Alvarado-Montes}, {Archibald}, {B{\'o}di}, {Bapat}, {Barentsen},
  {Baz{\'a}n}, {Biswas}, {Boquien}, {Burke}, {Cara}, {Cara}, {Conroy},
  {Conseil}, {Craig}, {Cross}, {Cruz}, {D'Eugenio}, {Dencheva}, {Devillepoix},
  {Dietrich}, {Eigenbrot}, {Erben}, {Ferreira}, {Foreman-Mackey}, {Fox},
  {Freij}, {Garg}, {Geda}, {Glattly}, {Gondhalekar}, {Gordon}, {Grant},
  {Greenfield}, {Groener}, {Guest}, {Gurovich}, {Handberg}, {Hart},
  {Hatfield-Dodds}, {Homeier}, {Hosseinzadeh}, {Jenness}, {Jones}, {Joseph},
  {Kalmbach}, {Karamehmetoglu}, {Ka{\l}uszy{\'n}ski}, {Kelley}, {Kern},
  {Kerzendorf}, {Koch}, {Kulumani}, {Lee}, {Ly}, {Ma}, {MacBride}, {Maljaars},
  {Muna}, {Murphy}, {Norman}, {O'Steen}, {Oman}, {Pacifici}, {Pascual},
  {Pascual-Granado}, {Patil}, {Perren}, {Pickering}, {Rastogi}, {Roulston},
  {Ryan}, {Rykoff}, {Sabater}, {Sakurikar}, {Salgado}, {Sanghi}, {Saunders},
  {Savchenko}, {Schwardt}, {Seifert-Eckert}, {Shih}, {Jain}, {Shukla}, {Sick},
  {Simpson}, {Singanamalla}, {Singer}, {Singhal}, {Sinha}, {Sip{\H{o}}cz},
  {Spitler}, {Stansby}, {Streicher}, {{\v{S}}umak}, {Swinbank}, {Taranu},
  {Tewary}, {Tremblay}, {de Val-Borro}, {Van Kooten}, {Vasovi{\'c}}, {Verma},
  {de Miranda Cardoso}, {Williams}, {Wilson}, {Winkel}, {Wood-Vasey}, {Xue},
  {Yoachim}, {Zhang}, {Zonca}, \& {Astropy Project Contributors}}]{astropy2022}
---. 2022{\natexlab{b}}, \apj, 935, 167, \dodoi{10.3847/1538-4357/ac7c74}

\bibitem[{{Bacon} {et~al.}(2010){Bacon}, {Accardo}, {Adjali}, {Anwand},
  {Bauer}, {Biswas}, {Blaizot}, {Boudon}, {Brau-Nogue}, {Brinchmann},
  {Caillier}, {Capoani}, {Carollo}, {Contini}, {Couderc}, {Daguis{\'e}},
  {Deiries}, {Delabre}, {Dreizler}, {Dubois}, {Dupieux}, {Dupuy}, {Emsellem},
  {Fechner}, {Fleischmann}, {Fran{\c{c}}ois}, {Gallou}, {Gharsa}, {Glindemann},
  {Gojak}, {Guiderdoni}, {Hansali}, {Hahn}, {Jarno}, {Kelz}, {Koehler},
  {Kosmalski}, {Laurent}, {Le Floch}, {Lilly}, {Lizon}, {Loupias}, {Manescau},
  {Monstein}, {Nicklas}, {Olaya}, {Pares}, {Pasquini}, {P{\'e}contal-Rousset},
  {Pell{\'o}}, {Petit}, {Popow}, {Reiss}, {Remillieux}, {Renault}, {Roth},
  {Rupprecht}, {Serre}, {Schaye}, {Soucail}, {Steinmetz}, {Streicher}, {Stuik},
  {Valentin}, {Vernet}, {Weilbacher}, {Wisotzki}, \& {Yerle}}]{Bacon2010}
{Bacon}, R., {Accardo}, M., {Adjali}, L., {et~al.} 2010, in Society of
  Photo-Optical Instrumentation Engineers (SPIE) Conference Series, Vol. 7735,
  Ground-based and Airborne Instrumentation for Astronomy III, ed. I.~S.
  {McLean}, S.~K. {Ramsay}, \& H.~{Takami}, 773508, \dodoi{10.1117/12.856027}

\bibitem[{{Balestra} {et~al.}(2010){Balestra}, {Mainieri}, {Popesso},
  {Dickinson}, {Nonino}, {Rosati}, {Teimoorinia}, {Vanzella}, {Cristiani},
  {Cesarsky}, {Fosbury}, {Kuntschner}, \& {Rettura}}]{Balestra2010}
{Balestra}, I., {Mainieri}, V., {Popesso}, P., {et~al.} 2010, \aap, 512, A12,
  \dodoi{10.1051/0004-6361/200913626}

\bibitem[{{Barro} {et~al.}(2014){Barro}, {Faber}, {P{\'e}rez-Gonz{\'a}lez},
  {Pacifici}, {Trump}, {Koo}, {Wuyts}, {Guo}, {Bell}, {Dekel}, {Porter},
  {Primack}, {Ferguson}, {Ashby}, {Caputi}, {Ceverino}, {Croton}, {Fazio},
  {Giavalisco}, {Hsu}, {Kocevski}, {Koekemoer}, {Kurczynski}, {Kollipara},
  {Lee}, {McIntosh}, {McGrath}, {Moody}, {Somerville}, {Papovich}, {Salvato},
  {Santini}, {Tal}, {van der Wel}, {Williams}, {Willner}, \&
  {Zolotov}}]{Barro2014}
{Barro}, G., {Faber}, S.~M., {P{\'e}rez-Gonz{\'a}lez}, P.~G., {et~al.} 2014,
  \apj, 791, 52, \dodoi{10.1088/0004-637X/791/1/52}

\bibitem[{{Beichman} {et~al.}(2012){Beichman}, {Rieke}, {Eisenstein}, {Greene},
  {Krist}, {McCarthy}, {Meyer}, \& {Stansberry}}]{Beichman2012}
{Beichman}, C.~A., {Rieke}, M., {Eisenstein}, D., {et~al.} 2012, in Society of
  Photo-Optical Instrumentation Engineers (SPIE) Conference Series, Vol. 8442,
  Space Telescopes and Instrumentation 2012: Optical, Infrared, and Millimeter
  Wave, ed. M.~C. {Clampin}, G.~G. {Fazio}, H.~A. {MacEwen}, \& J.~{Oschmann},
  Jacobus~M., 84422N, \dodoi{10.1117/12.925447}

\bibitem[{{Bennert} {et~al.}(2002){Bennert}, {Falcke}, {Schulz}, {Wilson}, \&
  {Wills}}]{Bennert2002}
{Bennert}, N., {Falcke}, H., {Schulz}, H., {Wilson}, A.~S., \& {Wills}, B.~J.
  2002, \apjl, 574, L105, \dodoi{10.1086/342420}

\bibitem[{{Bennert} {et~al.}(2006){Bennert}, {Jungwiert}, {Komossa}, {Haas}, \&
  {Chini}}]{Bennert2006}
{Bennert}, N., {Jungwiert}, B., {Komossa}, S., {Haas}, M., \& {Chini}, R. 2006,
  \aap, 456, 953, \dodoi{10.1051/0004-6361:20065319}

\bibitem[{{Bower} {et~al.}(2006){Bower}, {Benson}, {Malbon}, {Helly}, {Frenk},
  {Baugh}, {Cole}, \& {Lacey}}]{Bower2006}
{Bower}, R.~G., {Benson}, A.~J., {Malbon}, R., {et~al.} 2006, \mnras, 370, 645,
  \dodoi{10.1111/j.1365-2966.2006.10519.x}

\bibitem[{{Brammer} {et~al.}(2008){Brammer}, {van Dokkum}, \&
  {Coppi}}]{Brammer2008}
{Brammer}, G.~B., {van Dokkum}, P.~G., \& {Coppi}, P. 2008, \apj, 686, 1503,
  \dodoi{10.1086/591786}

\bibitem[{{Brusa} {et~al.}(2009){Brusa}, {Fiore}, {Santini}, {Grazian},
  {Comastri}, {Zamorani}, {Hasinger}, {Merloni}, {Civano}, {Fontana}, \&
  {Mainieri}}]{Brusa2009}
{Brusa}, M., {Fiore}, F., {Santini}, P., {et~al.} 2009, \aap, 507, 1277,
  \dodoi{10.1051/0004-6361/200912261}

\bibitem[{{Calvi} {et~al.}(2014){Calvi}, {Stiavelli}, {Bradley}, {Pizzella}, \&
  {Kim}}]{Calvi2014}
{Calvi}, V., {Stiavelli}, M., {Bradley}, L., {Pizzella}, A., \& {Kim}, S. 2014,
  \apj, 796, 102, \dodoi{10.1088/0004-637X/796/2/102}

\bibitem[{{Cano-D{\'\i}az} {et~al.}(2012){Cano-D{\'\i}az}, {Maiolino},
  {Marconi}, {Netzer}, {Shemmer}, \& {Cresci}}]{Cano-diaz2012}
{Cano-D{\'\i}az}, M., {Maiolino}, R., {Marconi}, A., {et~al.} 2012, \aap, 537,
  L8, \dodoi{10.1051/0004-6361/201118358}

\bibitem[{{Ciesla} {et~al.}(2018){Ciesla}, {Elbaz}, {Schreiber}, {Daddi}, \&
  {Wang}}]{Ciesla2018}
{Ciesla}, L., {Elbaz}, D., {Schreiber}, C., {Daddi}, E., \& {Wang}, T. 2018,
  \aap, 615, A61, \dodoi{10.1051/0004-6361/201832715}

\bibitem[{{Circosta} {et~al.}(2018){Circosta}, {Mainieri}, {Padovani},
  {Lanzuisi}, {Salvato}, {Harrison}, {Kakkad}, {Puglisi}, {Vietri}, {Zamorani},
  {Cicone}, {Husemann}, {Vignali}, {Balmaverde}, {Bischetti}, {Bongiorno},
  {Brusa}, {Carniani}, {Civano}, {Comastri}, {Cresci}, {Feruglio}, {Fiore},
  {Fotopoulou}, {Karim}, {Lamastra}, {Magnelli}, {Mannucci}, {Marconi},
  {Merloni}, {Netzer}, {Perna}, {Piconcelli}, {Rodighiero}, {Schinnerer},
  {Schramm}, {Schulze}, {Silverman}, \& {Zappacosta}}]{Circosta2018}
{Circosta}, C., {Mainieri}, V., {Padovani}, P., {et~al.} 2018, \aap, 620, A82,
  \dodoi{10.1051/0004-6361/201833520}

\bibitem[{{Cresci} {et~al.}(2023){Cresci}, {Tozzi}, {Perna}, {Brusa},
  {Marconcini}, {Marconi}, {Carniani}, {Brienza}, {Giroletti}, {Belfiore},
  {Ginolfi}, {Mannucci}, {Ulivi}, {Scholtz}, {Venturi}, {Arribas}, {{\"U}bler},
  {D'Eugenio}, {Mingozzi}, {Balmaverde}, {Capetti}, {Parlanti}, \&
  {Zana}}]{Cresci2023}
{Cresci}, G., {Tozzi}, G., {Perna}, M., {et~al.} 2023, \aap, 672, A128,
  \dodoi{10.1051/0004-6361/202346001}

\bibitem[{{Croton} {et~al.}(2006){Croton}, {Springel}, {White}, {De Lucia},
  {Frenk}, {Gao}, {Jenkins}, {Kauffmann}, {Navarro}, \& {Yoshida}}]{Croton2006}
{Croton}, D.~J., {Springel}, V., {White}, S. D.~M., {et~al.} 2006, \mnras, 365,
  11, \dodoi{10.1111/j.1365-2966.2005.09675.x}

\bibitem[{{Deconto-Machado} {et~al.}(2022){Deconto-Machado}, {Riffel}, {Ilha},
  {Rembold}, {Storchi-Bergmann}, {Riffel}, {Schimoia}, {Schneider}, {Bizyaev},
  {Feng}, {Wylezalek}, {da Costa}, {do Nascimento}, \& {Maia}}]{Machado2022}
{Deconto-Machado}, A., {Riffel}, R.~A., {Ilha}, G.~S., {et~al.} 2022, \aap,
  659, A131, \dodoi{10.1051/0004-6361/202140613}

\bibitem[{{Dempsey} \& {Zakamska}(2018)}]{Dempsey2018}
{Dempsey}, R., \& {Zakamska}, N.~L. 2018, \mnras, 477, 4615,
  \dodoi{10.1093/mnras/sty941}

\bibitem[{{Di Matteo} {et~al.}(2005){Di Matteo}, {Springel}, \&
  {Hernquist}}]{DiMatteo2005}
{Di Matteo}, T., {Springel}, V., \& {Hernquist}, L. 2005, \nat, 433, 604,
  \dodoi{10.1038/nature03335}

\bibitem[{{Dunlop} {et~al.}(2017){Dunlop}, {McLure}, {Biggs}, {Geach},
  {Micha{\l}owski}, {Ivison}, {Rujopakarn}, {van Kampen}, {Kirkpatrick},
  {Pope}, {Scott}, {Swinbank}, {Targett}, {Aretxaga}, {Austermann}, {Best},
  {Bruce}, {Chapin}, {Charlot}, {Cirasuolo}, {Coppin}, {Ellis}, {Finkelstein},
  {Hayward}, {Hughes}, {Ibar}, {Jagannathan}, {Khochfar}, {Koprowski},
  {Narayanan}, {Nyland}, {Papovich}, {Peacock}, {Rieke}, {Robertson},
  {Vernstrom}, {Werf}, {Wilson}, \& {Yun}}]{Dunlop2017}
{Dunlop}, J.~S., {McLure}, R.~J., {Biggs}, A.~D., {et~al.} 2017, \mnras, 466,
  861, \dodoi{10.1093/mnras/stw3088}

\bibitem[{{Durr{\'e}} \& {Mould}(2018)}]{Durre&Mould2018}
{Durr{\'e}}, M., \& {Mould}, J. 2018, \apj, 867, 149,
  \dodoi{10.3847/1538-4357/aae68e}

\bibitem[{{Durr{\'e}} \& {Mould}(2019)}]{Durre&Mould2019}
---. 2019, \apj, 870, 37, \dodoi{10.3847/1538-4357/aaf000}

\bibitem[{{Eisenstein} {et~al.}(2023{\natexlab{a}}){Eisenstein}, {Willott},
  {Alberts}, {Arribas}, {Bonaventura}, {Bunker}, {Cameron}, {Carniani},
  {Charlot}, {Curtis-Lake}, {D'Eugenio}, {Endsley}, {Ferruit}, {Giardino},
  {Hainline}, {Hausen}, {Jakobsen}, {Johnson}, {Maiolino}, {Rieke}, {Rieke},
  {Rix}, {Robertson}, {Stark}, {Tacchella}, {Williams}, {Willmer}, {Baker},
  {Baum}, {Bhatawdekar}, {Boyett}, {Chen}, {Chevallard}, {Circosta}, {Curti},
  {Danhaive}, {DeCoursey}, {de Graaff}, {Dressler}, {Egami}, {Helton},
  {Hviding}, {Ji}, {Jones}, {Kumari}, {L{\"u}tzgendorf}, {Laseter}, {Looser},
  {Lyu}, {Maseda}, {Nelson}, {Parlanti}, {Perna}, {Pusk{\'a}s}, {Rawle},
  {Rodr{\'\i}guez Del Pino}, {Sandles}, {Saxena}, {Scholtz}, {Sharpe},
  {Shivaei}, {Silcock}, {Simmonds}, {Skarbinski}, {Smit}, {Stone}, {Suess},
  {Sun}, {Tang}, {Topping}, {{\"U}bler}, {Villanueva}, {Wallace}, {Whitler},
  {Witstok}, \& {Woodrum}}]{Eisenstein2023a}
{Eisenstein}, D.~J., {Willott}, C., {Alberts}, S., {et~al.} 2023{\natexlab{a}},
  arXiv e-prints, arXiv:2306.02465, \dodoi{10.48550/arXiv.2306.02465}

\bibitem[{{Eisenstein} {et~al.}(2023{\natexlab{b}}){Eisenstein}, {Johnson},
  {Robertson}, {Tacchella}, {Hainline}, {Jakobsen}, {Maiolino}, {Bonaventura},
  {Bunker}, {Cameron}, {Cargile}, {Curtis-Lake}, {Hausen}, {Pusk{\'a}s},
  {Rieke}, {Sun}, {Willmer}, {Willott}, {Alberts}, {Arribas}, {Baker}, {Baum},
  {Bhatawdekar}, {Carniani}, {Charlot}, {Chen}, {Chevallard}, {Curti},
  {DeCoursey}, {D'Eugenio}, {de Graaff}, {Egami}, {Helton}, {Ji}, {Jones},
  {Kumari}, {L{\"u}tzgendorf}, {Laseter}, {Looser}, {Lyu}, {Maseda}, {Nelson},
  {Parlanti}, {Rauscher}, {Rawle}, {Rieke}, {Rix}, {Rujopakarn}, {Sandles},
  {Saxena}, {Scholtz}, {Sharpe}, {Shivaei}, {Simmonds}, {Smit}, {Topping},
  {{\"U}bler}, {Venturi}, {Williams}, {Witstok}, \&
  {Woodrum}}]{Eisenstein2023b}
{Eisenstein}, D.~J., {Johnson}, B.~D., {Robertson}, B., {et~al.}
  2023{\natexlab{b}}, arXiv e-prints, arXiv:2310.12340,
  \dodoi{10.48550/arXiv.2310.12340}

\bibitem[{{Fabian}(2012)}]{Fabian2012}
{Fabian}, A.~C. 2012, \araa, 50, 455,
  \dodoi{10.1146/annurev-astro-081811-125521}

\bibitem[{{Falcke} {et~al.}(1998){Falcke}, {Wilson}, \& {Simpson}}]{Falcke1998}
{Falcke}, H., {Wilson}, A.~S., \& {Simpson}, C. 1998, \apj, 502, 199,
  \dodoi{10.1086/305886}

\bibitem[{{Ferrarese} \& {Merritt}(2000)}]{Ferrarese2000}
{Ferrarese}, L., \& {Merritt}, D. 2000, \apjl, 539, L9, \dodoi{10.1086/312838}

\bibitem[{{Fiore} {et~al.}(2012){Fiore}, {Puccetti}, {Grazian}, {Menci},
  {Shankar}, {Santini}, {Piconcelli}, {Koekemoer}, {Fontana}, {Boutsia},
  {Castellano}, {Lamastra}, {Malacaria}, {Feruglio}, {Mathur}, {Miller}, \&
  {Pannella}}]{Fiore2012}
{Fiore}, F., {Puccetti}, S., {Grazian}, A., {et~al.} 2012, \aap, 537, A16,
  \dodoi{10.1051/0004-6361/201117581}

\bibitem[{{Fischer} {et~al.}(2010){Fischer}, {Sturm}, {Gonz{\'a}lez-Alfonso},
  {Graci{\'a}-Carpio}, {Hailey-Dunsheath}, {Poglitsch}, {Contursi}, {Lutz},
  {Genzel}, {Sternberg}, {Verma}, \& {Tacconi}}]{Fischer2010}
{Fischer}, J., {Sturm}, E., {Gonz{\'a}lez-Alfonso}, E., {et~al.} 2010, \aap,
  518, L41, \dodoi{10.1051/0004-6361/201014676}

\bibitem[{Fitzpatrick {et~al.}(2014)Fitzpatrick, Olsen, Economou, Stobie,
  Beers, Dickinson, Norris, Saha, Seaman, Silva, Swaters, Thomas, \&
  Valdes}]{Fitzpatrick2014}
Fitzpatrick, M.~J., Olsen, K., Economou, F., {et~al.} 2014, in Observatory
  Operations: Strategies, Processes, and Systems V, ed. A.~B. Peck, C.~R. Benn,
  \& R.~L. Seaman, Vol. 9149, International Society for Optics and Photonics
  (SPIE), 91491T, \dodoi{10.1117/12.2057445}

\bibitem[{{F{\"o}rster Schreiber} \& {Wuyts}(2020)}]{Schreiber2020}
{F{\"o}rster Schreiber}, N.~M., \& {Wuyts}, S. 2020, \araa, 58, 661,
  \dodoi{10.1146/annurev-astro-032620-021910}

\bibitem[{{Franco} {et~al.}(2018){Franco}, {Elbaz}, {B{\'e}thermin},
  {Magnelli}, {Schreiber}, {Ciesla}, {Dickinson}, {Nagar}, {Silverman},
  {Daddi}, {Alexander}, {Wang}, {Pannella}, {Le Floc'h}, {Pope}, {Giavalisco},
  {Maury}, {Bournaud}, {Chary}, {Demarco}, {Ferguson}, {Finkelstein}, {Inami},
  {Iono}, {Juneau}, {Lagache}, {Leiton}, {Lin}, {Magdis}, {Messias},
  {Motohara}, {Mullaney}, {Okumura}, {Papovich}, {Pforr}, {Rujopakarn},
  {Sargent}, {Shu}, \& {Zhou}}]{Franco2018}
{Franco}, M., {Elbaz}, D., {B{\'e}thermin}, M., {et~al.} 2018, \aap, 620, A152,
  \dodoi{10.1051/0004-6361/201832928}

\bibitem[{{Franco} {et~al.}(2020){Franco}, {Elbaz}, {Zhou}, {Magnelli},
  {Schreiber}, {Ciesla}, {Dickinson}, {Nagar}, {Magdis}, {Alexander},
  {B{\'e}thermin}, {Demarco}, {Daddi}, {Wang}, {Mullaney}, {Sargent}, {Inami},
  {Shu}, {Bournaud}, {Chary}, {Coogan}, {Ferguson}, {Finkelstein},
  {Giavalisco}, {G{\'o}mez-Guijarro}, {Iono}, {Juneau}, {Lagache}, {Lin},
  {Motohara}, {Okumura}, {Pannella}, {Papovich}, {Pope}, {Rujopakarn},
  {Silverman}, \& {Xiao}}]{Franco2020}
{Franco}, M., {Elbaz}, D., {Zhou}, L., {et~al.} 2020, \aap, 643, A30,
  \dodoi{10.1051/0004-6361/202038312}

\bibitem[{{Fraquelli} {et~al.}(2003){Fraquelli}, {Storchi-Bergmann}, \&
  {Levenson}}]{Fraquelli2003}
{Fraquelli}, H.~A., {Storchi-Bergmann}, T., \& {Levenson}, N.~A. 2003, \mnras,
  341, 449, \dodoi{10.1046/j.1365-8711.2003.06397.x}

\bibitem[{{Friedrich} {et~al.}(2010){Friedrich}, {Davies}, {Hicks}, {Engel},
  {M{\"u}ller-S{\'a}nchez}, {Genzel}, \& {Tacconi}}]{Friedrich}
{Friedrich}, S., {Davies}, R.~I., {Hicks}, E.~K.~S., {et~al.} 2010, \aap, 519,
  A79, \dodoi{10.1051/0004-6361/200913924}

\bibitem[{{Fujimoto} {et~al.}(2018){Fujimoto}, {Ouchi}, {Kohno}, {Yamaguchi},
  {Hatsukade}, {Ueda}, {Shibuya}, {Inoue}, {Oogi}, {Toft},
  {G{\'o}mez-Guijarro}, {Wang}, {Espada}, {Nagao}, {Tanaka}, {Ao}, {Umehata},
  {Taniguchi}, {Nakanishi}, {Rujopakarn}, {Ivison}, {Wang}, {Lee}, {Tadaki},
  {Tamura}, \& {Dunlop}}]{Fujimoto2018}
{Fujimoto}, S., {Ouchi}, M., {Kohno}, K., {et~al.} 2018, \apj, 861, 7,
  \dodoi{10.3847/1538-4357/aac6c4}

\bibitem[{{Garilli} {et~al.}(2021){Garilli}, {McLure}, {Pentericci},
  {Franzetti}, {Gargiulo}, {Carnall}, {Cucciati}, {Iovino}, {Amorin},
  {Bolzonella}, {Bongiorno}, {Castellano}, {Cimatti}, {Cirasuolo}, {Cullen},
  {Dunlop}, {Elbaz}, {Finkelstein}, {Fontana}, {Fontanot}, {Fumana}, {Guaita},
  {Hartley}, {Jarvis}, {Juneau}, {Maccagni}, {McLeod}, {Nandra}, {Pompei},
  {Pozzetti}, {Scodeggio}, {Talia}, {Calabr{\`o}}, {Cresci}, {Fynbo}, {Hathi},
  {Hibon}, {Koekemoer}, {Magliocchetti}, {Salvato}, {Vietri}, {Zamorani},
  {Almaini}, {Balestra}, {Bardelli}, {Begley}, {Brammer}, {Bell}, {Bowler},
  {Brusa}, {Buitrago}, {Caputi}, {Cassata}, {Charlot}, {Citro}, {Cristiani},
  {Curtis-Lake}, {Dickinson}, {Fazio}, {Ferguson}, {Fiore}, {Franco},
  {Georgakakis}, {Giavalisco}, {Grazian}, {Hamadouche}, {Jung}, {Kim},
  {Khusanova}, {Le F{\`e}vre}, {Longhetti}, {Lotz}, {Mannucci}, {Maltby},
  {Matsuoka}, {Mendez-Hernandez}, {Mendez-Abreu}, {Mignoli}, {Moresco},
  {Nonino}, {Pannella}, {Papovich}, {Popesso}, {Roberts-Borsani}, {Rosario},
  {Saldana-Lopez}, {Santini}, {Saxena}, {Schaerer}, {Schreiber}, {Stark},
  {Tasca}, {Thomas}, {Vanzella}, {Wild}, {Williams}, \& {Zucca}}]{Garilli2021}
{Garilli}, B., {McLure}, R., {Pentericci}, L., {et~al.} 2021, \aap, 647, A150,
  \dodoi{10.1051/0004-6361/202040059}

\bibitem[{{Gebhardt} {et~al.}(2000){Gebhardt}, {Bender}, {Bower}, {Dressler},
  {Faber}, {Filippenko}, {Green}, {Grillmair}, {Ho}, {Kormendy}, {Lauer},
  {Magorrian}, {Pinkney}, {Richstone}, \& {Tremaine}}]{Gebhardt2000}
{Gebhardt}, K., {Bender}, R., {Bower}, G., {et~al.} 2000, \apjl, 539, L13,
  \dodoi{10.1086/312840}

\bibitem[{{Giavalisco} {et~al.}(2004){Giavalisco}, {Ferguson}, {Koekemoer},
  {Dickinson}, {Alexander}, {Bauer}, {Bergeron}, {Biagetti}, {Brandt},
  {Casertano}, {Cesarsky}, {Chatzichristou}, {Conselice}, {Cristiani}, {Da
  Costa}, {Dahlen}, {de Mello}, {Eisenhardt}, {Erben}, {Fall}, {Fassnacht},
  {Fosbury}, {Fruchter}, {Gardner}, {Grogin}, {Hook}, {Hornschemeier}, {Idzi},
  {Jogee}, {Kretchmer}, {Laidler}, {Lee}, {Livio}, {Lucas}, {Madau},
  {Mobasher}, {Moustakas}, {Nonino}, {Padovani}, {Papovich}, {Park},
  {Ravindranath}, {Renzini}, {Richardson}, {Riess}, {Rosati}, {Schirmer},
  {Schreier}, {Somerville}, {Spinrad}, {Stern}, {Stiavelli}, {Strolger},
  {Urry}, {Vandame}, {Williams}, \& {Wolf}}]{Giavalisco2004}
{Giavalisco}, M., {Ferguson}, H.~C., {Koekemoer}, A.~M., {et~al.} 2004, \apjl,
  600, L93, \dodoi{10.1086/379232}

\bibitem[{{Gim} {et~al.}(2019){Gim}, {Yun}, {Owen}, {Momjian}, {Miller},
  {Giavalisco}, {Wilson}, {Lowenthal}, {Aretxaga}, {Hughes}, {Morrison}, \&
  {Kawabe}}]{Gim2019}
{Gim}, H.~B., {Yun}, M.~S., {Owen}, F.~N., {et~al.} 2019, \apj, 875, 80,
  \dodoi{10.3847/1538-4357/ab1011}

\bibitem[{{G{\'o}mez-Guijarro} {et~al.}(2022){G{\'o}mez-Guijarro}, {Elbaz},
  {Xiao}, {B{\'e}thermin}, {Franco}, {Magnelli}, {Daddi}, {Dickinson},
  {Demarco}, {Inami}, {Rujopakarn}, {Magdis}, {Shu}, {Chary}, {Zhou},
  {Alexander}, {Bournaud}, {Ciesla}, {Ferguson}, {Finkelstein}, {Giavalisco},
  {Iono}, {Juneau}, {Kartaltepe}, {Lagache}, {Le Floc'h}, {Leiton}, {Lin},
  {Motohara}, {Mullaney}, {Okumura}, {Pannella}, {Papovich}, {Pope}, {Sargent},
  {Silverman}, {Treister}, \& {Wang}}]{Gomez-Guijarro2022}
{G{\'o}mez-Guijarro}, C., {Elbaz}, D., {Xiao}, M., {et~al.} 2022, \aap, 658,
  A43, \dodoi{10.1051/0004-6361/202141615}

\bibitem[{{Gonzalez-Martin} {et~al.}(2010){Gonzalez-Martin}, {Acosta-Pulido},
  {Perez Garcia}, \& {Ramos Almeida}}]{Gonzalez-Martin}
{Gonzalez-Martin}, O., {Acosta-Pulido}, J.~A., {Perez Garcia}, A.~M., \& {Ramos
  Almeida}, C. 2010, \apj, 723, 1748, \dodoi{10.1088/0004-637X/723/2/1748}

\bibitem[{{Greene} {et~al.}(2011){Greene}, {Zakamska}, {Ho}, \&
  {Barth}}]{Greene2011}
{Greene}, J.~E., {Zakamska}, N.~L., {Ho}, L.~C., \& {Barth}, A.~J. 2011, \apj,
  732, 9, \dodoi{10.1088/0004-637X/732/1/9}

\bibitem[{{Guo} {et~al.}(2020){Guo}, {Gu}, {Ding}, {Contini}, \&
  {Chen}}]{Guo2020}
{Guo}, X., {Gu}, Q., {Ding}, N., {Contini}, E., \& {Chen}, Y. 2020, \mnras,
  492, 1887, \dodoi{10.1093/mnras/stz3589}

\bibitem[{{Guo} {et~al.}(2021){Guo}, {Gu}, {Ding}, {Yu}, \& {Chen}}]{Guo2021}
{Guo}, X., {Gu}, Q., {Ding}, N., {Yu}, X., \& {Chen}, Y. 2021, \apj, 908, 169,
  \dodoi{10.3847/1538-4357/abd0f5}

\bibitem[{{Hainline} {et~al.}(2013){Hainline}, {Hickox}, {Greene}, {Myers}, \&
  {Zakamska}}]{Hainline2013}
{Hainline}, K.~N., {Hickox}, R., {Greene}, J.~E., {Myers}, A.~D., \&
  {Zakamska}, N.~L. 2013, \apj, 774, 145, \dodoi{10.1088/0004-637X/774/2/145}

\bibitem[{{Hainline} {et~al.}(2014){Hainline}, {Hickox}, {Greene}, {Myers},
  {Zakamska}, {Liu}, \& {Liu}}]{Haineline2014}
{Hainline}, K.~N., {Hickox}, R.~C., {Greene}, J.~E., {et~al.} 2014, \apj, 787,
  65, \dodoi{10.1088/0004-637X/787/1/65}

\bibitem[{{Hainline} {et~al.}(2023){Hainline}, {Johnson}, {Robertson},
  {Tacchella}, {Helton}, {Sun}, {Eisenstein}, {Simmonds}, {Topping}, {Whitler},
  {Willmer}, {Rieke}, {Suess}, {Hviding}, {Cameron}, {Alberts}, {Baker},
  {Bhatawdekar}, {Boyett}, {Bunker}, {Carniani}, {Charlot}, {Chen}, {Curti},
  {Curtis-Lake}, {D'Eugenio}, {Egami}, {Endsley}, {Hausen}, {Ji}, {Looser},
  {Lyu}, {Maiolino}, {Nelson}, {Puskas}, {Rawle}, {Sandles}, {Saxena}, {Smit},
  {Stark}, {Williams}, {Willott}, \& {Witstok}}]{Hainline2023}
{Hainline}, K.~N., {Johnson}, B.~D., {Robertson}, B., {et~al.} 2023, arXiv
  e-prints, arXiv:2306.02468, \dodoi{10.48550/arXiv.2306.02468}

\bibitem[{{Harris} {et~al.}(2020){Harris}, {Millman}, {van der Walt},
  {Gommers}, {Virtanen}, {Cournapeau}, {Wieser}, {Taylor}, {Berg}, {Smith},
  {Kern}, {Picus}, {Hoyer}, {van Kerkwijk}, {Brett}, {Haldane}, {del R{\'\i}o},
  {Wiebe}, {Peterson}, {G{\'e}rard-Marchant}, {Sheppard}, {Reddy}, {Weckesser},
  {Abbasi}, {Gohlke}, \& {Oliphant}}]{harris2020}
{Harris}, C.~R., {Millman}, K.~J., {van der Walt}, S.~J., {et~al.} 2020, \nat,
  585, 357, \dodoi{10.1038/s41586-020-2649-2}

\bibitem[{{Harrison} {et~al.}(2016){Harrison}, {Alexander}, {Mullaney},
  {Stott}, {Swinbank}, {Arumugam}, {Bauer}, {Bower}, {Bunker}, \&
  {Sharples}}]{Harrison2016b}
{Harrison}, C.~M., {Alexander}, D.~M., {Mullaney}, J.~R., {et~al.} 2016,
  \mnras, 456, 1195, \dodoi{10.1093/mnras/stv2727}

\bibitem[{{He} {et~al.}(2018){He}, {Sun}, {Zakamska}, {Wylezalek}, {Kelly},
  {Greene}, {Rembold}, {Riffel}, \& {Riffel}}]{He2018}
{He}, Z., {Sun}, A.-L., {Zakamska}, N.~L., {et~al.} 2018, \mnras, 478, 3614,
  \dodoi{10.1093/mnras/sty1322}

\bibitem[{{Heckman} \& {Best}(2014)}]{Heckman&Best2014}
{Heckman}, T.~M., \& {Best}, P.~N. 2014, \araa, 52, 589,
  \dodoi{10.1146/annurev-astro-081913-035722}

\bibitem[{{Hickox} \& {Alexander}(2018)}]{Hickox&Alexander2018}
{Hickox}, R.~C., \& {Alexander}, D.~M. 2018, \araa, 56, 625,
  \dodoi{10.1146/annurev-astro-081817-051803}

\bibitem[{{Hopkins} \& {Elvis}(2010)}]{Hopkins&Elvis2010}
{Hopkins}, P.~F., \& {Elvis}, M. 2010, \mnras, 401, 7,
  \dodoi{10.1111/j.1365-2966.2009.15643.x}

\bibitem[{{Hopkins} {et~al.}(2007){Hopkins}, {Richards}, \&
  {Hernquist}}]{Hopkins2007}
{Hopkins}, P.~F., {Richards}, G.~T., \& {Hernquist}, L. 2007, \apj, 654, 731,
  \dodoi{10.1086/509629}

\bibitem[{{Hsu} {et~al.}(2014){Hsu}, {Salvato}, {Nandra}, {Brusa}, {Bender},
  {Buchner}, {Donley}, {Kocevski}, {Guo}, {Hathi}, {Rangel}, {Willner},
  {Brightman}, {Georgakakis}, {Budav{\'a}ri}, {Szalay}, {Ashby}, {Barro},
  {Dahlen}, {Faber}, {Ferguson}, {Galametz}, {Grazian}, {Grogin}, {Huang},
  {Koekemoer}, {Lucas}, {McGrath}, {Mobasher}, {Peth}, {Rosario}, \&
  {Trump}}]{Hsu2014}
{Hsu}, L.-T., {Salvato}, M., {Nandra}, K., {et~al.} 2014, \apj, 796, 60,
  \dodoi{10.1088/0004-637X/796/1/60}

\bibitem[{{Humphrey} {et~al.}(2010){Humphrey}, {Villar-Mart{\'\i}n},
  {S{\'a}nchez}, {Mart{\'\i}nez-Sansigre}, {Delgado}, {P{\'e}rez}, {Tadhunter},
  \& {P{\'e}rez-Torres}}]{Humphrey2010}
{Humphrey}, A., {Villar-Mart{\'\i}n}, M., {S{\'a}nchez}, S.~F., {et~al.} 2010,
  \mnras, 408, L1, \dodoi{10.1111/j.1745-3933.2010.00906.x}

\bibitem[{{Hunter}(2007)}]{Hunter2007}
{Hunter}, J.~D. 2007, Computing in Science and Engineering, 9, 90,
  \dodoi{10.1109/MCSE.2007.55}

\bibitem[{{Husemann} {et~al.}(2013){Husemann}, {Wisotzki}, {S{\'a}nchez}, \&
  {Jahnke}}]{Husemann2013}
{Husemann}, B., {Wisotzki}, L., {S{\'a}nchez}, S.~F., \& {Jahnke}, K. 2013,
  \aap, 549, A43, \dodoi{10.1051/0004-6361/201220076}

\bibitem[{{Joh} {et~al.}(2021){Joh}, {Nagao}, {Wada}, {Terao}, \&
  {Yamashita}}]{Joh2021}
{Joh}, K., {Nagao}, T., {Wada}, K., {Terao}, K., \& {Yamashita}, T. 2021,
  \pasj, 73, 1152, \dodoi{10.1093/pasj/psab065}

\bibitem[{Juneau {et~al.}(2021)Juneau, Olsen, Nikutta, Jacques, \&
  Bailey}]{Juneau2021_jupyter}
Juneau, S., Olsen, K., Nikutta, R., Jacques, A., \& Bailey, S. 2021, Computing
  in Science \& Engineering, 23, 15, \dodoi{10.1109/MCSE.2021.3057097}

\bibitem[{{Juneau} {et~al.}(2022){Juneau}, {Goulding}, {Banfield}, {Bianchi},
  {Duc}, {Ho}, {Dopita}, {Scharw{\"a}chter}, {Bauer}, {Groves}, {Alexander},
  {Davies}, {Elbaz}, {Freeland}, {Hampton}, {Kewley}, {Nikutta}, {Shastri},
  {Shu}, {Vogt}, {Wang}, {Wong}, \& {Woo}}]{Juneau}
{Juneau}, S., {Goulding}, A.~D., {Banfield}, J., {et~al.} 2022, \apj, 925, 203,
  \dodoi{10.3847/1538-4357/ac425f}

\bibitem[{{Kakkad} {et~al.}(2020){Kakkad}, {Mainieri}, {Vietri}, {Carniani},
  {Harrison}, {Perna}, {Scholtz}, {Circosta}, {Cresci}, {Husemann},
  {Bischetti}, {Feruglio}, {Fiore}, {Marconi}, {Padovani}, {Brusa}, {Cicone},
  {Comastri}, {Lanzuisi}, {Mannucci}, {Menci}, {Netzer}, {Piconcelli},
  {Puglisi}, {Salvato}, {Schramm}, {Silverman}, {Vignali}, {Zamorani}, \&
  {Zappacosta}}]{Kakkad2020}
{Kakkad}, D., {Mainieri}, V., {Vietri}, G., {et~al.} 2020, \aap, 642, A147,
  \dodoi{10.1051/0004-6361/202038551}

\bibitem[{{Kakkad} {et~al.}(2022){Kakkad}, {Sani}, {Rojas}, {Mallmann},
  {Veilleux}, {Bauer}, {Ricci}, {Mushotzky}, {Koss}, {Ricci}, {Treister},
  {Privon}, {Nguyen}, {B{\"a}r}, {Harrison}, {Oh}, {Powell}, {Riffel}, {Stern},
  {Trakhtenbrot}, \& {Urry}}]{Kakkad2022}
{Kakkad}, D., {Sani}, E., {Rojas}, A.~F., {et~al.} 2022, \mnras, 511, 2105,
  \dodoi{10.1093/mnras/stac103}

\bibitem[{{Keel} {et~al.}(2022){Keel}, {Moiseev}, {Kozlova}, {Ikhsanova},
  {Oparin}, {Uklein}, {Smirnova}, \& {Eselevich}}]{Keel2021}
{Keel}, W.~C., {Moiseev}, A., {Kozlova}, D.~V., {et~al.} 2022, \mnras, 510,
  4608, \dodoi{10.1093/mnras/stab3656}

\bibitem[{{Kollatschny} {et~al.}(2022){Kollatschny}, {Ochmann}, {Kaspi},
  {Schumacher}, {Behar}, {Chelouche}, {Horne}, {M{\"u}ller}, {Rafter}, {Chini},
  {Haas}, \& {Probst}}]{Kollatschny2022}
{Kollatschny}, W., {Ochmann}, M.~W., {Kaspi}, S., {et~al.} 2022, \aap, 657,
  A122, \dodoi{10.1051/0004-6361/202142007}

\bibitem[{{Kormendy} \& {Richstone}(1995)}]{Kormendy1995}
{Kormendy}, J., \& {Richstone}, D. 1995, \araa, 33, 581,
  \dodoi{10.1146/annurev.aa.33.090195.003053}

\bibitem[{{Laha} {et~al.}(2021){Laha}, {Reynolds}, {Reeves}, {Kriss},
  {Guainazzi}, {Smith}, {Veilleux}, \& {Proga}}]{Laha2021}
{Laha}, S., {Reynolds}, C.~S., {Reeves}, J., {et~al.} 2021, Nature Astronomy,
  5, 13, \dodoi{10.1038/s41550-020-01255-2}

\bibitem[{{Lamastra} {et~al.}(2009){Lamastra}, {Bianchi}, {Matt}, {Perola},
  {Barcons}, \& {Carrera}}]{Lamastra2009}
{Lamastra}, A., {Bianchi}, S., {Matt}, G., {et~al.} 2009, \aap, 504, 73,
  \dodoi{10.1051/0004-6361/200912023}

\bibitem[{{Leisenring} {et~al.}(2016){Leisenring}, {Rieke}, {Misselt}, \&
  {Robberto}}]{Leisenring2016}
{Leisenring}, J.~M., {Rieke}, M., {Misselt}, K., \& {Robberto}, M. 2016, in
  Society of Photo-Optical Instrumentation Engineers (SPIE) Conference Series,
  Vol. 9915, High Energy, Optical, and Infrared Detectors for Astronomy VII,
  ed. A.~D. {Holland} \& J.~{Beletic}, 99152N, \dodoi{10.1117/12.2233917}

\bibitem[{{Li} {et~al.}(2019){Li}, {Xue}, {Sun}, {Liu}, {Vito}, {Brandt},
  {Hughes}, {Yang}, {Tozzi}, {Zhu}, {Zheng}, {Luo}, {Chen}, {Vignali}, {Gilli},
  \& {Shu}}]{Li2019}
{Li}, J., {Xue}, Y., {Sun}, M., {et~al.} 2019, \apj, 877, 5,
  \dodoi{10.3847/1538-4357/ab184b}

\bibitem[{{Liu} {et~al.}(2014){Liu}, {Zakamska}, \& {Greene}}]{Liu2014}
{Liu}, G., {Zakamska}, N.~L., \& {Greene}, J.~E. 2014, \mnras, 442, 1303,
  \dodoi{10.1093/mnras/stu974}

\bibitem[{{Liu} {et~al.}(2013{\natexlab{a}}){Liu}, {Zakamska}, {Greene},
  {Nesvadba}, \& {Liu}}]{Liu2013a}
{Liu}, G., {Zakamska}, N.~L., {Greene}, J.~E., {Nesvadba}, N. P.~H., \& {Liu},
  X. 2013{\natexlab{a}}, \mnras, 430, 2327, \dodoi{10.1093/mnras/stt051}

\bibitem[{{Liu} {et~al.}(2013{\natexlab{b}}){Liu}, {Zakamska}, {Greene},
  {Nesvadba}, \& {Liu}}]{Liu2013b}
---. 2013{\natexlab{b}}, \mnras, 436, 2576, \dodoi{10.1093/mnras/stt1755}

\bibitem[{{L{\'o}pez-Cob{\'a}} {et~al.}(2020){L{\'o}pez-Cob{\'a}},
  {S{\'a}nchez}, {Anderson}, {Cruz-Gonz{\'a}lez}, {Galbany}, {Ruiz-Lara},
  {Barrera-Ballesteros}, {Prieto}, \& {Kuncarayakti}}]{Lopez-coba}
{L{\'o}pez-Cob{\'a}}, C., {S{\'a}nchez}, S.~F., {Anderson}, J.~P., {et~al.}
  2020, \aj, 159, 167, \dodoi{10.3847/1538-3881/ab7848}

\bibitem[{{Luo} {et~al.}(2017){Luo}, {Brandt}, {Xue}, {Lehmer}, {Alexander},
  {Bauer}, {Vito}, {Yang}, {Basu-Zych}, {Comastri}, {Gilli}, {Gu},
  {Hornschemeier}, {Koekemoer}, {Liu}, {Mainieri}, {Paolillo}, {Ranalli},
  {Rosati}, {Schneider}, {Shemmer}, {Smail}, {Sun}, {Tozzi}, {Vignali}, \&
  {Wang}}]{Luo2017}
{Luo}, B., {Brandt}, W.~N., {Xue}, Y.~Q., {et~al.} 2017, \apjs, 228, 2,
  \dodoi{10.3847/1538-4365/228/1/2}

\bibitem[{{Lyu} {et~al.}(2022){Lyu}, {Alberts}, {Rieke}, \&
  {Rujopakarn}}]{Lyu2022}
{Lyu}, J., {Alberts}, S., {Rieke}, G.~H., \& {Rujopakarn}, W. 2022, \apj, 941,
  191, \dodoi{10.3847/1538-4357/ac9e5d}

\bibitem[{{Madau} \& {Dickinson}(2014)}]{Madau2014}
{Madau}, P., \& {Dickinson}, M. 2014, \araa, 52, 415,
  \dodoi{10.1146/annurev-astro-081811-125615}

\bibitem[{{Magorrian} {et~al.}(1998){Magorrian}, {Tremaine}, {Richstone},
  {Bender}, {Bower}, {Dressler}, {Faber}, {Gebhardt}, {Green}, {Grillmair},
  {Kormendy}, \& {Lauer}}]{Magorrian1998}
{Magorrian}, J., {Tremaine}, S., {Richstone}, D., {et~al.} 1998, \aj, 115,
  2285, \dodoi{10.1086/300353}

\bibitem[{{Martin}(2005)}]{Martin2005}
{Martin}, C.~L. 2005, \apj, 621, 227, \dodoi{10.1086/427277}

\bibitem[{{May} {et~al.}(2018){May}, {Rodr{\'\i}guez-Ardila}, {Prieto},
  {Fern{\'a}ndez-Ontiveros}, {Diaz}, \& {Mazzalay}}]{May2018}
{May}, D., {Rodr{\'\i}guez-Ardila}, A., {Prieto}, M.~A., {et~al.} 2018, \mnras,
  481, L105, \dodoi{10.1093/mnrasl/sly155}

\bibitem[{{McLure} {et~al.}(2017){McLure}, {Pentericci}, \& {VANDELS
  Team}}]{McLure2017}
{McLure}, R., {Pentericci}, L., \& {VANDELS Team}. 2017, The Messenger, 167, 31

\bibitem[{{Mingozzi} {et~al.}(2019){Mingozzi}, {Cresci}, {Venturi}, {Marconi},
  {Mannucci}, {Perna}, {Belfiore}, {Carniani}, {Balmaverde}, {Brusa}, {Cicone},
  {Feruglio}, {Gallazzi}, {Mainieri}, {Maiolino}, {Nagao}, {Nardini}, {Sani},
  {Tozzi}, \& {Zibetti}}]{Mingozzi}
{Mingozzi}, M., {Cresci}, G., {Venturi}, G., {et~al.} 2019, \aap, 622, A146,
  \dodoi{10.1051/0004-6361/201834372}

\bibitem[{{Momcheva} {et~al.}(2016){Momcheva}, {Brammer}, {van Dokkum},
  {Skelton}, {Whitaker}, {Nelson}, {Fumagalli}, {Maseda}, {Leja}, {Franx},
  {Rix}, {Bezanson}, {Da Cunha}, {Dickey}, {F{\"o}rster Schreiber},
  {Illingworth}, {Kriek}, {Labb{\'e}}, {Ulf Lange}, {Lundgren}, {Magee},
  {Marchesini}, {Oesch}, {Pacifici}, {Patel}, {Price}, {Tal}, {Wake}, {van der
  Wel}, \& {Wuyts}}]{Momcheva2016}
{Momcheva}, I.~G., {Brammer}, G.~B., {van Dokkum}, P.~G., {et~al.} 2016, \apjs,
  225, 27, \dodoi{10.3847/0067-0049/225/2/27}

\bibitem[{{Morganti}(2017)}]{Morganti2017}
{Morganti}, R. 2017, Frontiers in Astronomy and Space Sciences, 4, 42,
  \dodoi{10.3389/fspas.2017.00042}

\bibitem[{{Morganti} {et~al.}(1998){Morganti}, {Oosterloo}, \&
  {Tsvetanov}}]{Morganti1998}
{Morganti}, R., {Oosterloo}, T., \& {Tsvetanov}, Z. 1998, \aj, 115, 915,
  \dodoi{10.1086/300236}

\bibitem[{{Nesvadba} {et~al.}(2008){Nesvadba}, {Lehnert}, {De Breuck},
  {Gilbert}, \& {van Breugel}}]{Nesvadba2008}
{Nesvadba}, N.~P.~H., {Lehnert}, M.~D., {De Breuck}, C., {Gilbert}, A.~M., \&
  {van Breugel}, W. 2008, \aap, 491, 407, \dodoi{10.1051/0004-6361:200810346}

\bibitem[{{Netzer}(2015)}]{Netzer2015}
{Netzer}, H. 2015, \araa, 53, 365, \dodoi{10.1146/annurev-astro-082214-122302}

\bibitem[{Nikutta {et~al.}(2020)Nikutta, Fitzpatrick, Scott, \&
  Weaver}]{Nikutta2020}
Nikutta, R., Fitzpatrick, M., Scott, A., \& Weaver, B. 2020, Astronomy and
  Computing, 33, 100411, \dodoi{https://doi.org/10.1016/j.ascom.2020.100411}

\bibitem[{{Oesch} {et~al.}(2023){Oesch}, {Brammer}, {Naidu}, {Bouwens},
  {Chisholm}, {Illingworth}, {Matthee}, {Nelson}, {Qin}, {Reddy}, {Shapley},
  {Shivaei}, {van Dokkum}, {Weibel}, {Whitaker}, {Wuyts}, {Covelo-Paz},
  {Endsley}, {Fudamoto}, {Giovinazzo}, {Herard-Demanche}, {Kerutt},
  {Kramarenko}, {Labbe}, {Leonova}, {Lin}, {Magee}, {Marchesini}, {Maseda},
  {Mason}, {Matharu}, {Meyer}, {Neufeld}, {Prieto Lyon}, {Schaerer}, {Sharma},
  {Shuntov}, {Smit}, {Stefanon}, {Wyithe}, \& {Xiao}}]{Oesch2023}
{Oesch}, P.~A., {Brammer}, G., {Naidu}, R.~P., {et~al.} 2023, \mnras, 525,
  2864, \dodoi{10.1093/mnras/stad2411}

\bibitem[{{Oesch, Pascal} \& {Magee, Dan}(2023)}]{FRESCO_doi}
{Oesch, Pascal}, \& {Magee, Dan}. 2023, The JWST FRESCO Survey,  STScI/MAST,
  \dodoi{10.17909/GDYC-7G80}

\bibitem[{{Pantoni} {et~al.}(2021){Pantoni}, {Lapi}, {Massardi}, {Donevski},
  {Bressan}, {Silva}, {Pozzi}, {Vignali}, {Talia}, {Cimatti}, {Ronconi}, \&
  {Danese}}]{Pantoni2021}
{Pantoni}, L., {Lapi}, A., {Massardi}, M., {et~al.} 2021, \mnras, 504, 928,
  \dodoi{10.1093/mnras/stab674}

\bibitem[{{Pei}(1995)}]{Pei1995}
{Pei}, Y.~C. 1995, \apj, 438, 623, \dodoi{10.1086/175105}

\bibitem[{{Peimbert} {et~al.}(2017){Peimbert}, {Peimbert}, \&
  {Delgado-Inglada}}]{Peimbert2017}
{Peimbert}, M., {Peimbert}, A., \& {Delgado-Inglada}, G. 2017, \pasp, 129,
  082001, \dodoi{10.1088/1538-3873/aa72c3}

\bibitem[{{Perrin} {et~al.}(2014){Perrin}, {Sivaramakrishnan}, {Lajoie},
  {Elliott}, {Pueyo}, {Ravindranath}, \& {Albert}}]{Perrin2014}
{Perrin}, M.~D., {Sivaramakrishnan}, A., {Lajoie}, C.-P., {et~al.} 2014, in
  Society of Photo-Optical Instrumentation Engineers (SPIE) Conference Series,
  Vol. 9143, Space Telescopes and Instrumentation 2014: Optical, Infrared, and
  Millimeter Wave, ed. J.~{Oschmann}, Jacobus~M., M.~{Clampin}, G.~G. {Fazio},
  \& H.~A. {MacEwen}, 91433X, \dodoi{10.1117/12.2056689}

\bibitem[{{Puchwein} \& {Springel}(2013)}]{Puchwein2013}
{Puchwein}, E., \& {Springel}, V. 2013, \mnras, 428, 2966,
  \dodoi{10.1093/mnras/sts243}

\bibitem[{{Rauch} {et~al.}(2013){Rauch}, {Becker}, {Haehnelt}, {Carswell}, \&
  {Gauthier}}]{Rauch2013}
{Rauch}, M., {Becker}, G.~D., {Haehnelt}, M.~G., {Carswell}, R.~F., \&
  {Gauthier}, J.~R. 2013, \mnras, 431, L68, \dodoi{10.1093/mnrasl/slt010}

\bibitem[{{Revalski}(2019)}]{Revalski2019}
{Revalski}, M. 2019, PhD thesis, Georgia State University

\bibitem[{{Revalski} {et~al.}(2021){Revalski}, {Meena}, {Martinez}, {Polack},
  {Crenshaw}, {Kraemer}, {Collins}, {Fischer}, {Schmitt}, {Schmidt}, {Maksym},
  \& {Rafelski}}]{Revalski2021}
{Revalski}, M., {Meena}, B., {Martinez}, F., {et~al.} 2021, \apj, 910, 139,
  \dodoi{10.3847/1538-4357/abdcad}

\bibitem[{{Rieke} {et~al.}(2005){Rieke}, {Kelly}, \& {Horner}}]{Rieke2005}
{Rieke}, M.~J., {Kelly}, D., \& {Horner}, S. 2005, in Society of Photo-Optical
  Instrumentation Engineers (SPIE) Conference Series, Vol. 5904, Cryogenic
  Optical Systems and Instruments XI, ed. J.~B. {Heaney} \& L.~G. {Burriesci},
  1--8, \dodoi{10.1117/12.615554}

\bibitem[{{Rieke} {et~al.}(2003){Rieke}, {Baum}, {Beichman}, {Crampton},
  {Doyon}, {Eisenstein}, {Greene}, {Hodapp}, {Horner}, {Johnstone}, {Lesyna},
  {Lilly}, {Meyer}, {Martin}, {McCarthy}, {Rieke}, {Roellig}, {Stauffer},
  {Trauger}, \& {Young}}]{Rieke2003}
{Rieke}, M.~J., {Baum}, S.~A., {Beichman}, C.~A., {et~al.} 2003, in Society of
  Photo-Optical Instrumentation Engineers (SPIE) Conference Series, Vol. 4850,
  IR Space Telescopes and Instruments, ed. J.~C. {Mather}, 478--485,
  \dodoi{10.1117/12.489103}

\bibitem[{{Rieke} {et~al.}(2023{\natexlab{a}}){Rieke}, {Kelly}, {Misselt},
  {Stansberry}, {Boyer}, {Beatty}, {Egami}, {Florian}, {Greene}, {Hainline},
  {Leisenring}, {Roellig}, {Schlawin}, {Sun}, {Tinnin}, {Williams}, {Willmer},
  {Wilson}, {Clark}, {Rohrbach}, {Brooks}, {Canipe}, {Correnti}, {DiFelice},
  {Gennaro}, {Girard}, {Hartig}, {Hilbert}, {Koekemoer}, {Nikolov}, {Pirzkal},
  {Rest}, {Robberto}, {Sunnquist}, {Telfer}, {Wu}, {Ferry}, {Lewis}, {Baum},
  {Beichman}, {Doyon}, {Dressler}, {Eisenstein}, {Ferrarese}, {Hodapp},
  {Horner}, {Jaffe}, {Johnstone}, {Krist}, {Martin}, {McCarthy}, {Meyer},
  {Rieke}, {Trauger}, \& {Young}}]{Rieke2023a}
{Rieke}, M.~J., {Kelly}, D.~M., {Misselt}, K., {et~al.} 2023{\natexlab{a}},
  \pasp, 135, 028001, \dodoi{10.1088/1538-3873/acac53}

\bibitem[{{Rieke} {et~al.}(2023{\natexlab{b}}){Rieke}, {Robertson},
  {Tacchella}, {Hainline}, {Johnson}, {Hausen}, {Ji}, {Willmer}, {Eisenstein},
  {Pusk{\'a}s}, {Alberts}, {Arribas}, {Baker}, {Baum}, {Bhatawdekar},
  {Bonaventura}, {Boyett}, {Bunker}, {Cameron}, {Carniani}, {Charlot},
  {Chevallard}, {Chen}, {Curti}, {Curtis-Lake}, {Danhaive}, {DeCoursey},
  {Dressler}, {Egami}, {Endsley}, {Helton}, {Hviding}, {Kumari}, {Looser},
  {Lyu}, {Maiolino}, {Maseda}, {Nelson}, {Rieke}, {Rix}, {Sandles}, {Saxena},
  {Sharpe}, {Shivaei}, {Skarbinski}, {Smit}, {Stark}, {Stone}, {Suess}, {Sun},
  {Topping}, {{\"U}bler}, {Villanueva}, {Wallace}, {Williams}, {Willott},
  {Whitler}, {Witstok}, \& {Woodrum}}]{Rieke2023b}
{Rieke}, M.~J., {Robertson}, B., {Tacchella}, S., {et~al.} 2023{\natexlab{b}},
  \apjs, 269, 16, \dodoi{10.3847/1538-4365/acf44d}

\bibitem[{{Rieke, Marcia} {et~al.}(2023{\natexlab{a}}){Rieke, Marcia},
  {Robertson, Brant}, {Tacchella, Sandro}, {Willmer, Christopher}, {Johnson,
  Ben}, {Carniani, Stefano}, {Bunker, Andy}, \& {Willott, Chris}}]{JADES_doi}
{Rieke, Marcia}, {Robertson, Brant}, {Tacchella, Sandro}, {et~al.}
  2023{\natexlab{a}}, Data from the JWST Advanced Deep Extragalactic Survey
  (JADES),  STScI/MAST, \dodoi{10.17909/8TDJ-8N28}

\bibitem[{{Rieke, Marcia} {et~al.}(2023{\natexlab{b}}){Rieke, Marcia},
  {Robertson, Brant}, {Tacchella, Sandro}, {Willmer, Christopher}, {Johnson,
  Ben}, {Carniani, Stefano}, {Bunker, Andy}, \& {Willott,
  Chris}}]{JADES_DR2_doi}
---. 2023{\natexlab{b}}, Data from the JWST Advanced Deep Extragalactic Survey
  (JADES),  STScI/MAST, \dodoi{10.17909/8TDJ-8N28}

\bibitem[{{Riffel} {et~al.}(2009){Riffel}, {Storchi-Bergmann}, {Dors}, \&
  {Winge}}]{Riffel2009}
{Riffel}, R.~A., {Storchi-Bergmann}, T., {Dors}, O.~L., \& {Winge}, C. 2009,
  \mnras, 393, 783, \dodoi{10.1111/j.1365-2966.2008.14250.x}

\bibitem[{{Rigby} {et~al.}(2023){Rigby}, {Perrin}, {McElwain}, {Kimble},
  {Friedman}, {Lallo}, {Doyon}, {Feinberg}, {Ferruit}, {Glasse}, {Rieke},
  {Rieke}, {Wright}, {Willott}, {Colon}, {Milam}, {Neff}, {Stark}, {Valenti},
  {Abell}, {Abney}, {Abul-Huda}, {Acton}, {Adams}, {Adler}, {Aguilar}, {Ahmed},
  {Albert}, {Alberts}, {Aldridge}, {Allen}, {Altenburg},
  {{\'A}lvarez-M{\'a}rquez}, {Alves de Oliveira}, {Andersen}, {Anderson},
  {Anderson}, {Argyriou}, {Armstrong}, {Arribas}, {Artigau}, {Arvai},
  {Atkinson}, {Bacon}, {Bair}, {Banks}, {Barrientes}, {Barringer}, {Bartosik},
  {Bast}, {Baudoz}, {Beatty}, {Bechtold}, {Beck}, {Bergeron}, {Bergkoetter},
  {Bhatawdekar}, {Birkmann}, {Blazek}, {Blome}, {Boccaletti}, {B{\"o}ker},
  {Boia}, {Bonaventura}, {Bond}, {Bosley}, {Boucarut}, {Bourque}, {Bouwman},
  {Bower}, {Bowers}, {Boyer}, {Bradley}, {Brady}, {Braun}, {Breda},
  {Bresnahan}, {Bright}, {Britt}, {Bromenschenkel}, {Brooks}, {Brooks},
  {Brown}, {Brown}, {Brown}, {Bunker}, {Burger}, {Bushouse}, {Cale}, {Cameron},
  {Cameron}, {Canipe}, {Caplinger}, {Caputo}, {Cara}, {Carey}, {Carniani},
  {Carrasquilla}, {Carruthers}, {Case}, {Catherine}, {Chance}, {Chapman},
  {Charlot}, {Charlow}, {Chayer}, {Chen}, {Cherinka}, {Chichester}, {Chilton},
  {Chonis}, {Clampin}, {Clark}, {Clark}, {Coe}, {Coleman}, {Comber}, {Comeau},
  {Connolly}, {Cooper}, {Cooper}, {Coppock}, {Correnti}, {Cossou}, {Coulais},
  {Coyle}, {Cracraft}, {Curti}, {Cuturic}, {Davis}, {Davis}, {Dean}, {DeLisa},
  {deMeester}, {Dencheva}, {Dencheva}, {DePasquale}, {Deschenes}, {Hunor
  Detre}, {Diaz}, {Dicken}, {DiFelice}, {Dillman}, {Dixon}, {Doggett},
  {Donaldson}, {Douglas}, {DuPrie}, {Dupuis}, {Durning}, {Easmin}, {Eck},
  {Edeani}, {Egami}, {Ehrenwinkler}, {Eisenhamer}, {Eisenhower}, {Elie},
  {Elliott}, {Elliott}, {Ellis}, {Engesser}, {Espinoza}, {Etienne}, {Etxaluze},
  {Falini}, {Feeney}, {Ferry}, {Filippazzo}, {Fincham}, {Fix}, {Flagey},
  {Florian}, {Flynn}, {Fontanella}, {Ford}, {Forshay}, {Fox}, {Franz}, {Fu},
  {Fullerton}, {Galkin}, {Galyer}, {Garc{\'\i}a Mar{\'\i}n}, {Gardner},
  {Gardner}, {Garland}, {Garrett}, {Gasman}, {Gaspar}, {Gaudreau}, {Gauthier},
  {Geers}, {Geithner}, {Gennaro}, {Giardino}, {Girard}, {Giuliano},
  {Glassmire}, {Glauser}, {Glazer}, {Godfrey}, {Golimowski}, {Gollnitz},
  {Gong}, {Gonzaga}, {Gordon}, {Gordon}, {Goudfrooij}, {Greene}, {Greenhouse},
  {Grimaldi}, {Groebner}, {Grundy}, {Guillard}, {Gutman}, {Ha}, {Haderlein},
  {Hagedorn}, {Hainline}, {Haley}, {Hami}, {Hamilton}, {Hammel}, {Hansen},
  {Harkins}, {Harr}, {Hart}, {Hart}, {Hartig}, {Hashimoto}, {Haskins},
  {Hathaway}, {Havey}, {Hayden}, {Hecht}, {Heller-Boyer}, {Henriques}, {Henry},
  {Hermann}, {Hernandez}, {Hesman}, {Hicks}, {Hilbert}, {Hines}, {Hoffman},
  {Holfeltz}, {Holler}, {Hoppa}, {Hott}, {Howard}, {Howard}, {Hunter},
  {Hunter}, {Hurst}, {Husemann}, {Hustak}, {Ilinca Ignat}, {Illingworth},
  {Irish}, {Jackson}, {Jahromi}, {Jakobsen}, {James}, {James}, {Januszewski},
  {Jenkins}, {Jirdeh}, {Johnson}, {Johnson}, {Jones}, {Jones}, {Jones},
  {Jones}, {Jordan}, {Jordan}, {Jurczyk}, {Jurling}, {Kaleida}, {Kalmanson},
  {Kammerer}, {Kang}, {Kao}, {Karakla}, {Kavanagh}, {Kelly}, {Kendrew},
  {Kennedy}, {Kenny}, {Keski-kuha}, {Keyes}, {Kidwell}, {Kinzel}, {Kirk},
  {Kirkpatrick}, {Kirshenblat}, {Klaassen}, {Knapp}, {Knight}, {Knollenberg},
  {Koehler}, {Koekemoer}, {Kovacs}, {Kulp}, {Kumari}, {Kyprianou}, {La Massa},
  {Labador}, {Labiano}, {Lagage}, {Lajoie}, {Lallo}, {Lam}, {Lamb}, {Lambros},
  {Lampenfield}, {Langston}, {Larson}, {Law}, {Lawrence}, {Lee}, {Leisenring},
  {Lepo}, {Leveille}, {Levenson}, {Levine}, {Levy}, {Lewis}, {Lewis},
  {Libralato}, {Lightsey}, {Link}, {Liu}, {Lo}, {Lockwood}, {Logue}, {Long},
  {Long}, {Loomis}, {Lopez-Caniego}, {Lorenzo Alvarez}, {Love-Pruitt}, {Lucy},
  {Luetzgendorf}, {Maghami}, {Maiolino}, {Major}, {Malla}, {Malumuth},
  {Manjavacas}, {Mannfolk}, {Marrione}, {Marston}, {Martel}, {Maschmann},
  {Masci}, {Masciarelli}, {Maszkiewicz}, {Mather}, {McKenzie}, {McLean},
  {McMaster}, {Melbourne}, {Mel{\'e}ndez}, {Menzel}, {Merz}, {Meyett}, {Meza},
  {Miskey}, {Misselt}, {Moller}, {Morrison}, {Morse}, {Moseley}, {Mosier},
  {Mountain}, {Mueckay}, {Mueller}, {Mullally}, {Murphy}, {Murray}, {Murray},
  {Mustelier}, {Muzerolle}, {Mycroft}, {Myers}, {Myrick}, {Nanavati}, {Nance},
  {Nayak}, {Naylor}, {Nelan}, {Nickson}, {Nielson}, {Nieto-Santisteban},
  {Nikolov}, {Noriega-Crespo}, {O'Shaughnessy}, {O'Sullivan}, {Ochs}, {Ogle},
  {Oleszczuk}, {Olmsted}, {Osborne}, {Ottens}, {Owens}, {Pacifici}, {Pagan},
  {Page}, {Park}, {Parrish}, {Patapis}, {Paul}, {Pauly}, {Pavlovsky}, {Pedder},
  {Peek}, {Pena-Guerrero}, {Penanen}, {Perez}, {Perna}, {Perriello},
  {Phillips}, {Pietraszkiewicz}, {Pinaud}, {Pirzkal}, {Pitman}, {Piwowar},
  {Platais}, {Player}, {Plesha}, {Pollizi}, {Polster}, {Pontoppidan},
  {Porterfield}, {Proffitt}, {Pueyo}, {Pulliam}, {Quirt}, {Quispe Neira},
  {Ramos Alarcon}, {Ramsay}, {Rapp}, {Rapp}, {Rauscher}, {Ravindranath},
  {Rawle}, {Regan}, {Reichard}, {Reis}, {Ressler}, {Rest}, {Reynolds}, {Rhue},
  {Richon}, {Rickman}, {Ridgaway}, {Ritchie}, {Rix}, {Robberto}, {Robinson},
  {Robinson}, {Robinson}, {Rock}, {Rodriguez}, {Rodriguez Del Pino}, {Roellig},
  {Rohrbach}, {Roman}, {Romelfanger}, {Rose}, {Roteliuk}, {Roth}, {Rothwell},
  {Rowlands}, {Roy}, {Royer}, {Royle}, {Rui}, {Rumler}, {Runnels}, {Russ},
  {Rustamkulov}, {Ryden}, {Ryer}, {Sabata}, {Sabatke}, {Sabbi}, {Samuelson},
  {Sapp}, {Sappington}, {Sargent}, {Sauer}, {Scheithauer}, {Schlawin},
  {Schlitz}, {Schmitz}, {Schneider}, {Schreiber}, {Schulze}, {Schwab}, {Scott},
  {Sembach}, {Shanahan}, {Shaughnessy}, {Shaw}, {Shawger}, {Shay}, {Sheehan},
  {Shen}, {Sherman}, {Shiao}, {Shih}, {Shivaei}, {Sienkiewicz}, {Sing},
  {Sirianni}, {Sivaramakrishnan}, {Skipper}, {Sloan}, {Slocum}, {Slowinski},
  {Smith}, {Smith}, {Smith}, {Smith}, {Snyder}, {Soh}, {Sohn}, {Soto},
  {Spencer}, {Stallcup}, {Stansberry}, {Starr}, {Starr}, {Stewart},
  {Stiavelli}, {Straughn}, {Strickland}, {Stys}, {Summers}, {Sun}, {Sunnquist},
  {Swade}, {Swam}, {Swaters}, {Swoish}, {Taylor}, {Taylor}, {Te Plate}, {Tea},
  {Teague}, {Telfer}, {Temim}, {Thatte}, {Thompson}, {Thompson}, {Thomson},
  {Tikkanen}, {Tippet}, {Todd}, {Toolan}, {Tran}, {Trejo}, {Truong},
  {Tsukamoto}, {Tustain}, {Tyra}, {Ubeda}, {Underwood}, {Uzzo}, {Van Campen},
  {Vandal}, {Vandenbussche}, {Vila}, {Volk}, {Wahlgren}, {Waldman}, {Walker},
  {Wander}, {Warfield}, {Warner}, {Wasiak}, {Watkins}, {Weaver}, {Weilert},
  {Weiser}, {Weiss}, {Weissman}, {Welty}, {West}, {Wheate}, {Wheatley},
  {Wheeler}, {White}, {Whiteaker}, {Whitehouse}, {Whiteleather}, {Whitman},
  {Williams}, {Willmer}, {Willoughby}, {Wilson}, {Wirth}, {Wislowski}, {Wolf},
  {Wolfe}, {Wolff}, {Workman}, {Wright}, {Wu}, {Wu}, {Wymer}, {Yates},
  {Yeager}, {Yeates}, {Yerger}, {Yoon}, {Young}, {Yu}, {Zak}, {Zeidler},
  {Zhou}, {Zielinski}, {Zincke}, \& {Zonak}}]{Rigby2023}
{Rigby}, J., {Perrin}, M., {McElwain}, M., {et~al.} 2023, \pasp, 135, 048001,
  \dodoi{10.1088/1538-3873/acb293}

\bibitem[{{Rinaldi} {et~al.}(2023){Rinaldi}, {Caputi}, {Costantin}, {Gillman},
  {Iani}, {P{\'e}rez-Gonz{\'a}lez}, {{\"O}stlin}, {Colina}, {Greve},
  {Noorgard-Nielsen}, {Wright}, {Alonso-Herrero}, {{\'A}lvarez-M{\'a}rquez},
  {Eckart}, {Garc{\'\i}a-Mar{\'\i}n}, {Hjorth}, {Ilbert}, {Kendrew}, {Labiano},
  {Le F{\`e}vre}, {Pye}, {Tikkanen}, {Walter}, {van der Werf}, {Ward},
  {Annunziatella}, {Azzollini}, {Bik}, {Boogaard}, {Bosman}, {Crespo
  G{\'o}mez}, {Jermann}, {Langeroodi}, {Melinder}, {Meyer}, {Moutard},
  {Peissker}, {Topinka}, {van Dishoeck}, {G{\"u}del}, {Henning}, {Lagage},
  {Ray}, {Vandenbussche}, {Waelkens}, {Navarro-Carrera}, \&
  {Kokorev}}]{Rinaldi2023}
{Rinaldi}, P., {Caputi}, K.~I., {Costantin}, L., {et~al.} 2023, \apj, 952, 143,
  \dodoi{10.3847/1538-4357/acdc27}

\bibitem[{{Robberto}(2010)}]{Robberto2010}
{Robberto}, M. 2010, {NIRCam Optimal Readout II: General Case (Including Photon
  Noise)}, Technical Report JWST-STScI-002100

\bibitem[{{Romaniello}(2022)}]{Romaniello2022}
{Romaniello}, M. 2022, in Society of Photo-Optical Instrumentation Engineers
  (SPIE) Conference Series, Vol. 12186, Observatory Operations: Strategies,
  Processes, and Systems IX, ed. D.~S. {Adler}, R.~L. {Seaman}, \& C.~R.
  {Benn}, 121860D, \dodoi{10.1117/12.2628253}

\bibitem[{{Romaniello} {et~al.}(2023){Romaniello}, {Arnaboldi}, {Barbieri},
  {Delmotte}, {Dobrzycki}, {Fourniol}, {Freudling}, {Grave}, {Mascetti},
  {Micol}, {Retzlaff}, {Rosse}, {Tax}, {Vuong}, {Hainaut}, {Rejkuba}, \&
  {Sterzik}}]{Romaniello2023}
{Romaniello}, M., {Arnaboldi}, M., {Barbieri}, M., {et~al.} 2023, The
  Messenger, 191, 29, \dodoi{10.18727/0722-6691/5338}

\bibitem[{{Rupke} {et~al.}(2005){Rupke}, {Veilleux}, \& {Sanders}}]{Rupke2005}
{Rupke}, D.~S., {Veilleux}, S., \& {Sanders}, D.~B. 2005, \apj, 632, 751,
  \dodoi{10.1086/444451}

\bibitem[{{Sarajedini} {et~al.}(2011){Sarajedini}, {Koo}, {Klesman}, {Laird},
  {Perez Gonzalez}, \& {Mozena}}]{Sarajedini2011}
{Sarajedini}, V.~L., {Koo}, D.~C., {Klesman}, A.~J., {et~al.} 2011, \apj, 731,
  97, \dodoi{10.1088/0004-637X/731/2/97}

\bibitem[{{Scharr{\'e}} {et~al.}(2024){Scharr{\'e}}, {Sorini}, \&
  {Dav{\'e}}}]{Scharre2024}
{Scharr{\'e}}, L., {Sorini}, D., \& {Dav{\'e}}, R. 2024, arXiv e-prints,
  arXiv:2404.07252, \dodoi{10.48550/arXiv.2404.07252}

\bibitem[{{Sijacki} {et~al.}(2007){Sijacki}, {Springel}, {Di Matteo}, \&
  {Hernquist}}]{Sijacki2007}
{Sijacki}, D., {Springel}, V., {Di Matteo}, T., \& {Hernquist}, L. 2007,
  \mnras, 380, 877, \dodoi{10.1111/j.1365-2966.2007.12153.x}

\bibitem[{{Solimano} {et~al.}(2025){Solimano}, {Gonz{\'a}lez-L{\'o}pez},
  {Aravena}, {Alcalde Pampliega}, {Assef}, {B{\'e}thermin}, {Boquien},
  {Bovino}, {Casey}, {Cassata}, {da Cunha}, {Davies}, {De Looze}, {Ding},
  {D{\'\i}az-Santos}, {Faisst}, {Ferrara}, {Fisher}, {F{\"o}rster-Schreiber},
  {Fujimoto}, {Ginolfi}, {Gruppioni}, {Guaita}, {Hathi}, {Herrera-Camus},
  {Ibar}, {Inami}, {Jones}, {Koekemoer}, {Lee}, {Li}, {Liu}, {Liu}, {Molina},
  {Ogle}, {Posses}, {Pozzi}, {Rela{\~n}o}, {Riechers}, {Romano}, {Spilker},
  {Sulzenauer}, {Telikova}, {Vallini}, {Vasan}, {Veilleux}, {Vergani},
  {Villanueva}, {Wang}, {Yan}, \& {Zamorani}}]{Solimano2025}
{Solimano}, M., {Gonz{\'a}lez-L{\'o}pez}, J., {Aravena}, M., {et~al.} 2025,
  \aap, 693, A70, \dodoi{10.1051/0004-6361/202451551}

\bibitem[{{Soltan}(1982)}]{Soltan1982}
{Soltan}, A. 1982, \mnras, 200, 115, \dodoi{10.1093/mnras/200.1.115}

\bibitem[{{Springel} {et~al.}(2005){Springel}, {Di Matteo}, \&
  {Hernquist}}]{Springel2005}
{Springel}, V., {Di Matteo}, T., \& {Hernquist}, L. 2005, \mnras, 361, 776,
  \dodoi{10.1111/j.1365-2966.2005.09238.x}

\bibitem[{{Straatman} {et~al.}(2016){Straatman}, {Spitler}, {Quadri},
  {Labb{\'e}}, {Glazebrook}, {Persson}, {Papovich}, {Tran}, {Brammer},
  {Cowley}, {Tomczak}, {Nanayakkara}, {Alcorn}, {Allen}, {Broussard}, {van
  Dokkum}, {Forrest}, {van Houdt}, {Kacprzak}, {Kawinwanichakij}, {Kelson},
  {Lee}, {McCarthy}, {Mehrtens}, {Monson}, {Murphy}, {Rees}, {Tilvi}, \&
  {Whitaker}}]{Straatman2016}
{Straatman}, C. M.~S., {Spitler}, L.~R., {Quadri}, R.~F., {et~al.} 2016, \apj,
  830, 51, \dodoi{10.3847/0004-637X/830/1/51}

\bibitem[{{Stutz} {et~al.}(2008){Stutz}, {Papovich}, \&
  {Eisenstein}}]{Stutz2008}
{Stutz}, A.~M., {Papovich}, C., \& {Eisenstein}, D.~J. 2008, \apj, 677, 828,
  \dodoi{10.1086/529114}

\bibitem[{{Sun} {et~al.}(2017){Sun}, {Greene}, \& {Zakamska}}]{Sun2017}
{Sun}, A.-L., {Greene}, J.~E., \& {Zakamska}, N.~L. 2017, \apj, 835, 222,
  \dodoi{10.3847/1538-4357/835/2/222}

\bibitem[{{Szokoly} {et~al.}(2004){Szokoly}, {Bergeron}, {Hasinger}, {Lehmann},
  {Kewley}, {Mainieri}, {Nonino}, {Rosati}, {Giacconi}, {Gilli}, {Gilmozzi},
  {Norman}, {Romaniello}, {Schreier}, {Tozzi}, {Wang}, {Zheng}, \&
  {Zirm}}]{Szokoly2004}
{Szokoly}, G.~P., {Bergeron}, J., {Hasinger}, G., {et~al.} 2004, \apjs, 155,
  271, \dodoi{10.1086/424707}

\bibitem[{{Szomoru} {et~al.}(2012){Szomoru}, {Franx}, \& {van
  Dokkum}}]{Szomoru2012}
{Szomoru}, D., {Franx}, M., \& {van Dokkum}, P.~G. 2012, \apj, 749, 121,
  \dodoi{10.1088/0004-637X/749/2/121}

\bibitem[{{Tasca} {et~al.}(2017){Tasca}, {Le F{\`e}vre}, {Ribeiro}, {Thomas},
  {Moreau}, {Cassata}, {Garilli}, {Le Brun}, {Lemaux}, {Maccagni},
  {Pentericci}, {Schaerer}, {Vanzella}, {Zamorani}, {Zucca}, {Amorin},
  {Bardelli}, {Cassar{\`a}}, {Castellano}, {Cimatti}, {Cucciati}, {Durkalec},
  {Fontana}, {Giavalisco}, {Grazian}, {Hathi}, {Ilbert}, {Paltani}, {Pforr},
  {Scodeggio}, {Sommariva}, {Talia}, {Tresse}, {Vergani}, {Capak}, {Charlot},
  {Contini}, {de la Torre}, {Dunlop}, {Fotopoulou}, {Guaita}, {Koekemoer},
  {L{\'o}pez-Sanjuan}, {Mellier}, {Salvato}, {Scoville}, {Taniguchi}, \&
  {Wang}}]{Tasca2017}
{Tasca}, L.~A.~M., {Le F{\`e}vre}, O., {Ribeiro}, B., {et~al.} 2017, \aap, 600,
  A110, \dodoi{10.1051/0004-6361/201527963}

\bibitem[{{Tozzi} {et~al.}(2024){Tozzi}, {Cresci}, {Perna}, {Mainieri},
  {Mannucci}, {Marconi}, {Kakkad}, {Marasco}, {Brusa}, {Bertola}, {Bischetti},
  {Carniani}, {Cicone}, {Circosta}, {Fiore}, {Feruglio}, {Harrison},
  {Lamperti}, {Netzer}, {Piconcelli}, {Puglisi}, {Scholtz}, {Vietri},
  {Vignali}, \& {Zamorani}}]{Tozzi2024}
{Tozzi}, G., {Cresci}, G., {Perna}, M., {et~al.} 2024, \aap, 690, A141,
  \dodoi{10.1051/0004-6361/202450162}

\bibitem[{{Vayner} {et~al.}(2023){Vayner}, {Zakamska}, {Ishikawa}, {Sankar},
  {Wylezalek}, {Rupke}, {Veilleux}, {Bertemes}, {Barrera-Ballesteros}, {Chen},
  {Diachenko}, {Goulding}, {Greene}, {Hainline}, {Hamann}, {Heckman},
  {Johnson}, {Grace Lim}, {Liu}, {Lutz}, {L{\"u}tzgendorf}, {Mainieri},
  {McCrory}, {Murphree}, {Nesvadba}, {Ogle}, {Sturm}, \&
  {Whitesell}}]{Vayner2023}
{Vayner}, A., {Zakamska}, N.~L., {Ishikawa}, Y., {et~al.} 2023, \apj, 955, 92,
  \dodoi{10.3847/1538-4357/ace784}

\bibitem[{{Vayner} {et~al.}(2024){Vayner}, {Zakamska}, {Ishikawa}, {Sankar},
  {Wylezalek}, {Rupke}, {Veilleux}, {Bertemes}, {Barrera-Ballesteros}, {Chen},
  {Diachenko}, {Goulding}, {Greene}, {Hainline}, {Hamann}, {Heckman},
  {Johnson}, {Grace Lim}, {Liu}, {Lutz}, {L{\"u}tzgendorf}, {Mainieri},
  {McCrory}, {Murphree}, {Nesvadba}, {Ogle}, {Sturm}, \&
  {Whitesell}}]{Vayner2024}
---. 2024, \apj, 960, 126, \dodoi{10.3847/1538-4357/ad0be9}

\bibitem[{{Veilleux} {et~al.}(2005){Veilleux}, {Cecil}, \&
  {Bland-Hawthorn}}]{Veilleux2005}
{Veilleux}, S., {Cecil}, G., \& {Bland-Hawthorn}, J. 2005, \araa, 43, 769,
  \dodoi{10.1146/annurev.astro.43.072103.150610}

\bibitem[{{Veilleux} {et~al.}(2001){Veilleux}, {Shopbell}, \&
  {Miller}}]{Veilleux2001}
{Veilleux}, S., {Shopbell}, P.~L., \& {Miller}, S.~T. 2001, \aj, 121, 198,
  \dodoi{10.1086/318046}

\bibitem[{{Venturi} \& {Marconi}(2021)}]{Ventur&Marconi2021}
{Venturi}, G., \& {Marconi}, A. 2021, in Galaxy Evolution and Feedback across
  Different Environments, ed. T.~{Storchi Bergmann}, W.~{Forman},
  R.~{Overzier}, \& R.~{Riffel}, Vol. 359, 212--220,
  \dodoi{10.1017/S1743921320002203}

\bibitem[{{V{\'e}ron-Cetty} \& {V{\'e}ron}(2010)}]{Veron-Cetty2010}
{V{\'e}ron-Cetty}, M.~P., \& {V{\'e}ron}, P. 2010, \aap, 518, A10,
  \dodoi{10.1051/0004-6361/201014188}

\bibitem[{{Villar-Mart{\'\i}n} {et~al.}(2016){Villar-Mart{\'\i}n}, {Arribas},
  {Emonts}, {Humphrey}, {Tadhunter}, {Bessiere}, {Cabrera Lavers}, \& {Ramos
  Almeida}}]{Villar2016}
{Villar-Mart{\'\i}n}, M., {Arribas}, S., {Emonts}, B., {et~al.} 2016, \mnras,
  460, 130, \dodoi{10.1093/mnras/stw901}

\bibitem[{{Weilbacher} {et~al.}(2020){Weilbacher}, {Palsa}, {Streicher},
  {Bacon}, {Urrutia}, {Wisotzki}, {Conseil}, {Husemann}, {Jarno}, {Kelz},
  {P{\'e}contal-Rousset}, {Richard}, {Roth}, {Selman}, \&
  {Vernet}}]{Weilbacher2020}
{Weilbacher}, P.~M., {Palsa}, R., {Streicher}, O., {et~al.} 2020, \aap, 641,
  A28, \dodoi{10.1051/0004-6361/202037855}

\bibitem[{{Williams} {et~al.}(2023){Williams}, {Tacchella}, {Maseda},
  {Robertson}, {Johnson}, {Willott}, {Eisenstein}, {Willmer}, {Ji}, {Hainline},
  {Helton}, {Alberts}, {Baum}, {Bhatawdekar}, {Boyett}, {Bunker}, {Carniani},
  {Charlot}, {Chevallard}, {Curtis-Lake}, {de Graaff}, {Egami}, {Franx},
  {Kumari}, {Maiolino}, {Nelson}, {Rieke}, {Sandles}, {Shivaei}, {Simmonds},
  {Smit}, {Suess}, {Sun}, {{\"U}bler}, \& {Witstok}}]{Williams2023}
{Williams}, C.~C., {Tacchella}, S., {Maseda}, M.~V., {et~al.} 2023, \apjs, 268,
  64, \dodoi{10.3847/1538-4365/acf130}

\bibitem[{{Williams, Christina} {et~al.}(2023){Williams, Christina},
  {Tacchella, Sandro}, \& {Maseda, Michael}}]{JEMS_doi}
{Williams, Christina}, {Tacchella, Sandro}, \& {Maseda, Michael}. 2023, Data
  from the JWST Extragalactic Medium-band Survey (JEMS),  STScI/MAST,
  \dodoi{10.17909/FSC4-DT61}

\bibitem[{{Wylezalek} {et~al.}(2022){Wylezalek}, {Vayner}, {Rupke}, {Zakamska},
  {Veilleux}, {Ishikawa}, {Bertemes}, {Liu}, {Barrera-Ballesteros}, {Chen},
  {Goulding}, {Greene}, {Hainline}, {Hamann}, {Heckman}, {Johnson}, {Lutz},
  {L{\"u}tzgendorf}, {Mainieri}, {Maiolino}, {Nesvadba}, {Ogle}, \&
  {Sturm}}]{Wylezalek2022}
{Wylezalek}, D., {Vayner}, A., {Rupke}, D. S.~N., {et~al.} 2022, \apjl, 940,
  L7, \dodoi{10.3847/2041-8213/ac98c3}

\bibitem[{{Xue} {et~al.}(2011){Xue}, {Luo}, {Brandt}, {Bauer}, {Lehmer},
  {Broos}, {Schneider}, {Alexander}, {Brusa}, {Comastri}, {Fabian}, {Gilli},
  {Hasinger}, {Hornschemeier}, {Koekemoer}, {Liu}, {Mainieri}, {Paolillo},
  {Rafferty}, {Rosati}, {Shemmer}, {Silverman}, {Smail}, {Tozzi}, \&
  {Vignali}}]{Xue2011}
{Xue}, Y.~Q., {Luo}, B., {Brandt}, W.~N., {et~al.} 2011, \apjs, 195, 10,
  \dodoi{10.1088/0067-0049/195/1/10}

\bibitem[{{Zakamska} \& {Greene}(2014)}]{Zakamska2014}
{Zakamska}, N.~L., \& {Greene}, J.~E. 2014, \mnras, 442, 784,
  \dodoi{10.1093/mnras/stu842}

\bibitem[{{Zanchettin} {et~al.}(2023){Zanchettin}, {Feruglio}, {Massardi},
  {Lapi}, {Bischetti}, {Cantalupo}, {Fiore}, {Bongiorno}, {Malizia},
  {Marinucci}, {Molina}, {Piconcelli}, {Tombesi}, {Travascio}, {Tozzi}, \&
  {Tripodi}}]{Zanchettin2023}
{Zanchettin}, M.~V., {Feruglio}, C., {Massardi}, M., {et~al.} 2023, \aap, 679,
  A88, \dodoi{10.1051/0004-6361/202245729}

\end{thebibliography}
\bibliographystyle{aasjournal}

\end{document}